\documentclass[twocolumn,astrosymb,twocolappendix]{aastex631}

\newcommand{\aicmod}{\mathrm{AIC_{p}}}

\newcommand{\rhodm}{\mathrm{\rho_{DM}}}
\newcommand{\mbh}{\mathrm{M_{\bullet}}}
\newcommand{\rhonorm}{\mathrm{\rho_{0}}}
\newcommand{\rhokpc}{\mathrm{\rho_{1kpc}}}
\newcommand{\slpin}{\mathrm{\gamma_{in}}}
\newcommand{\slpout}{\mathrm{\gamma_{out}}}
\newcommand{\sclrad}{\mathrm{r_{s}}}
\newcommand{\qdm}{\mathrm{q_{DM}}}
\newcommand{\reff}{r_\mathrm{e}}
\newcommand{\reval}{r_\mathrm{ev}}
\newcommand{\MDM}{\mathrm{M_{DM}\left(r<\reval \right)}}
\newcommand{\centredis}{\Delta r_\mathrm{sky,M87}}
\newcommand{\dmfrac}{\mathrm{f_{DM}}}
\newcommand{\paperrefeSTARS}{VW--\MakeUppercase{\romannumeral 1}}
\newcommand{\paperrefeSTARSSP}{VW--\MakeUppercase{\romannumeral 1} } 
\newcommand{\paperrefe}{LT}
\newcommand{\paperrefeSP}{LT } 
\newcommand{\rsoi}{\mathrm{r_{SOI}}}
\newcommand{\MTOT}{\mathrm{M_{tot}\left(<r \right)}}
\newcommand{\dmgrad}{\mathrm{\frac{\partial \rhodm}{\partial r}}}

\shortauthors{Lipka et al.}

\begin{document}

\title{The VIRUS-dE Survey II: Cuspy and round halos in dwarf ellipticals - A result of early assembly?}

\correspondingauthor{Mathias Lipka}
\email{mlipka@mpe.mpg.de}

\author[0000-0002-0730-0351]{Mathias Lipka}
\affiliation{Max-Planck-Institut für extraterrestrische Physik, Giessenbachstrasse, D-85748 Garching}
\affiliation{Universitäts-Sternwarte München, Scheinerstrasse 1, D-81679 München, Germany}

\author[0000-0003-2868-9244]{Jens Thomas}
\affiliation{Max-Planck-Institut für extraterrestrische Physik, Giessenbachstrasse, D-85748 Garching}
\affiliation{Universitäts-Sternwarte München, Scheinerstrasse 1, D-81679 München, Germany}

\author[0000-0003-0378-7032]{Roberto Saglia}
\affiliation{Max-Planck-Institut für extraterrestrische Physik, Giessenbachstrasse, D-85748 Garching}
\affiliation{Universitäts-Sternwarte München, Scheinerstrasse 1, D-81679 München, Germany}

\author[0000-0001-7179-0626]{Ralf Bender}
\affiliation{Max-Planck-Institut für extraterrestrische Physik, Giessenbachstrasse, D-85748 Garching}
\affiliation{Universitäts-Sternwarte München, Scheinerstrasse 1, D-81679 München, Germany}

\author[0000-0002-7025-6058]{Maximilian Fabricius}
\affiliation{Max-Planck-Institut für extraterrestrische Physik, Giessenbachstrasse, D-85748 Garching}
\affiliation{Universitäts-Sternwarte München, Scheinerstrasse 1, D-81679 München, Germany}

\author[0009-0003-9985-0012]{Christian Partmann}
\affiliation{Max-Planck-Institut für Astrophysik, Karl-Schwarzschild-Str. 1, D-85748, Garching, Germany}

\begin{abstract}
We analyze the dark matter (DM) halos of a sample of dwarf Ellitpicals (dE) and discuss cosmological and evolutionary implications. Using orbit modeling we recover their density slopes and, for the first time, the halo flattening. We find the `cusp-core' tension is mild, on average dEs have central slopes slightly below the Navarro–Frenk–White (NFW) predictions. However, the measured flattenings are still more spherical than cosmological simulations predict. Unlike brighter ETGs the total density slopes of dEs are shallower, and their average DM density does not follow their scaling relation with luminosity. Conversely, dE halos are denser and the densities steeper than in LTGs. We find average DM density and slope are strongly correlated with the environment and moderately with the angular momentum. Central, non-rotating dEs have dense and cuspy halos, whereas rotating dEs in Virgo's outskirts are more cored and less dense. This can be explained by a delayed formation of the dEs in the cluster outskirts, or alternatively, by the accumulated baryonic feedback the dEs in the outskirts have experienced during their very different star formation history. Our results suggest halo profiles are not universal (they depend on assembly conditions) and they evolve only mildly due to internal feedback. We conclude dEs in the local Universe have assembled at a higher redshift than local spirals. In these extreme conditions (e.g. star-formation, halo assembly) were very different, suggesting no new dEs are formed at present. 
\end{abstract}


\section{Introduction}
\label{sec:intro}
In the standard cosmological model ($\Lambda$CDM) structures like galaxies have assembled from collapsing dark matter (DM) over-densities as baryons followed them to build the galaxies observed today. In the hierarchical formation scenario the \textit{largest} galaxies are thought to have formed via mergers of \textit{smaller} galaxies, i.e. smaller DM halos, which has dramatically changed their mass and kinematic structure. The smaller dwarf galaxies on the other hand, which avoided merging to this day, should have pristine halos which makes them an ideal probe of the initial DM over-density collapse. 

In this framework, dwarf ellipticals (dEs) that inhabit an \textit{intermediate} mass regime ($\log_{10}(M_{*}/M_{\sun})\approx 7-9$) are often thought to be the largest \textit{fundamental} building blocks within the sequence of the quiescent early-type galaxies (ETGs) that have not formed via merging. However, while they may have avoided merging to this day, they are also much more exposed to their environment due to their much shallower potential wells. Over time \textit{internal} feedback processes like star-formation, and/or \textit{environmental} processes like ram-pressure stripping RMS \citep[][]{Gunn_1972,Lin_1983} and galaxy harassment \citep[][]{Moore_1998}, could have modified the distribution of baryons and as such also that of the DM. Therefore the present-day structure of dE dark matter halos may not only be an excellent probe of the underlying cosmology but also an avenue to investigate the effect cluster environments and/or internal feedback has had on the dark matter as time passed. 

While overall quite successful in explaining the observed clustering of mass on large scales \citep[][]{Croft_2002,Spergel_2003,Springel_2006}, $\Lambda$CDM predictions are much harder to reconcile with observational evidence on the smaller galaxy scales. Firstly, the statistical occurrence and distribution of the dwarf galaxies in the local universe differ from cosmological predictions, which is known as the `missing satellite-' and the `too big to fail'-problem \citep[e.g.][]{Klypin_1999,Boylan_Kolchin_2011,Boylan_Kolchin_2012}. Secondly, observational constraints on the individual DM halo distributions suggest they are closer to spherical than $\Lambda$CDM simulations anticipated \citep[e.g.][]{Allgood_2006,Hayashi_2007,Chua_2019,Bovy_2016,Wegg_2019}. And thirdly, the steepness of the inner DM distribution conflicts with predictions, which is known as the `cusp-core problem' and was first reported by \citet{Moore_1994} \citep[For a review see][]{de_Blok2010,Del_Popolo_2021}. Simulations of halo formation suggest \textit{cuspy} central density profiles for the halos of dwarf galaxies. Examples are the Navarro–Frenk–White profile (NFW), which has a central logarithmic slope of $-1$, or even cuspier halos with slopes of $\sim -1.5$ \citep[e.g.][]{Diemand_2004,Diemand_2011,Moore_1998,Moore_2001,Klypin_2001,Klypin_2011,Navarro_2010}. This stands in contrast to the majority of observational findings, which often find cored density distributions. For example, the rotation curve modeling of HI disks suggests that at least the smaller dwarf galaxies have a strong preference towards \textit{cored} halos \citep[e.g.][]{de_Blok_2002,de_Blok_2008,Donato_2009,Oh_2011_a,Plana_2010}.  

If these discrepancies between $\Lambda$CDM predictions and observations are quantified accurately, then one may be able to identify the reason behind it: be it an exotic nature of dark matter particles \citep[][]{Spergel_2000,Marsh_2014,Elbert_2015} or the feedback of baryonic physics on the dark matter \citep[][]{de_Souza_2011,Gnedin_2002,Governato_2010,Madau_2014,Navarro_1996,Oh_2011_b}. However, the majority of observational evidence for cored halos stems from dynamical modeling using gas as tracer of the gravitational potential and is thus mostly restricted to late-type galaxies. Observational constraints on the degree of sphericity of halos are even more scarce, and the majority of evidence comes from Milky Way studies.

For \textit{early-type} galaxies, i.e. for galaxies without significant amounts of gas, stellar-based dynamical models can be employed to infer the structure of their DM halos. However, the existing literature is mostly restricted to the very small but near-by dwarf spheroidals (dSphs) within the Local Group or to very massive ETGs ($\log_{10}(M_{*}/M_{\sun})\gtrsim10$) that inhabit the more distant massive galaxy clusters (Coma, Virgo, Fornax ...). In contrast, the DM structure of dEs (i.e. the intermediate mass regime) is scarcely probed even though they are by far the most common type of galaxy found in the nearby clusters. This deficiency of observational constraints in the dE regime is mostly because the Local Group only has a few dEs, while the dEs in the nearby clusters are faint and require a very high spectral resolution to be analyzed using stellar dynamical models. 

This paper is part of a series aimed at studying the mass distribution, stellar populations and dynamical composition by analyzing a sample of 9 such dEs in the Virgo cluster. The first paper, Lipka et al. (2024), in the following \paperrefeSTARS, is a comprehensive analysis of the \textit{stellar} structure of the dEs. There we also discuss basic properties of the dE sample, all the data sets we obtained, and describe the \textit{dynamical} and {population} modeling techniques we employed to infer the intrinsic 3D structure. The current paper is focused on the dark matter structure and its interpretation in the broader cosmological and galaxy evolutionary context. It is aimed to fill the gap in our understanding of dark matter in the intermediate mass regime of early-type galaxies. 

The current paper is organized as follows: In Section~\ref{sec:sample} we briefly describe the dE sample we obtained and recapitulate some of the main findings of \paperrefeSTARS. In Section~\ref{sec:technique} we explain how we specifically modified our dynamical modeling technique to optimally recover the 3D density distributions of the dark matter halos. In preparation for this study we stress-tested this modeling approach by applying it to an N-body simulation, the results of which are discussed in App.~\ref{append:simulation}. Section~\ref{sec:dwarf_modelling} shows the dynamical constraints and recovered dark mass distributions of the dEs sample. Using these modeling results we then examine whether the slopes and flattening of the halos of dEs are in tension with $\Lambda$CDM predictions or not (Section~\ref{sec:CDM-Discussion}). Under the umbrella of the $\Lambda$CDM paradigm we then discuss what our results imply regarding the formation and evolution of dEs and how they are related to other galaxy types (Section~\ref{sec:formation_evolution}). The paper concludes with a summary in Section~\ref{sec:conclusions}.

\section{The VIRUS-W dwarf sample}
\label{sec:sample}
The sample we analyze in this paper consists of 9 dEs with stellar masses $\log_{10}(M_{*})\in [8.5,9.5]$ which inhabit different environments within the Virgo cluster, ranging from its center to just beyond its virial radius. The basic properties (like distance) of the galaxies we adopted for the dynamical modeling can be found in Table 2 of \paperrefeSTARS. To distinguish the individual galaxies in our sample, we keep the same color-coding we used in \paperrefeSTARS. The VCC-Catalog ID of each galaxy and its corresponding color can be inferred from Fig.~\ref{fig:halo_sampling} or Tab.~\ref{tab:data_table} which shows some of the most important quantities we measured in this paper.

\definecolor{royalblue}{HTML}{4169E1}
\definecolor{darkorange}{HTML}{FF8C00}
\definecolor{fuchsia}{HTML}{FF00FF}
\definecolor{lime}{HTML}{00FF00}
\definecolor{black}{HTML}{000000}
\definecolor{gold}{HTML}{FFD700}
\definecolor{red}{HTML}{FF0000}
\definecolor{slategrey}{HTML}{708090}
\definecolor{aqua}{HTML}{00FFFF}

\begin{table*}
	\centering
	\caption{Table with some of the important quantities we measured. From left to right: The average DM density within $1\reff$ in $M_{\sun}/\rm kpc^3$, the mass-weighted slope (eq.~\ref{eq:MW_slope}) of the total density, the volume averaged slope $\eta_{DM}$  within $0.8 \rm kpc$ (eq.~\ref{eq:mean_slope}), the axis ratio of the halo, the (mean) axis ratio of stars, the dark matter fraction within $1\reff$, the SSP age in Gyrs, metallicity, [Mg/Fe] ratio at $r=2.5\arcsec$, stellar angular momentum in $[\mathrm{kpc} \cdot \mathrm{km/s}]$, the angular momentum parameter. Total stellar mass and distance to M87 are tabulated in \paperrefeSTARS.}
	\label{tab:data_table}
	\begin{tabular}{ccccccccccccc} 
     VCC ID   & Color& $\log_{10}(\overline{\rho_{\rm DM}})$ & $\gamma_{\mathrm{MW}}$ & $\eta_{\rm DM}$ & $q_{*}$ & $q_{\rm DM}$ & $f_{\rm DM}$ & Age [Gyr] & $[Z/\rm H]$ & [Mg/Fe] & $\log_{10}(j_{*})$ &$\lambda_{e/2}$\\ \hline
     VCC 200  & \cellcolor{royalblue}  & 7.324 & -1.944 & -1.113 & 0.895 & 0.9 & 0.33 &  11.2 & -0.63 & 0.28 & 0.957 & 0.208 \\
     VCC 308  & \cellcolor{darkorange} & 7.295 & -1.540 & -0.521 & 0.826 & 1.0 & 0.24 &  2.8  & -0.24 & 0.08 & 1.014 & 0.222 \\
     VCC 543  & \cellcolor{fuchsia}    & 7.138 & -1.791 & -0.715 & 0.612 & 0.9 & 0.17 &  6.3  & -0.34 & 0.21 & 1.488 & 0.375 \\ 
     VCC 856  & \cellcolor{lime}       & 7.724 & -1.282 & -0.713 & 0.609 & 1.0 & 0.56 &  7.8  & -0.49 & 0.23 & 1.719 & 0.330 \\
     VCC 1261 & \cellcolor{black}      & 7.756 & -1.660 & -1.426 & 0.713 & 1.0 & 0.53 &  6.6  & -0.30 & 0.16 & 1.085 & 0.046 \\
     VCC 1528 & \cellcolor{gold}       & 8.048 & -1.815 & -1.414 & 0.839 & 0.7 & 0.35 &  6.7  & -0.28 & 0.19 & 0.307 & 0.028 \\
     VCC 1861 & \cellcolor{red}        & 7.616 & -1.388 & -0.719 & 0.897 & 0.9 & 0.54 &  10.0 & -0.24 & 0.18 & 1.176 & 0.109 \\
     VCC 1910 & \cellcolor{slategrey}  & 7.591 & -2.067 & -1.111 & 0.865 & 1.0 & 0.25 &  2.0  &  0.35 & 0.19 & 1.020 & 0.126 \\
     VCC 2048 & \cellcolor{aqua}       & 7.199 & -1.805 & -0.421 & 0.423 & 0.5 & 0.07 &  3.5  & -0.20 & 0.17 & 1.366 & 0.256 \\
     \hline
	\end{tabular}
\end{table*}

The main data set we obtained with the integral-field-unit (IFU) spectrograph VIRUS-W \citep[][]{Fabricius_2008,Fabricius_2012} at the Harlan J. Smith Telescope (McDonald Observatory). VIRUS-W's very high spectral resolution ($R=7900$ to $9000$) is essential to measure the low velocity dispersions of the stars in dEs, because spectrographs with a lower resolution tend to overestimate $\sigma$ by a significant margin (cf. \paperrefeSTARS) which would inevitably bias any dynamical mass reconstruction. As one of the first IFU studies of dEs the VIRUS-W data allow us to access their full 2D spatially-resolved on-sky kinematic which vastly improves the constraints of the dynamical modeling required to infer the 3D density distribution of the dark matter halos analyzed here.   

To obtain the line-of-sight-velocity-distributions (LOSVDs) of the stars we employed the spectral-fitting code WINGFIT (Thomas et al. in prep.) which allows us to retrieve the full extent of the information contained in the spectra well beyond just the mean velocity and dispersion. Higher moment information is crucial to break the mass-anisotropy degeneracy \citep[e.g.][]{Merrifield_1990,Marel_1993} and conversely enable a robust recovery of the dark matter distribution. To ensure a robust recovery of the LOSVDs we binned the spectra with the Voronoi tesselation method \citep[][]{Cappellari_2003} and excluded any bins that did not fulfil our $S/N$ requirement. The remaining data covers the kinematic out to approximately 1 effective radius. 

For a detailed description of the data reduction/preparation and the final resulting LOSVDs we refer to \paperrefeSTARS, in it we also show our results from stellar population modeling and the \textit{stellar} density reconstruction we obtained from our dynamical modeling implementation (which is also used here). In short we find: stellar populations are spatially homogeneous, but display a larger variety in age with some having stopped forming stars only recently while others did so 12 Gyrs ago. Dynamically, the dEs have a more isotropic orbit structure and a suppressed angular momentum compared to other galaxy types \citep[see also][]{Scott_2020}. We find that their mass-to-light ratios are anti-correlated with their single stellar population (SSP) age and we identify two possible explanations: i) In the Virgo cluster dEs have been formed continuously starting 12 Gyrs ago until now and their initial mass function (IMF) changed with their formation epoch, ii) or the bulk of Virgo dEs has formed early on in the same epoch, but (subject to internal/external influences) have experienced varying degrees of extended star formation history (SFH) with some dE being quenched shortly after gravitational collapse while others were able to sustain several Gyrs of continuous or bursty star formation until they were eventually quenched. These processes could have left imprints in the distribution of the dark matter, which we will investigate in this work. 

\section{Recovering dark matter with dynamical models}
\label{sec:technique}
The modeling code we employ is a state-of-the-art axisymmetric implementation of the Schwarzschild orbit superposition technique \citep[][]{Schwarzschild_1979,Thomas_2004}. In short, the modeling principle is simple: For a given galaxy a number of \textit{candidate} mass models is established, and then a set of representative orbits in each of the corresponding potentials is integrated. Each of the orbits is given a weight, and the weighted superposition determines the stellar phase-space density of the orbit model. Given a set of observed data, e.g. LOSVDs, an optimal set of orbit weights can be determined by fitting a candidate model's LOSVDs to the data\footnote{This involves a regularization to avoid overfitting since the number of weights is typically larger than the number of data constraints. In \citet{Lipka_2021} and \citet{Thomas_2022} we introduced a novel data-driven approach that allows the determination of the optimum amount of regularization to avoid both over- and under-fitting.}. To find the candidate mass model that best represents a given galaxy, the fit of each candidate model to the data is then compared using an evaluation statistic. A popular choice for this statistic is $\chi^2$, however, in \citet{Lipka_2021} we demonstrated that $\chi^2$ is biased due to the varying fit flexibility of different candidate models. We developed a new model selection approach \citep{Thomas_2022} that takes this flexibility into account in its evaluation. The corresponding evaluation statistic is an extension of the Akaike information \citep[][]{Akaike_73,Akaike_74} criterion and called $\aicmod$. Analogous to a $\chi^2$ approach, the mass model with the minimum $\aicmod$ is deemed to be the best representation of the galaxy under investigation. Beyond the $\aicmod$ which improves modeling constraints in general, we made some adjustments in the setup of the candidate models and the analysis of results with the specific goal to ensure an unbiased and accurate recovery of the dark matter. In the following section we discuss those modifications. A stress-test of this entire modeling procedure applied to an N-body simulation that is placed under similar conditions as the VIRUS-W dE observations can be found in App.~\ref{append:simulation}.

\subsection{A flexible halo mass model}
\label{subsec:halo_model}
The choice which candidate mass models are being probed in the first place lies in the modellers choice, but it is crucial as it may distort the dynamical constraints. We plan to investigate this on a general methodical level in Lipka et al. (in prep.). In the following this paper will be referenced as \paperrefe, but it is not part of the VIRUS-W dE survey as it is not concerned with dEs in particular. 

For early-type galaxies where gas contribution is negligible, it is common to describe the models with a 3-component density distribution: 
\begin{equation}
\rho(r)=\Upsilon_{*} \cdot \nu+\rhodm+\mbh\cdot \delta(r)
	\label{eq:Total_mass_model}
\end{equation}
Here the stellar component is determined by the stellar mass-to-light ratio $\Upsilon_{*}$ and the 3D luminosity distribution $\nu$. The latter is obtained from the deprojection of the observed photometry (see \paperrefeSTARS). The mass $\mbh$ of the supermassive central black hole (SMBH) and the dark matter halo density $\rhodm$ form the non-visible components of the system.

We also followed this 3 component approach for the modeling of the dE sample. However, for our dE sample an extension of e.g. the $\mbh-\sigma$ relation \citep[][]{Ferrarese_2000,Gebhardt_2000,Hu_2008} suggest black hole masses with a sphere of influence that are well below the spatial resolution of our data which would imply the black holes are undetectable. Nevertheless, in the very-low mass regime of spheroidal galaxies some observations \citep{Bustamante-Rosell_2021} and simulations \citep{weller2023_arxiv} suggest the possibility of a presence of `over-massive' black holes that significantly exceed typical relations. Stripping of baryonic mass could move galaxies above the $\mbh-M_{*}$ relation \citep{pacucci_2023_arxiv}. The degree to which such effects play a role in the more massive dEs is yet to be determined as little is known about black holes in this galaxy regime. Therefore we still decided to equip our candidate models with a variable black hole mass, even though we expect to only find an upper limit. 

The stellar densities we probed allow for different (axisymmetric) \textit{flattenings} and a \textit{radially variable} stellar mass-to-light ratio $\Upsilon_{*}$. We achieve this by sampling different viewing angles in the deprojection, and probing models with different inner and outer mass-to-light ratios ($\Upsilon_{i}$ at radius $r=r_{i}$ and $\Upsilon_{o}$ at radius $r=r_{o}$). Within one effective radius we find a sample mean of the stellar mass-to-light ratio $\Upsilon_{*}$ of $\sim1.3\pm0.4$ ($z$-band), a value fairly consistent with color-based stellar mass-to-light estimates for dEs \citep[e.g.][]{Eftekhari_2022}. For details on the implementation of the stellar mass-to-light ratio gradients and a detailed analysis of the stellar mass-to-light results see \paperrefeSTARS.

The halo density $\rhodm$ is most susceptible to the modeler's choice as it is usually obtained by probing a parametric description motivated by simulation results such as for example a Navarro-Frenk-White profile \citep[][]{Navarro_1996_B,Navarro_1997}. In \paperrefeSP we will argue that it is important to adopt a description that is highly flexible. Halo parametrizations that are too restrictive in their profiles can bias the results towards specific configuration inherited from said halo model. Only a flexible halo allows one to probe a variety of different halo densities and to accurately gauge the actual constraining strength of the dynamical models in an unbiased manner. Therefore, for the modeling of the dEs our fiducial halo model is:
\begin{equation}
\rhodm(m,\theta)=\frac{\rhonorm}{\left(\frac{m}{\sclrad}\right)^{\slpin} \cdot \left(1+\frac{m}{\sclrad}\right)^{\slpout-\slpin}}
	\label{eq:dark_matter_model}
\end{equation}
where $m$ and $\theta$ are elliptical coordinates. $\slpin$ and $\slpout$ are the inner and outer logarithmic slopes separated by the scale radius $\sclrad$. The halo normalization $\rhonorm$ is sampled by probing different values of $\rhokpc$ which is the density of the halo at $m=1.0 \rm kpc$. They are related to each other by: 
\begin{equation}
\rhonorm=\rhokpc \left(\frac{1\rm kpc}{\sclrad}\right)^{\slpin} \cdot \left(1+\frac{1 \mathrm{kpc}}{\sclrad}\right)^{\slpout-\slpin}
	\label{eq:density_norm}
\end{equation}
Equation~\ref{eq:dark_matter_model} is essentially a Zhao-profile \citep[][]{Zhao_1996} in elliptical coordinates where the parameter that describes the transition width is fixed to unity. To probe the \textit{flattening} of the halo component (independently of the stellar flattening) we additionally endowed the above halo model with a (globally constant) spheroidal flattening parameter $\qdm$.

\subsection{Nuisance parameters and sampling strategy}
\label{subsec:sampling_choice}
A naive interpretation of the parameters that establish \textit{parametric} model descriptions can be very deceiving and biased as they are often correlated with each other which makes a comparison of models with different sets of parameters non-trivial (see \paperrefeSP for a comprehensive discussion). For example, for our choice of the halo model the nominal values of the \textit{asymptotic} slopes $\slpin$ and $\slpout$ describe the density gradients at $r \rightarrow 0$ and $r \rightarrow \infty$. But these are not actually the radii responsible for constraining the values of these two parameters. This is because they are \textit{global} parameters, and it is the combination of $\slpin$, $\slpout$ and $\sclrad$ that fully determine the slope at \textit{any} radius of the halo model. Therefore the constraints on the density slopes at every radius indirectly constrain the value of said model parameters. Instead of interpreting the halo parameters as physically meaningful parameters we treat them as \textit{nuisance parameters} that simply serve us in setting up different trial mass distributions. Consequently, instead of interpreting and evaluating the nuisance parameters we should focus on the evaluation of the actual mass distributions they generate.

With this is mind we can also optimize the parameter sampling. The parameters in our mass model that have the strongest inter-correlations are those describing the DM halo. Inter-correlations imply that one may probe nominally very different halo parameters without actually changing the mass distribution in a meaningful way, essentially probing the same dynamical model multiple times. Therefore to keep the available grid of candidate models efficiently small and avoid repeated sampling of essentially identical mass distributions we set up the halo parameter grid as follows: i) $\qdm$ is sampled from spherical $\qdm=1.0$ down to $0.6$ and $0.7$, thus covering the range of intrinsic flattenings observed in the respective stellar systems. Only for VCC 2048 we extended the range down to $0.3$ as the $\aicmod$ showed improvements down to $\qdm=0.5$. ii) The scale radius $\sclrad$ is probed between $0.1 \rm kpc$ and $5.0 \rm kpc$, i.e. starting from a fraction of the spatial resolution out to radii several times the size of the FoV. iii) The asymptotic inner slope $\slpin$ is varied between 0.0 and 1.5, whereas $\slpout$ is sampled for a broader range from -1.0 to 3.0. This broadens the space of possible models, allowing even unrealistic halo distribution with a positive radial density gradient (e.g. if $\sclrad \to 0$ and $\slpout=-1$) or a sign reversal in the mass gradient. The goal of this sampling choice is to allow as many different mass distributions as possible, retaining generality, all the while maximizing the differences between individual candidate models as much as possible to avoid redundant sampling. 

\begin{figure*}
	\centering
	\includegraphics[width=1.0\textwidth]{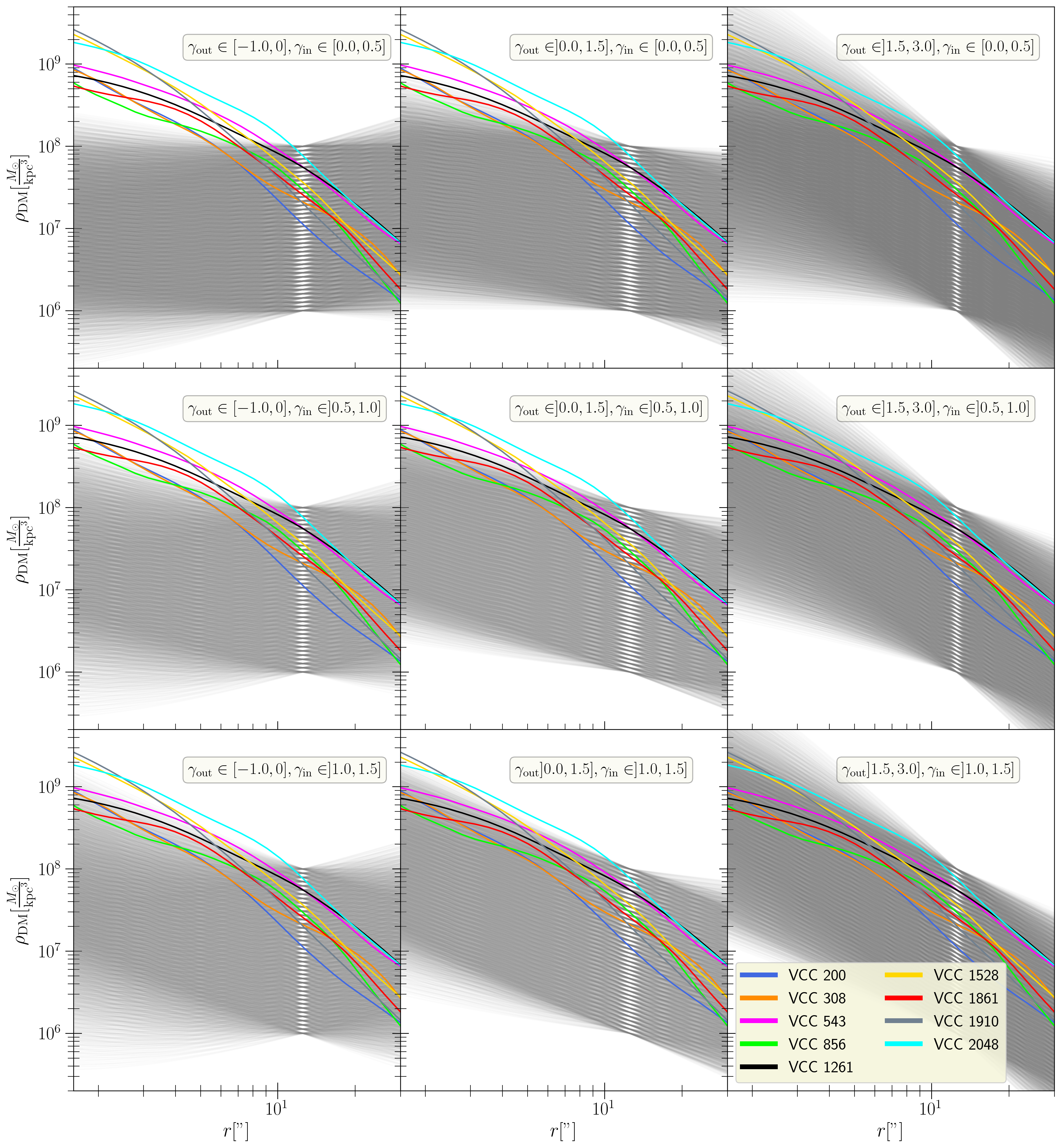}
    \caption{Representative illustration of all the candidate dark matter profiles (\textit{gray}) we probed for each of the dEs in the sample. The color-coding for each dE we us in this study is indicated in the bottom right panel. The candidate models are obtained by sampling the Zhao-like mass description (eq.~\ref{eq:dark_matter_model}) on the parameter grid discussed in Sec.~\ref{subsec:sampling_choice}. In detail, the exact value of the sampling steps differed slightly from galaxy to galaxy, which is not shown here. To show the diversity in allowed profiles, we separated the candidate models into the different panels shown here according to their respective $\slpin$ and $\slpout$. For simplicity we only plotted the \textit{spherical} models ($\qdm=1.0$). For comparison, we also overlay the \textit{stellar} density distribution that we found for each of the dEs in our sample (\textit{colored} lines).} 
    \label{fig:halo_sampling}
\end{figure*}

An illustration of the variety of halo models that are probed by this sampling choice is shown in Fig.~\ref{fig:halo_sampling}. The profiles and shapes range from galaxies with halos much more massive than their baryonic component to galaxies with virtually no Dark matter contribution. Including the flattening $\qdm$, the space of probed Zhao halos encompasses $10^{5}-10^{6}$ different halos, densely covering the entire physically plausible density space $\rhodm$. 

The entire space of candidate models (eq.~\ref{eq:Total_mass_model}) is even larger and spans a 9D grid where each parameter ($\Upsilon_{i}$, $\Upsilon_{o}$, $i$ , $\mbh$ , $\rhonorm$, $\sclrad$, $\slpin$, $\slpout$, $\qdm$) is probed with 5–20 values. A full grid search is obviously unfeasible, which is why we search the grid by employing the Nonlinear Optimization by Mesh Adaptive Direct search NOMAD \citep{Audet_2006,Digabel_2022}. We estimate the errors of all dynamically derived galaxy properties by evaluating the scatter between the best 25 $\aicmod$ models, which is roughly equivalent to a $\Delta \aicmod \lesssim 10$ criterion\footnote{In statistical modeling the models with a $\Delta \mathrm{AIC}>10$ are considered extremely unlikely \citep[][]{burnham_2002}.}. For details regarding this choice of error estimation see \paperrefeSTARS.  

\subsection{Where are dark matter halo constrained the best?}
\label{subsec:recovery_evaluation}
We expect the constraints on the mass models to be most robust in the regions where the data coverage is dense, conversely the mass recovery is uncertain and possibly biased at the smallest scales where we lack spatial resolution and at the largest scales where no data is available \citep[cf.][]{Gerhard_1998,Thomas_2005}. This is also observed in the simulated stress-test in App.~\ref{append:simulation}, which is why we decided for the more cautious approach to trust and analyze the mass recovery only in the regions where data coverage is good. This implies for the dEs in our sample we should trust the mass recovery mostly in between $\sim 3\arcsec$ and $\sim15\arcsec$ (see data coverage in \paperrefeSTARS), which at the distance of the Virgo cluster corresponds to $\sim 0.25 \rm kpc$ and $\sim1.2 \rm kpc$.

This relation of constraining power and data coverage applies to the distribution of total mass in general. However, the goal of this paper is to determine the distribution of the DM halos which requires a \textit{decomposition} of the individual mass components that make up the total mass (eq.~\ref{eq:Total_mass_model}). For a mass component to be dynamically detectable, this requires a significant contribution of the individual mass component to the \textit{total} enclosed mass (see \paperrefe). For the DM component this implies the constraining power scales with the enclosed dark matter fraction $\dmfrac$. If the galaxy has a high dark matter fraction it means the halo may well be constrained by the data because the contribution to the total dynamical mass is an essential requirement for the models to emulate the dynamics of the whole system. In contrast, if the dark matter fraction is negligible (e.g. in the center) the \textit{relative} uncertainty of the DM profile becomes very large. For example, one could easily double the DM density without changing the total mass and gravitational potential of the model in a dynamically detectable manner. In other words: the exact shape and profile of the halo gets very uncertain within the regions where the models suggest $\dmfrac \approx 0$. 

Unfortunately for the majority of galaxies the very central parts are expected to be dominated by the baryonic/luminous matter (and/or $\mbh$) such that $\dmfrac$ is a monotonically rising function eventually dominating the total mass at larger radii. If that is indeed the case for our dEs, a dynamical measurement of their \textit{central} dark matter density and their slope becomes extremely ambiguous. Instead it is preferable to focus the analysis on the DM properties near the edge of the FoV where $\dmfrac$ is significant and the total mass is still well constrained by the data coverage (see also \paperrefe). As discussed in \paperrefeSTARS, the contribution of the blue central nuclei to the central VIRUS-W bins is dynamically negligible, such that it is not worth treating them as a separate dynamical component and trying to recover their properties.

\section{Modeling results}
\label{sec:dwarf_modelling}
In the following we present the halo density structure that we were able to find with the dynamical modeling setup as described above. We continue with the discussion and interpretation of these dark matter results in the subsequent sections. We discussed the \textit{stellar} and \textit{kinematic} structure in a detailed manner in \paperrefeSTARS. There we also showed how well the dynamical models are able to reproduce the observations: the spatially resolved features in the mean velocity, dispersion but also the higher-order Gauss--Hermite moments of the dEs are well reproduced by the best axisymmetric model found. The 3D deprojected luminosity density is a boundary constraint in the orbit modelling technique we use \citep[cf.][]{Thomas_2004}, i.e. it must be reproduced by each orbit model we probe. This ensures that the 2D surface brightness is reproduced as well. 

\subsection{Constraints on the (nuisance) parameters}
\label{subsec:nuisance}
\begin{figure*}
	\centering
	\includegraphics[width=1.0\textwidth]{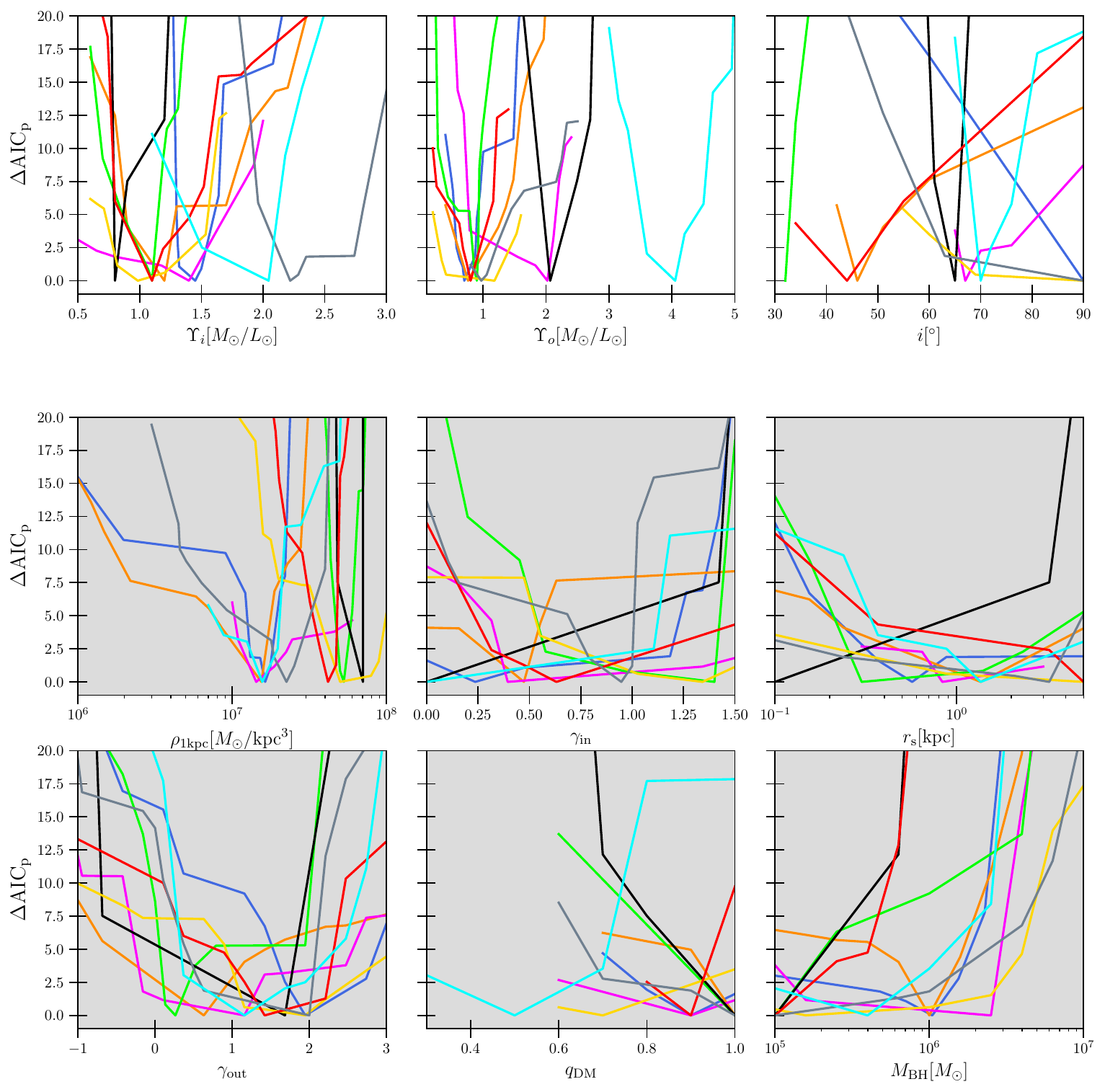}
    \caption{The $\Delta \aicmod$ envelopes of the orbit models probed for each dE. The upper 3 panels that are concerned with the luminous model components were discussed in \paperrefeSTARS. The gray-colored panels show the parameters that describe the `dark' components of the orbits models, i.e. the black hole and dark matter. As argued in Sec.~\ref{subsec:sampling_choice} the parameters that generate the halo distribution should be treated as nuisance parameters, specifically $\slpin$, $\slpout$, and $\sclrad$ are very inter-correlated when the FoV is limited (as is always the case).} 
    \label{fig:dE_AIC_curves}
\end{figure*}

Fig.~\ref{fig:dE_AIC_curves} shows the $\aicmod$-constraints we obtained from all the orbit models that were probed on the 9D-parameter mass model grid. The stellar mass-to-light ratios, the dark matter normalization $\rhokpc$, and the inclination appear to be the most strongly constrained as indicated by sharp lower and upper boundaries in $\aicmod$. These parameters exhibit the sharpest constraints, because they directly dictate the \textit{scale} of the global mass distribution. The asymptotic slopes and the scale radius of the halo profile show much more diversity and scatter, with some $\aicmod$ constraints even reaching the edges of the explored parameter space. 
These parameters ($\sclrad$, $\slpin$, $\slpout$) are strongly inter-correlated and change the radial behavior of the mass distribution within the dynamically relevant range more indirectly. 

In \paperrefeSP we demonstrate that, because the FoV with the kinematic constraints is always limited in radius, the values of the halo parameters themselves should not be interpreted physically. Instead these parameters should be viewed as nuisance parameters only, that allow us to construct flexible, yet smooth, halo mass distributions (cf. Fig.~\ref{fig:halo_sampling}). Neither the larger scatter in $\aicmod$ in these parameters nor the fact that some of these parameters can reach grid sampling limits is necessarily an issue. The halo of VCC 1261, for example, reaches the grid sampling limits at $\slpin=0.0$ and $\sclrad=0.1 \rm kpc$. However, an even lower $\sclrad$ or $\slpin$ would have essentially no impact on the total mass distribution at the scales where we can probe it with the models, because that would change the mass and slope predominantly at radii that are much smaller than the resolution of the VIRUS-W data ($\sim$$0.13 \rm kpc$). In that case the inner slope $\slpin=0.0$ becomes obsolete as $\sclrad$ shrinks and instead the outer $\slpout \sim 1.7$ (which is well constrained in $\aicmod$) is the important parameter that determines the actual slope behaviour of the halo mass distribution. An approximately identical profile could be generated by setting the $\slpin$$\sim$$1.7$ and moving the scale radius far outside, reaching the \textit{upper} limits $\sclrad$, and consequently rendering $\slpout$ irrelevant. Likewise, it is not very concerning that the outer asymptotic slopes $\slpout$ appear to be implausibly shallow: the best models we found have $\slpout$ $\in [0.2,2.0]$, and that would mean the models have an infinite total mass when integrated out to $r \rightarrow \infty$. However, due to the degradation of the constraining power outside the FoV, mass models that differ only far outside the FoV are virtually indistinguishable for dynamical models. The $\slpout$ constraints we measure instead describe, at best, the slope in the vicinity of the FoV edge, and even then only if the $\sclrad$ is small enough for $\slpout$ to be relevant. We will investigate the variability of the constraining power with radius on a methodological level in \paperrefe. To some degree this behaviour has been noted in the past several times, as it manifested as an overestimation of the total mass and uncertainty of the kinematic structure at large radii outside the FoV \citep[e.g.][]{Gerhard_1998,Thomas_2005}. 

\subsection{Halo densities and dark matter fraction}
\label{subsec:mass_results}
\begin{figure*}
	\centering
	\includegraphics[width=1.0\textwidth]{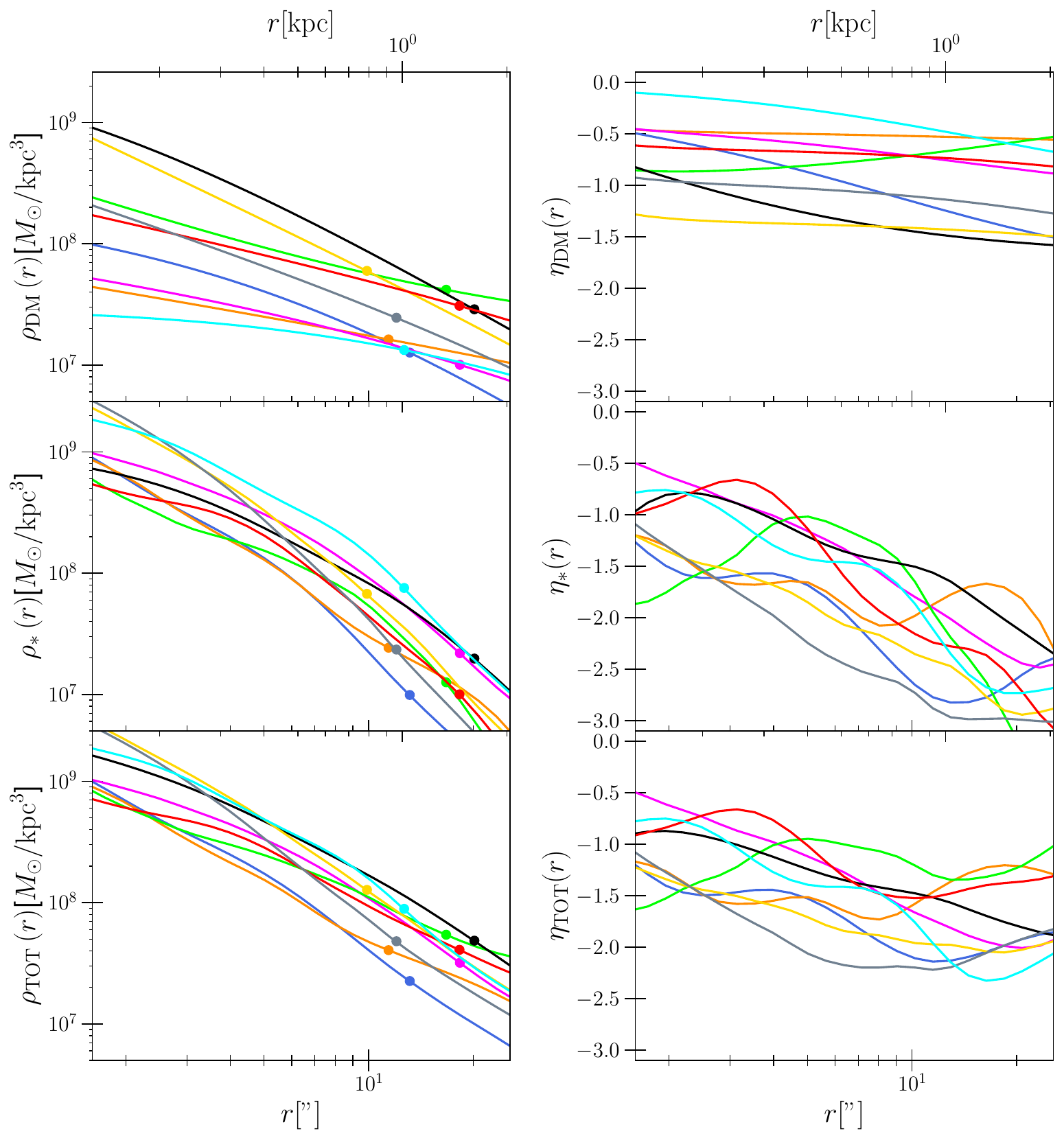}
    \caption{\textit{Left side:} The (spherically averaged) matter densities $\rho$ of each dE obtained from the best dynamical model we found. From top to bottom: dark matter density, stellar density (including the $\Upsilon_{*}(r)$), total mass density. We indicate the location of the (stellar) effective radius of each dE (cf. \paperrefeSTARS) by the point that overlap the corresponding curve. \textit{Right side:} The corresponding (volume) averaged slopes $\eta$ within spheres of radius r (see eq.~\ref{eq:mean_slope}) as a function of r. While the halo slopes stem from a parametric Zhao-description (i.e. the slopes are smooth by definition) the stellar component is non-parametric and is completely determined by the photometric deprojection, the inclination and the $\Upsilon_{*}$-gradient of the best fit model. At a radius of $10\arcsec$, i.e. within the region where we expect the DM to be constrained best by the data (Sec.~\ref{subsec:recovery_evaluation}) the typical statistical $1\sigma$-error of $\eta_{\mathrm{DM}}$ is $\sigma_{\eta}=0.26$. The individual errors resolved for each dE can be inferred from Fig.~\ref{fig:correlation_DM} and Fig.~\ref{fig:correlation_DM_POPULATION}.} 
    \label{fig:dE_avg_slopes}
\end{figure*}

\begin{figure*}
	\centering
	\includegraphics[width=1.0\textwidth]{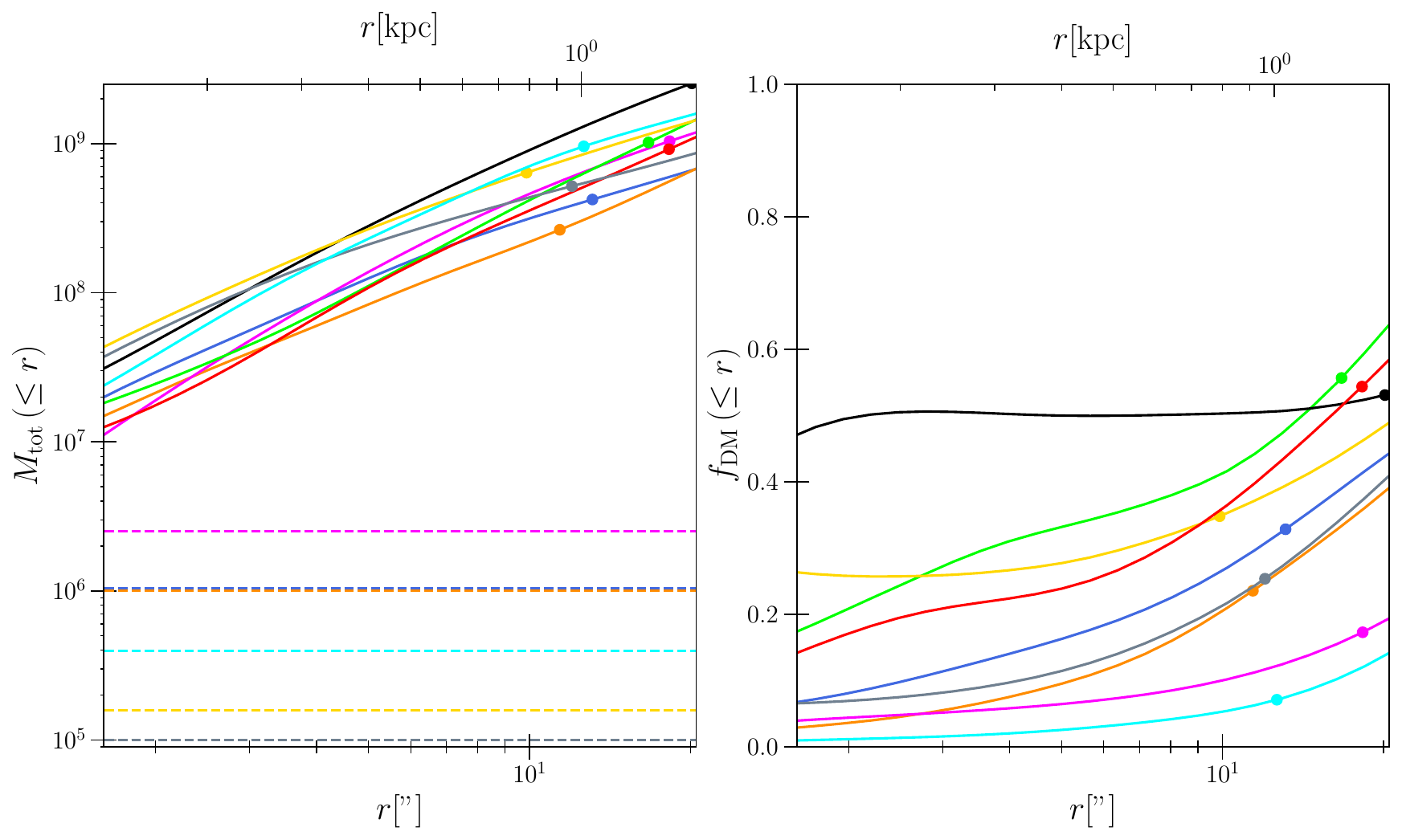}
    \caption{\textit{Left panel:} The total enclosed mass of each dE vs radius (solid lines). \textit{Dashed Lines:} The corresponding black hole masses of the best dynamical models. The spatial resolution of the VIRUS-W data is mostly limited by the fibre size and the seeing and is effectively of the order of $\sim $$2\arcsec$, hence, we do not expect to resolve central black holes. While our models (formally) recovered black hole masses (Fig.~\ref{fig:dE_AIC_curves}) as shown here, their contribution to the total enclosed mass within the spatial resolution limit is negligible. Therefore we consider the best models to be essentially degenerate with any models with black hole masses $\mbh\lesssim5\cdot10^6 M_{\sun}$. \textit{Right panel:} The cumulated dark matter fraction within radius r. With some exceptions, the potential within the central few arcseconds of the dEs is dominated by the stellar mass.} 
    \label{fig:dE_enclosedmass}
\end{figure*}

Instead of interpreting these nuisance parameters we now focus on directly evaluating the mass distributions they generate. The \textit{left} panels of Fig.~\ref{fig:dE_avg_slopes} show the (spherically averaged) mass densities of the dark matter halo, the stars, and their combined density of the \textit{best} dynamical models we found for each dE. In other words the Figure shows the actual mass distribution of the model with $\Delta \aicmod=0$ in the nuisance parameter grid (Fig.~\ref{fig:dE_AIC_curves}). The stellar density shown here incorporates the spatially variable stellar mass-to-light ratios $\Upsilon_{*}(r)$ we equipped the models with, meaning it's not merely a deprojection of the photometry at some viewing angle. Fig.~\ref{fig:dE_enclosedmass} displays the \textit{total} enclosed mass $M_{\mathrm{tot}}=M_{\mathrm{DM}}(<r)+M_{*}(<r)$ vs radius in the left panel and in the right panel we quantify the relative contribution of the dark matter by displaying the cumulated dark matter fraction $\dmfrac=\frac{M_{\mathrm{DM}}(<r)}{M_{\mathrm{tot}}}$ as a function of radius. As a visual guide we also indicate the location of the stellar effective radius $\reff$ in the mass profiles.  

Despite the huge variety in candidate halo profiles we probed (see Fig.~\ref{fig:halo_sampling}) the recovered density profiles $\rho_{\mathrm{DM}}$ show that the DM halos of the dEs behave qualitatively similar to each other, but, at the same time, are distinct from their baryonic counterparts. Dark matter is much less centrally concentrated and contributes only little to the total mass within the center, but its density profiles fall off less steeply. As such, the dark matter contribution to the total mass budget becomes increasingly more important with radius, as the dark matter fraction is surging outside radii greater than about $1\reff$. 

In and of itself this is of course nothing new as numerous studies in the past decades have found this to be the case \citep[starting from][]{Rubin_1970}. However, our results extend on the majority of previous studies in that we relax the assumption that the baryonic matter component follows the light distribution. Instead, we allow for a spatially variable \textit{stellar} mass-to-light ratios (similar as \citealt{Mehrgan_2023_b} did for massive ETGs). This has the advantage that `missing' mass, i.e. any difference between light and total dynamical mass is not necessarily counted as DM but that we can also account for changes in the stellar populations. 
The fact that we still see the self-similarity between the recovered halos and a systematic difference to the baryonic mass distributions is encouraging and increases the significance of the DM detection. If the stellar matter and dark matter components were completely \textit{degenerate}, or if the missing mass was caused solely by a complex variability of the underlying stellar population, then we would expect a larger variability in the recovered halo masses, given the huge variety of candidate models that we probed (Fig.~\ref{fig:halo_sampling}). For example, halos that increase in density with radius or halos that are congruent with the baryons (mass-follows-light) appear to be strongly ruled out by the dynamical modeling. All in all, the nuisance parameters that set up the halo may be fairly noisy (which is amplified by the large flexibility we allow in the mass models), but the need for an additional non-baryonic mass contribution that is distributed in a certain way is evident. 

\subsection{Density gradients - Dark matter vs Baryons}
\label{subsec:mass_slope_shape}

To quantify the degree to which the baryonic and dark matter differ we can evaluate their local density gradients as a function of radius. To this end we should focus on the density gradients in the regions where we believe the masses to be robustly constrained. As discussed in Sec.~\ref{subsec:mass_slope_shape} the dynamical constraints are strongest where data coverage is good, and the dark matter fraction is high (see also \paperrefe). Therefore, to measure the slope in the most robust way, one could measure it within the aperture defined by the FoV and then weigh the local density gradient by the local $\dmfrac$. However, this turns out to be dangerous because the true $\dmfrac$ itself is unknown and is a property that we estimate from the modeling itself. Therefore it relies on an accurate decomposition of baryonic and DM. Furthermore we may well expect that other galaxy types (e.g. `ordinary' ETGs or dwarf spirals) have intrinsically very different dark matter fractions than dEs. Therefore if we would link $\dmfrac$ and the definition of the slope we could find artificial correlations because both are not independent anymore.  

To overcome these issues, we decided to calculate the \textit{volume} averaged radial (logarithmic) slope within the sphere of radius $\reval$: 
\begin{equation}
\eta(<\reval)=\frac{3}{\reval^{3}} \cdot \int_{0}^{\reval} dr r^2 \frac{\partial \ln(\rho)}{\partial \ln(r)}
	\label{eq:mean_slope}
\end{equation}
where $\reval$ is the `evaluation' radius. Several similar but not identical definitions and aperture conventions have been used in the past \citep[e.g.][]{Poci_2017,Dutton_2014,Derkenne_2023} to estimate the slopes of \textit{total} matter densities. To place our results in a broader context and compare them to previous studies, we also analyzed the total slopes of our dE sample using some of the existing slope conventions (see Sec.~\ref{subsec:Total_slopes}). 

We plot $\eta$ as a function the $r=\reval$ in the \textit{right} panels of Fig.~\ref{fig:dE_avg_slopes} for the dark matter and stellar components, as well as for the combined mass distribution. For the latter two the volume averaged slope is not an urgently needed measure since the constraints of total mass and baryonic mass do not scale with $\dmfrac$ but we nonetheless display them here to enable a fair comparison between the different mass components\footnote{This line of argumentation is of course also true for the stellar component, i.e. we expect the constraints on the stars to become worse where $\dmfrac \sim 1$. However, for our dE study this is not relevant since the baryonic contribution is sufficiently large throughout the entire FoV.}. The local density gradients of the two components show very different radial behavior. While both components have similar gradients in the galaxy centers, the curves diverge with increasing radius as the dark component barely gets steeper at larger radii. This is dynamically required as the models prefer total density gradients within the effective radius which are relatively shallow (notably shallower than an isothermal profile). However, the models achieve this without forcing the halo profiles to be extremely flat/cored ($\eta_{\mathrm{DM}}\lesssim-0.5$).

\subsection{Black holes - Are the centers dominated by luminous or `dark' mass?}
\label{subsec:mbh}
For the majority of the galaxies in our sample the halo contribution in the center is small to insignificant (Fig.~\ref{fig:dE_enclosedmass}) which suggests that the luminous baryons dominate the dynamics within the center. Still, the models formally also include a second `dark' component, a central black hole that could affect the orbits of the stars tracing the center. The $\aicmod$ constraints for the black-hole mass (Fig.~\ref{fig:dE_AIC_curves}) suggest a strong upper limit, with $\mbh \lesssim$$10^{6}M_{\sun}$. The weaker constraints towards lower black hole masses are within our expectations since the spatial resolution we achieve with the VIRUS-W data would only allow us to resolve the sphere-of-influence of black holes with masses larger than $10^{7}M_{\sun}$. This is illustrated in Fig.~\ref{fig:dE_enclosedmass} which shows the enclosed mass of the total mass in comparison to the recovered black hole mass (horizontal lines). Adding or removing the recovered black hole masses to the total mass budget essentially has no impact on the total cumulated mass within our resolution limit ($\approx 2\arcsec$) that always exceeds $10^7M_{\sun}$.

Our results suggest that neither DM nor black holes are a particularly relevant contribution to the potential in the center of most dEs (with some exceptions, e.g. VCC 1261). Instead the luminous matter distribution dominates the inner dynamics where the mass distribution is well approximated by the light without any need for additional `dark' components (on the scales that we probed). Therefore we believe that the \textit{upper} limits we obtained for black hole masses are relatively accurate and robust. All in all the results seem to suggest that over-massive black holes in dEs with $\mbh \gtrsim$$5\cdot10^{6}M_{\sun}$ are strongly ruled out by the dynamical modeling. In the context of the cusp-core problem (see below, Sec.~\ref{subsec:core-cusp}) this already has implications since lower black hole masses are less likely to have affected the dark matter halos significantly, e.g. via Active galactic nuclei (AGN) feedback. Even though the fraction of dwarf galaxies that exhibit detectable activity is small in the local Universe, AGNs in dwarfs could have been more important at higher redshift \citep[][]{Mezcua_2019,Sharma_2022}. Future studies with significantly higher resolution may even better constrain how important black holes are in dwarf galaxies. Our upper limits $\mbh \approx(10^5-10^{6})M_{\sun}$ suggest that their influence on the DM distribution can not have been dramatic.

\section{Dark matter in quiescent dwarfs - In tension with standard cosmology?}
\label{sec:CDM-Discussion}
Apart from some notable exceptions \citep[][]{van_Dokkum_2018,Shen_2021,comeron_2023_arxiv} the vast majority of galaxies are known to require a dark matter halo (or a modification of gravity). In line with this, the dEs we investigated also exhibit a dynamical necessity for an additional dark component that increases the total mass budget (particularly at larger radii) beyond what the luminous mass distribution seems to suggest. However, with our advanced modeling technique, we attempt to go beyond merely demonstrating the dynamical necessity of dark matter in dEs: We try to accurately recover the exact profile, flattening and amount of dark matter in these galaxies. In principle, such distributions could even allow us to infer information on the underlying cosmology in which the halos have formed. Within standard cosmology model ($\Lambda$CDM) structure formation on large scales is generally explained well but several problems and tensions at smaller scales were reported that seem to conflict with it (cf. Sec.~\ref{sec:intro}), which will be investigated in the following. 

\subsection{Properties to analyze the DM structure}
\label{subsec:DM_tools}
To address whether our dynamical constraints for the dEs are in tension with the standard $\Lambda$CDM paradigm, we first need to establish a set of properties that describe the halo succinctly. As discussed in Sec.~\ref{sec:dwarf_modelling} the DM halo is best constrained at radii where $\dmfrac$ is large and within the FoV. For most of our sample this is roughly around $10\arcsec$ as the majority of dEs have negligible or low dark matter contribution in the center and our FoV covers radii out to $10\arcsec$-$15\arcsec$ (cf. \paperrefeSTARS). Since we intend to compare the halos of each dE to one another and interpret their physical origin and evolution consistently we want to measure the average density and slope evaluated within a distance-independent, \textit{identical} aperture for all galaxies. Therefore we decided to evaluate each halo within a circular aperture of the same intrinsic radius $\reval=0.8 \rm kpc\approx 10\arcsec$ which is at around one effective radius for the dEs.

Given such an aperture, we can then characterize the dark matter halos using the following \textbf{three quantities}: 

i) The \textbf{average density} $\overline{\rhodm}$ of the dark matter enclosed within the aperture:
\begin{equation}
\overline{\rhodm}=\frac{3}{4\pi}\cdot \frac{\MDM}{\reval^3}
	\label{eq:mean_density}
\end{equation}

ii) The volume \textbf{averaged logarithmic slope} $\eta_{\mathrm{DM}}$ of the halo within the aperture, as given by equation eq.~\ref{eq:mean_slope} evaluated at $\reval=0.8 \rm kpc$. 

iii) The halo \textbf{flattening} within the aperture $\reval=0.8 \rm kpc$ which, in our case, is aperture independent and identical to the nuisance parameter $\qdm$ because the models only allow for spatially constant axisymmetric dark matter flattenings.

\subsection{The cusp-core problem in dEs}
\label{subsec:core-cusp}
From Fig.~\ref{fig:dE_avg_slopes} it is evident that the dark matter slopes of the dEs exhibit a similar diversity (sample scatter) as the slopes of their luminous stellar distributions do. But the dark matter distribution is (on average) shallower and changes less drastically with increasing radius than the stars. First and foremost, this result highlights the well known requirement of galaxies having a total mass that decreases less steeply with radius than the luminosity distribution would suggest \citep[e.g.][]{Rubin_1970}. For many observed galaxies that are being studied with dynamical modeling this expresses itself in the need for an almost entirely flat/cored halo model component that flattens the overall mass gradient. The resulting halo models are often times so cored that they are in tension with the $\Lambda$CDM simulations of structure formation which predict cuspier distributions, particularly in the regime of small dwarf galaxies and late-type galaxies (see Sec.~\ref{sec:intro}). 

However, in the regime of dEs the question whether the cusp-core problem persists, and if so how strong, is far from settled. While the dark matter in our dE models falls off less steeply with radius than the baryons (as expected), the halo density models are not completely flat either. There is no unequivocal preference of the halos being either very cored, or particularly cuspy. Instead, we find the slopes $\eta_{\mathrm{DM}}$ to be mostly \textit{moderate} but with considerable scatter across the different dEs in our sample (Fig.~\ref{fig:dE_avg_slopes}). The most cored dEs have $\eta_{\mathrm{DM}}\sim -0.5$ while the most cuspy ones have $\eta_{\mathrm{DM}}\sim -1.4$. This sample scatter is larger than the typical $1\sigma$-error $\sigma_{\mathrm{DM}}=0.26$ we find for the dark matter slopes and larger than the radial change in DM slope within one effective radius. Therefore we argue this sample scatter of the halo slopes is not a measurement uncertainty but displays the real diversity of the halos. In fact, we may expect exactly this level of diversity considering the corresponding level of variety in slopes of the stellar matter (middle panel of Fig.~\ref{fig:dE_avg_slopes}) which stems from the deprojection of the observed photometry (and indirectly its gradients). If the luminous matter exhibits this level of diversity, then it may be reasonable to expect that the halos of dEs can also exhibit a similar diversity in slopes.

Compared to the sample scatter across different dEs the radial change in slope is slightly smaller within the investigated radial range. Within one effective radius the change in slope is similar in scale to the $1\sigma$ error and, thus, barely statistically significant. Conversely, the slopes of the baryonic matter display a significant and systematic decrease over the entire range covered by kinematic data. Under the \textit{assumption} that the true distribution of dark matter follows a profile with a \textit{single} distinct scale radius, these results imply that the \textit{scale radii} are dynamically unconstrained by our data. They must be located either far outside the FoV or in the very center where a negligible dark matter contribution (e.g. if $\dmfrac$$\sim$$0$) could prevent a change in slope to be dynamically detectable by the models as it barely changes the net potential. The former case is more likely because the total enclosed mass would be infinite otherwise. While the detailed size of halos relative to their stellar half-light radius may depend on various factors (e.g. redshift, angular momentum, etc.), we can anticipate the halo size to be generally much larger than the stellar half-light radius \citep[e.g.][]{Somerville_2018}. Consequently, we expect the scale radius to be far outside the $1\reff$ FoV, and we argue the nearly constant slopes we measure inside our FoV are mostly representative of the true \textit{inner} logarithmic halo slopes. Strictly speaking, of course, this holds only if the true dark matter distribution has only a \textit{single} scale radius. One could easily imagine a second or multiple slope transitions (i.e. scale radii). For example, a very small core within the regions where we measured, $\dmfrac \approx 0$ which we would not be able to detect dynamically with our current observations. 

Given the above arguments, we interpret the dark matter density slopes we measured within the $0.8 \rm kpc$ aperture to be representative of the \textit{central} slopes of the dark matter distributions. If we analyze the density slopes within this distance-independent aperture $\reval$ the sample average for the halo component is $\eta_{\mathrm{DM}}=-0.91\pm0.35$, which is contrasted by a much steeper luminous matter slope with a sample average of $\eta_{*}=-2.04\pm0.40$. While we believe the slope within $0.8 \rm kpc$ is the most robustly constrained measure of the central dark matter slopes we can obtain from the dynamical models, the values are not very sensitive to the exact aperture that we use. For example, if we were to evaluate the dark matter slopes within $\reval=0.2 \rm kpc$ (i.e. close to the spatial resolution limit of our data) we would obtain an average dark matter slope of $\eta_{\mathrm{DM}}(r=0.2 \rm kpc)=-0.73\pm0.38$.  

These slope estimates allow us to address the cusp-core problem from a new perspective, as detailed measurements of halos in this mass regime mostly stem from rotation curve modeling. This has a strong selection bias towards LTGs since it relies on the existence of a rotating HI disk. The dynamical models we employ here can complement these constraints, as they use stars as tracers of the gravitational potential instead. While stellar dynamical modeling has been employed to measure the halos of ETGs before, most studies are confined to more massive ETGs or smaller dwarf galaxies (dSphs) within the Local Group. Studies of early-types in the intermediate mass regime of the dEs are rare and, even then, do not attempt to explicitly measure the slope of the decomposed dark matter component but instead the \textit{total} density slope (which we will also compare in Sec.~\ref{subsec:Total_slopes}). While a direct comparison to ETGs in the same mass range of our dEs galaxy is difficult, we can place our results in the context of the measured DM slopes of LTGs and the smaller resolved early-type dSphs of the Local Group. 

While DM slopes for the most massive spirals are often ambiguous (some are cored other cuspy) the majority of studies of dwarf LTGs ($M_{B}\gtrsim-19\mathrm{mag}$) tends to favor cored dark matter distributions \citep[][]{McGaugh_1998,Cote_2000,de_Blok_2002,Gentile_2004,de_Blok_2008,Donato_2009,Plana_2010,Oh_2011_a}. For example, the dwarf irregulars (dIrr) analyzed by \citet{Oh_2015}, which are in the mass regime of our dEs, suggest a typical DM slope $\sim -0.29$, and the dwarfs sample of \citet{Adams_2014} has a mean DM slope of $\sim -0.58$. In contrast, in the \textit{quiescent} counterparts of the dIrr, the dSphs, the circumstances are much more ambiguous. First results of \citet{Walker_2011} using a chemo-dynamical approach seemed to have ruled out NFW-like halos for Fornax and Sculptor, though \citet[][]{Genina_2018} later showed that the models can possibly lead to mis-identification of cusps as cores. In a series of papers \citet[][]{Jardel_2012,Jardel_2013_a,Jardel_2013_b} have modelled dSphs of the Milky Way (Carina, Draco, Fornax, Sculptor, and Sextans) using Schwarzschild models. They found a diverse range of central DM profiles, with some being cuspy while others being cored. They concluded that (at least on average) the central halo slope scatters around an NFW-like profile. Similarly, \citet{Hayashi_2020} found that the halos of 8 dSphs show a significant variety but tend to favor more cuspy density distributions. Recently, \citet{De_Leo_2023}  analyzed resolved stellar motions in the Small Magellanic Cloud using Jeans modeling and found a cuspy structure that is more consistent with theoretical predictions. 

Taking into account the results of our dEs and the ambiguity in the dSphs samples, it appears the results for ETGs samples are not as suggestive of a core-cusp problem as the observational constraints for LTGs are. This disparity between LTG and ETG samples could be an artifact from \textit{systematic} differences between the modeling methods that are being employed. Many studies test parametric models such as NFW-profiles, or non-singular isothermal/logarithmic profiles, which may or may not be flexible enough to capture the structure of real DM halos. The effects of halo parametrization have not yet been investigated systematically, and it could be the case that some cored halo models nominally provide a better fit but underestimate the central DM density considerably. In \paperrefeSP we will investigate the effects of halo parametrization on a methodical level. Apart from this parametrization issue that likely extends to all dynamical modeling techniques, the observations of cores could also be facilitated by the fact that the majority of LTG studies are conducted using gas rotation curves whereas ETGs are constrained using stars as tracers. However, for a sample of \textit{late-type} dwarfs \citet{Adams_2014} investigated whether dynamical models that use gas as tracer for the gravitational potential recover different DM slopes than Jeans models (with stars as tracers). Using generalized NFW (gNFW) models they find both approaches to be fairly consistent with the gas models having a sample averaged central slope $\gamma=-0.67\pm0.1$ and the stellar models $\gamma=-0.58\pm0.24$. This suggests that the apparent difference in dark matter slopes of early- and late-types does not stem from an intrinsic bias/difference between the stellar vs gas dynamical models. Still, further comparisons of different modeling techniques applied to the same galaxy are required to judge about modeling systematics. 

If for now we assume systematics are negligible and results for LTGs and ETGs can be trusted equally, this slope-morphology dichotomy may well be genuine because the two classes can be expected to follow very different evolutionary paths. The correspondingly different baryonic feedback and/or environmental effects could have driven initially identical dark matter halos to diverge over time. In other words, even if all DM halos had initially NFW-like slopes, we may expect that LTGs observed today are on average more cored than the dEs. Since we argue the amount of diversity in measured slopes within our dE sample is not driven by noise or systematic errors but reflects the true diversity that dEs exhibit (see above), the even larger slope difference to the LTG studies strongly suggests that there must be mechanisms (ab initio or secular ones) that discriminate between different galaxies, leading to some halos being shallower than those of others (see also Sec.~\ref{sec:formation_evolution}). 

The situation whether dark matter in quiescent dwarfs is, in fact, cored enough to be considered in tension with standard cosmology appears to be much more ambiguous than it is for the LTGs. A considerable amount of ETGs studies suggest mildly cored and near NFW slopes. Still, the typical central slopes that we measured $\in[-0.7,-0.9]$ are in mild to moderate tension with the vast majority of DM-only simulations (DMO) of $\Lambda$CDM structure formation. Early DMO studies predicted (often universal) cuspy halo profiles with central slopes ranging from $-1$ to $-1.5$ \citep[][]{Navarro_1996,Navarro_1997,Moore_1998,Fukushige_2001}, and even more recent DMO simulations with higher resolution struggle to reach slopes much flatter than $-0.8$ \citep[][]{Gao_2008,Stadel_2009}. While the tension between the DM slopes of simulations and observations is significantly mellowed in the quiescent dwarfs, the existence of galaxies with DM slopes of $-0.5$ nonetheless rejects DMO simulations, suggesting at least some requirement for modification. Hydrodynamic simulations which include the response of the DM to baryon physics and friction appropriately may (possibly more comfortably than for LTGs) explain these remaining differences \citep[e.g.][]{El-Zant_2001,Mashchenko_2006,Del_Popolo_2009,Governato_2010,Cole_2011,Governato_2012,Nipoti_2015,Orkeny_2021}. 

All in all, neither the simulation nor the observation side seem to be entirely decided whether the halos of dwarf galaxies are cuspy or cored. In the last decades several solutions have been proposed to reconcile the observations with simulations \citep[for a review see][]{Del_Popolo_2021}. Broadly speaking, solutions can be categorized as follows: i) Baryonic feedback/outflows which transfer to the DM. For example, star formation bursts could have removed parts of the baryons, and in the course of that also some of the dark matter, leading to the observed cores \citep[e.g.][]{Navarro_1996,Gnedin_2002,Governato_2010,de_Souza_2011,Oh_2011_b,Madau_2014}. Similarly, interactions with the environment such as mergers, harassment or ram-pressure stripping (directly and indirectly by regulating star formation) could have affected the dark matter profiles \citep[e.g.][]{Del_Popolo_2012}. ii) Cosmological solutions: Theoretical concepts such as self-interacting dark matter, fuzzy dark matter, superfluid dark matter, could naturally produce cored halo profiles \citep[e.g.][]{Spergel_2000,Harko_2011,Robles_2012,Elbert_2015} iii) Modifications of Newtonian gravitation can make cuspy halos appear to be cored in a Newtonian analysis \citep[e.g.][]{Benetti_2023}. 

Overall we conclude that the cusp-core tension for dEs is not as severe and standard $\Lambda$CDM simulations that accurately model the effects of baryons may well be able to explain the halo distributions without the need for invoking more exotic physics. In Sec.~\ref{sec:formation_evolution} we will explore (under the umbrella of standard $\Lambda$CDM cosmology) whether our results point towards a specific baryonic mechanism that is driving the diversity in measured DM slopes of the dEs and the systematic difference to LTGs.  

\subsection{Are DM halos spherical?}
\label{subsec:halo_flattening}
Numerical Simulations of DM halo formation not just predict their central slopes but also their 3D shape. Therefore the axisymmetric flattening $\qdm$ we measured for the dEs (Fig.~\ref{fig:dE_AIC_curves}) puts us in the position to probe cosmological predictions and halo assembly further. The flattening of dark matter halos (other than the Milky Way's halo) is rarely probed by dynamical modeling studies. In App.~\ref{append:simulation} we demonstrate that dark matter flattening is detectable using our dynamical modeling setup and, in fact, plays a significant role in determining the overall quality of the model. Even though the N-body simulation in App.~\ref{append:simulation} is slightly triaxial and changes shape with radius the constant, axisymmetric flattening $\qdm=0.8$ we recovered approximates the average shape of the halo (to first order) well and unbiased. 

\begin{figure}
	\centering
	\includegraphics[width=1.0\columnwidth]{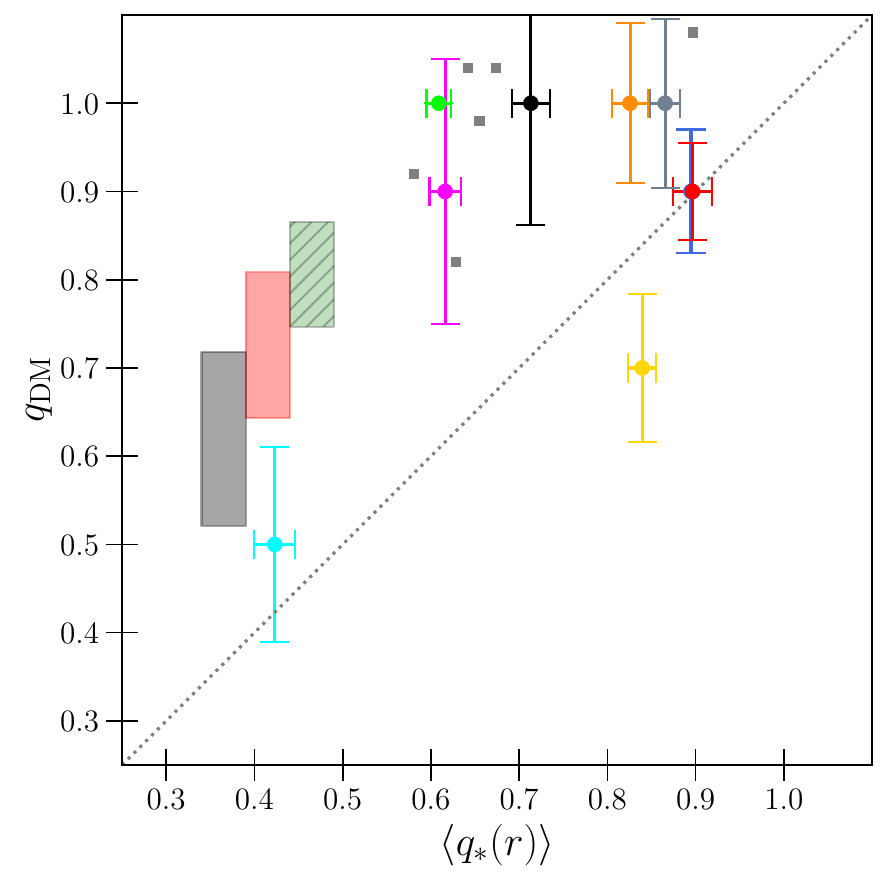}
    \caption{The \textit{intrinsic} axisymmetric flattening of the DM halo $\qdm$ vs the corresponding (radially averaged) $q_{*}$ of the stars for the dEs. Also included are the dynamically recovered axis ratios of the dSph of the Milky Way (\textit{gray squares}) from \citet{Hayashi_2020}. We indicate the \textit{expected} $\qdm$ from the Illustris simulations \citep[][]{Chua_2019} for their DMO runs (\textit{gray bar}), for the full hydrodynamical run in the dE regime $M_{*}\sim 10^9M_{\sun}$ (\textit{red bar}) and for higher masses ($10^{11}M_{\sun}$) at which their halos are roundest (\textit{green bar}). They describe the halo triaxial by minor/intermediate axis-ratios $q$ and $s$, to compare it to our axisymmetric values, the mean of $q$ and $s$ is shown here. The bar height represents the 25th-75th percentiles of their galaxies. The placement of the bars on the $q_{*}$-axis is arbitrary because they do not resolve $q_{*}$, but from the analysis of more massive ETGs in Illustris TNG100 by \citet{Pulsoni_2021} we may expect $\qdm \geq q_{*}$ with a large range of stellar flattening $0.2\leq q_{*}\leq 0.9$. Congruent with these simulations, our dEs have $\qdm \geq q_{*}$, but have even rounder halos than statistically expected.} 
    \label{fig:dE_flattening}
\end{figure}

For dEs we find that dark matter and stars are not only different with regard to their \textit{radial} distributions, but they also have systematically different flattenings. Fig.~\ref{fig:dE_flattening} compares the average \textit{intrinsic} axis ratios of the luminous matter $q_{*}$ with that of the dark matter $\qdm$. For all galaxies, but VCC 1528, the halo appears to be rounder than the stellar component it hosts. In fact, the majority of the DM halos are spherical or close to it. The only halo that exhibits a strong amount of halo flattening is that of VCC 2048. However, VCC 2048 also has the flattest stellar distribution ($q_{*}$$\simeq$$0.4$) and the lowest dark matter fraction with $\dmfrac<0.1$ throughout the entire investigated radial range (cf. Fig.~\ref{fig:dE_enclosedmass}), hence, the central halo shape may not be very well constrained. 

DMO simulations usually predict complex halo shapes that can be prolate or triaxial with significant shape variation at different radii \citep[e.g.][]{Frenk_1988,Allgood_2006,Hayashi_2007}, therefore it may be surprising that almost the entirety of the dE sample is fitted best with nearly spherical dark matter halos. However, these DMO predictions are also in tension with most observational results for the Milky Way's DM halo \citep[e.g.][]{Ibata_2001,Law_2010,Vera_ciro_2013,Bovy_2016,Wegg_2019}. These suggest a nearly spherical, oblate dark matter structure despite the Milky Way's much flatter stellar structure. For example, by analyzing the \textit{Gaia} proper motions of RR Lyrae stars \citet{Wegg_2019} have found an ellipsoidal flattening $\qdm=1.0\pm0.09$ of the Milky Way halo out to radii of $20 \rm kpc$. 

Similar to the cusp-core problem, this tension with the DMO simulations is significantly mellowed when baryonic interactions are included in the numerical simulations: the halo shapes are transformed by the interactions with baryons to be more spherical/oblate \citep[e.g.][]{Katz_1991,Chisari_2017,Chua_2019,Chua_2022,Cataldi_2023,Orkney_2023}. This reflects that DM particles that reach the center on box orbits are gravitationally affected by the condensating baryons such that their orbits change and become more tube-like and circular \citep[cf.][]{Debattista_2008}. This modification of the flattening caused by the baryons is most dominant in the center of the galaxies: while DMOs predict prolate distributions in the center which only become more spherical at large radii the baryon inclusion make the center more spherical which overall leads to a more radially constant and close-to-spherical halo distribution \citep[][]{Abadi_2010}. Hydrodynamical simulations not only predict that the resulting halo becomes more spherical, but also that they end up to be rounder (by about $\Delta q \sim 0.2$) than the baryonic distribution they host \citep[e.g.][]{Tenneti_2014,Pulsoni_2021}. This is congruent to our finding that the dark matter is more spherically distributed than the baryonic mass (Fig.~\ref{fig:dE_flattening}). 

In Fig.~\ref{fig:dE_flattening} we indicate the range of halo flattenings we may expect from numerical $\Lambda$CDM simulations by showing the median flattening of galaxies in the Illustris simulation analyzed by \citet{Chua_2019}. The inclusion of baryons typically changes the axis ratios of the simulated galaxies by about $+0.1$ to $+0.2$ which makes the average halo to have $\qdm \approx 0.75$. Like it is the case for the Milky Way's halo, the inclusion of baryons does lift some of the tension, but our observational constraints still suggest surprisingly spherical dark matter distributions. 7 out of 9 dEs have a $\qdm$ larger than the 25-75th percentile of the Illustris simulated galaxies. The analysis of \citet{Chua_2022} suggests that the effect of baryon-induced dark matter flattening becomes smaller as the total halo mass is decreased and/or as baryonic feedback parameters, such as stellar wind or black hole feedback strength, are increased. The former implies that we may expect galaxies in more massive halos to be even more spherical, while the latter implies that galaxies with stronger baryonic feedback are more prolate as they decrease the stellar mass fraction. Still, the observational constraints of our dE sample suggest that even halos in quiescent dwarfs are similar to the Milky Way in that they follow a nearly spherical distribution. Whether one can detect a dependency of the sphericity on the total mass or morphology (like it seems to be the case with the density slopes) remains to be seen at this point, as observational constraints need to be extended to more galaxies.  

The fact that some statistical tension between hydrodynamic simulations and our observation persists could be explained by the approximations and assumptions that were made to obtain the observational constraints. Firstly, our small sample of dEs may not be overly representative for the average galaxy in this mass regime. The majority of our sample is observed to have a fairly round light distribution already (cf. ellipticities in \paperrefeSTARS) which could also favor more spherical DM halos, i.e. $\qdm$ could be selection biased. Furthermore, considering baryons are a driving factor in determining the halo shape, we may also expect there to be significant differences between different galaxy morphologies\footnote{E.g. between ETGs and LTGs of the same mass. Though we may expect from the comparison of central slopes (Sec.~\ref{subsec:core-cusp}) that $\qdm$ of late-type dwarfs is even more spherical than early-type dwarfs.}. Secondly, the dynamical models assume an axisymmetric shape with a radially constant flattening, while real halos are probably triaxial with radially changing shape. This could bias our results for $\qdm$ to be more spherical, though the test on the N-body simulation (App.~\ref{append:simulation}) does not suggest so. Thirdly, especially for low mass halos, the baryon distribution and its embedding dark matter halo may be misaligned significantly \citep[][]{Chisari_2017}, an effect which our dynamical models can not emulate at this point. 

If, on the other hand, the observational constraints are robust and the hydrodynamical simulations accurately model baryonic interactions, the measured shapes of DM halos could be a strong indication of a deviation from $\Lambda$CDM cosmology. The excess in sphericity could be a direct consequence of non-CDM particles. For example, self interactions of the dark matter particles can heat up their orbits as they pass through the dense halo center and interact with each other \citep[e.g.][]{Peter_2013,Vogelsberger_2016,Brinckmann_2018}. As a result, the halos become rounder in the center, which is where we constrain the dE halos. Similarly to self-interacting dark matter, Fuzzy Dark matter particles can also produce more spherical halo shapes \citep[][]{Marsh_2014,Chowdhury_2023}. 

At this point in time, observational attempts at measuring the intrinsic shape of DM halos are still in the beginning stages. The few constraints that exists are for nearby spirals using \textit{edge-on} rotation curve fitting \citep[e.g.][]{Peters_2017}, for Local Group dSphs using Jeans modeling \citep[e.g.][]{Hayashi_2020}, or are for the Milky Way's halo inferred from globular cluster and stellar streams \citep[e.g.][]{Ibata_2001,Law_2010,Vera_ciro_2013,Bovy_2016,Posti_2019,Wegg_2019}. While gravitational lensing can provide independent shape estimates at larger distances, it can only directly constrain the \textit{projected} flattening of the mass distribution. Constraining the halo shapes with stellar dynamical modeling could therefore be an invaluable additional probe for early-type galaxies and further our understanding of cosmology.

Our dynamical constraints corroborate the observational results found for the Milky Way's halo in that the dark matter is distributed close to spherical even though the stars occupy orbits in a more flattened distribution. However, as outlined above, several simulations and theories may explain such observations. To draw definite physical conclusions on the implications of the measured dE halo shapes, larger sample sizes with FoVs beyond $1\reff$ and dynamical models that are triaxial and/or allow radially changing shapes are required. Similarly, galaxy samples at different mass and redshift scales could also be helpful in our understanding. We only probe dEs with $M_{*}\sim10^9{M_{\sun}}$ in this study, but the ability/efficiency, e.g. of baryonic feedback, to make the halos more spherical is expected to vary with total mass, stellar mass fraction and redshift \citep[e.g.][]{Allgood_2006,Chua_2019,Chua_2022}. Investigating these features of the halo shape may allow us to differentiate between the different cosmological scenarios in the future. 

\section{Halo structure - An imprint of galaxy formation and evolution?}
\label{sec:formation_evolution}
The discussion in Sec.~\ref{sec:CDM-Discussion} has shown that the $\Lambda$CDM cosmology is not strongly ruled out by the observational shape and slope constraints we found for the dark matter halos of dEs. While there remains some tension (particularly in the halo flattening) numerical simulations that accurately model baryonic and environmental effects may well be able to explain the observations without the need for invoking more exotic physics. In the following few sections, we will explore what the recovered dark matter distributions can tell us about the evolution and formation of dEs, given we assume $\Lambda$CDM is indeed an accurate description of underlying cosmology. We will preface this analysis by a discussion of the total densities (stars and DM combined) which complements the discussion of the individual DM halos and facilitates comparison with the existing literature.    

\subsection{The total density slopes of dEs}
\label{subsec:Total_slopes}
The DM density slopes we reported in Sec.~\ref{sec:CDM-Discussion} rely on a robust decomposition of the DM halo from the stars and stellar remnants. While we stress-tested this on a simulation (App.~\ref{append:simulation}) there is of course no guarantee that this always works for real galaxies. However, under the assumption the $\Lambda$CDM paradigm is correct, we can also investigate galaxy evolution and differences between different morphologies without the explicit requirement for an accurate decomposition if we analyze the \textit{total} mass distribution. The total mass is more strongly constrained by the dynamical models than the individual mass components are (see \paperrefe) and its recovery is not reliant on a successful decomposition. The same will hold true for the corresponding \textit{total} density slopes. Furthermore, compared to the analysis of decomposed DM, a manifold of published studies exist that investigate the total densities but not the dark matter component on its own. In the following, we complement our analysis of the DM slopes (Sec.~\ref{subsec:core-cusp}) by comparing the total density slopes of our dEs to those of published literature. This has the goal to place the dEs in a broader context, compare them to different galaxy classes and investigate possible formation mechanisms. Various conventions of measuring the total slopes (calculated within different apertures) are used in the literature. To make the comparison straightforward, we calculate the slopes for our dEs as they are defined in the studies that we compare them to in the following section. 

\subsubsection{Comparison with `ordinary' ETGs}
\label{subsubsec:tot_slope_massive_ETGs}
\begin{figure*}
	\centering
	\includegraphics[width=1.0\textwidth]{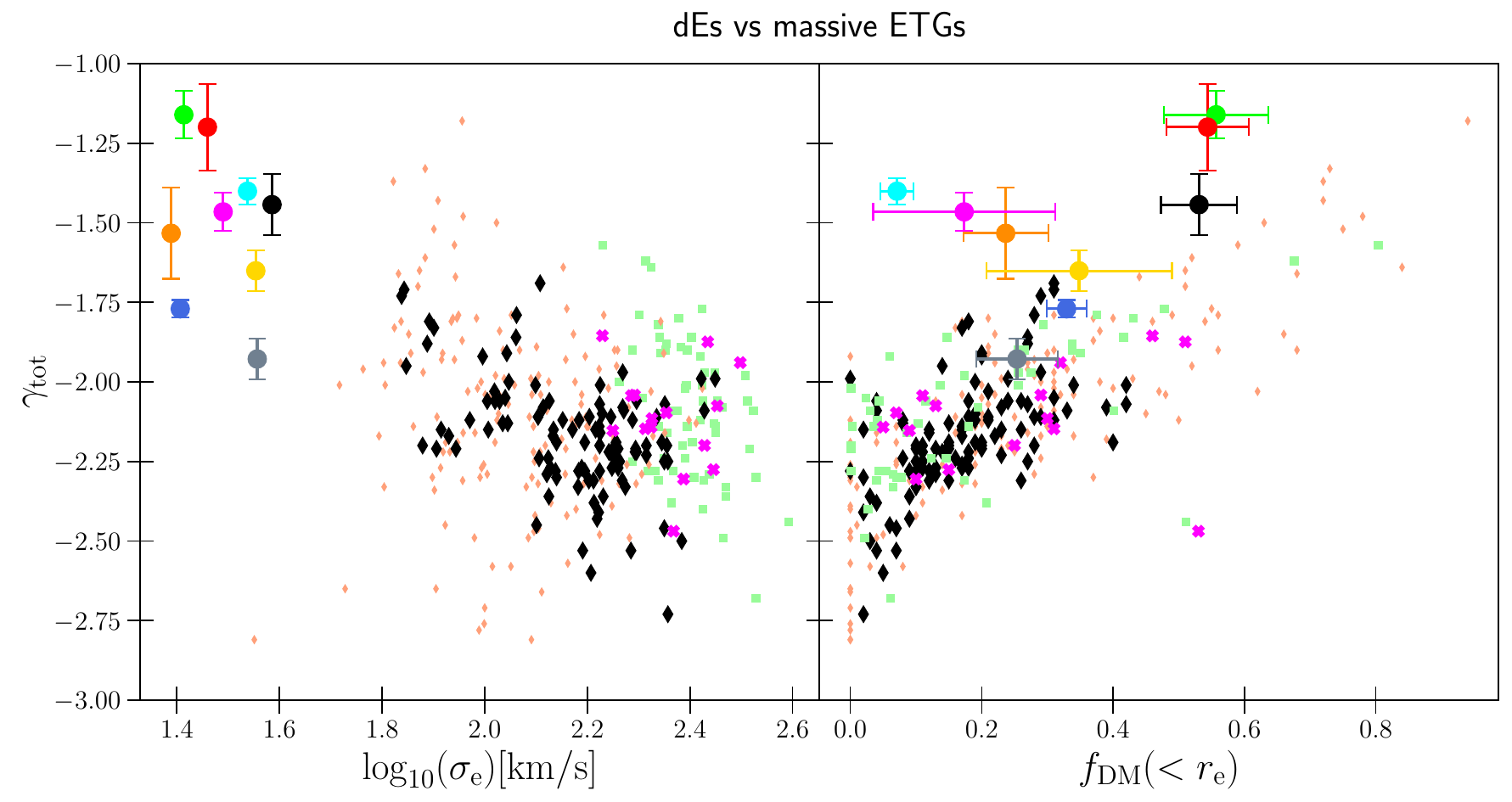}
    \caption{\textit{Left panel:} The average logarithmic power-law slope $\gamma_{\mathrm{tot}}$ of the \textit{total} mass distribution evaluated between $0.1\reff$ and $1\reff$ vs the luminosity-weighted velocity dispersion $\sigma_{\mathrm{e}}$ within one effective radius. \textit{Colored Dots:} Our dE sample. \textit{Diamonds:} The massive ETGs, part of the ATLAS$^{\rm 3D}$-survey \citep[][]{Cappellari_2011}, for which we obtained the slopes of model 1 in \citet{Poci_2017}. Their aperture varies slightly for each galaxy since it depends on the FoV coverage. But for the majority their $\gamma_{\mathrm{tot}}$ is essentially measured between $0.1\reff$ and $1\reff$. Following \citet{Poci_2017} we differentiate between ATLAS$^{\rm 3D}$ galaxies with good data quality (\textit{Black diamonds}) and poor quality (\textit{Red diamonds}). This data quality for each ETG was evaluated in \citet{Cappellari_2013}. \textit{Green squares:} The power-law slopes obtained from \citet{Auger_2010} for the strong lensing ETGs which are part of the SLACS survey \citep[][]{Bolton_2008,Auger_2009}. \textit{Magenta crosses:} The total slopes calculated from the mass profiles of \citet{Thomas_2007} which were obtained with Schwarzschild which are based on an earlier version of the axisymmetric orbit code we use in this study. \textit{Right panel:}. The total slope $\gamma_{\mathrm{tot}}$ vs the dark matter fraction $\dmfrac$ enclosed within $1\reff$ for the same data that was shown in the left panel. For the SLACS survey we obtained dark matter fraction from \citet{Posacki_2014}. For more published data of giant ETGs and comparison with cosmological simulations see also Fig.11 of \citet{Derkenne_2023}. } 
    \label{fig:Total_slope_vs_dispersion}
\end{figure*}

In Fig.~\ref{fig:Total_slope_vs_dispersion} we display $\gamma_{\mathrm{tot}}$, the mean logarithmic density slope of the total mass within the effective radius (see \citealt{Poci_2017}), vs the effective velocity dispersions $\sigma_{\mathrm{e}}$ and the dark matter fraction $\dmfrac$ for our dE sample and studies of `ordinary' ETGs ($M_{*}\gtrsim10^{10}M_{\sun}$). For the latter, we show three different samples, each based on a different modeling technique: i) Lensing models, ii) Schwarzschild models, and Jeans anisotropic modeling \citep[][]{Cappellari_2008}. Independently of the applied modeling method, the density slopes of the `ordinary' ETGs congregate at $\gamma_{\mathrm{tot}}\approx-2.1$ on average. In stark contrast, the slopes of our dEs are noticeably shallower (with a sample average of $\gamma_{\mathrm{tot}}=-1.51\pm0.24$) and concentrate outside the scatter of `ordinary' ETGs. 

In the high mass range $\sim 10^{10}-10^{12}M_{\sun}$ (i.e. $\log_{10}(\sigma_{\mathrm{e}})\gtrsim2$) it is well established that (with little scatter) galaxies have around isothermal total density profiles at one effective radius. Because the steeper baryonic and shallower dark matter components seemingly `know' of each other so as to always produce a nearly isothermal density together, this is also known as the `bulge--halo conspiracy' \citep[e.g.][]{Gerhard_2001,Thomas_2007,Koopmans_2009,Auger_2010,Dutton_2014}. However, this `bulge--halo conspiracy' actually only seems to hold in a limited mass regime. Several studies \citep[][]{Poci_2017,Tortora_2019,Li_2019} have noticed that below $\log_{10}(\sigma_\mathrm{e})\sim 2.1$ the total slope and velocity dispersion (or stellar mass) of a galaxy are anti-correlated (i.e. less massive ETGs are shallower). As the total mass is increased, the galaxy profiles become steeper until near $\log_{10}(\sigma_{\mathrm{e}})\sim 2.1$ the typical slope reaches a minimum at $\gamma_{\mathrm{tot}}\approx -2.3$, at which point the $\gamma_{\mathrm{tot}}$-$\sigma_{\mathrm{e}}$ breaks and flattens or arguably even reverts as galaxies with even higher masses become slightly more shallow again. 

This break point near $\log_{10}(\sigma_{\mathrm{e}})=2.1$ and $\gamma_{\mathrm{tot}}=-2.3$ can also be identified in Fig.~\ref{fig:Total_slope_vs_dispersion} as more massive galaxies do not become cuspier. If, on the other hand, the $\gamma_{\mathrm{tot}}$-$\sigma_{\mathrm{e}}$ correlation of the ETGs with $\log_{10}(\sigma_{\mathrm{e}})\leq2.1$ is extended to the lower dispersions of our dE sample, then, the dEs are arguably exactly where one would expect them to be. This may suggest that the dEs, together with slightly more massive ETGs ($\log_{10}(\sigma_{\mathrm{e}})\lesssim2.1$), form a continuous sequence for which the total slope $\gamma_{\mathrm{tot}}$ systematically decreases with increasing velocity dispersion $\sigma_{\mathrm{e}}$. Whether continuously connected or not, the dEs have much shallower density distributions than `ordinary' ETGs.

Conversely to the `ordinary' ETGs, the total slopes of dEs do not fall on the same correlation of the dark matter fraction $\dmfrac$ with the density slope (right panel of Fig.~\ref{fig:Total_slope_vs_dispersion}) as their shallower slopes would require a much higher dark matter fraction around $\dmfrac\sim0.8$ than what is measured. In other words, the shallow $\gamma_{\mathrm{tot}}$ of dEs is not \textit{solely} a result of an increased contribution of the halo to the total density within the effective radius. The strong correlation of total slope and DM fraction in `ordinary' ETGs may be expected. Simulations suggest both $\gamma_{\mathrm{tot}}$ and $\dmfrac$ increase with stellar mass \citep[e.g.][]{Lovell_2018,Mukherjee_2022} for galaxies with masses above $\gtrsim3\cdot10^{10}M_{\sun}$. If $\dmfrac$ and $\gamma_{\mathrm{tot}}$ both correlate with stellar mass then they will also correlate with each other as seen here. We conclude the total density slopes of dEs arguably lie on the extension of `ordinary' ETGs to lower masses, but their slopes and dark matter fractions are not as strongly correlated as those galaxies.

\subsubsection{Comparison with quiescent and star-forming dwarfs}
\label{subsubsec:tot_slope_dwarfs}
\begin{figure*}
	\centering
	\includegraphics[width=1.0\textwidth]{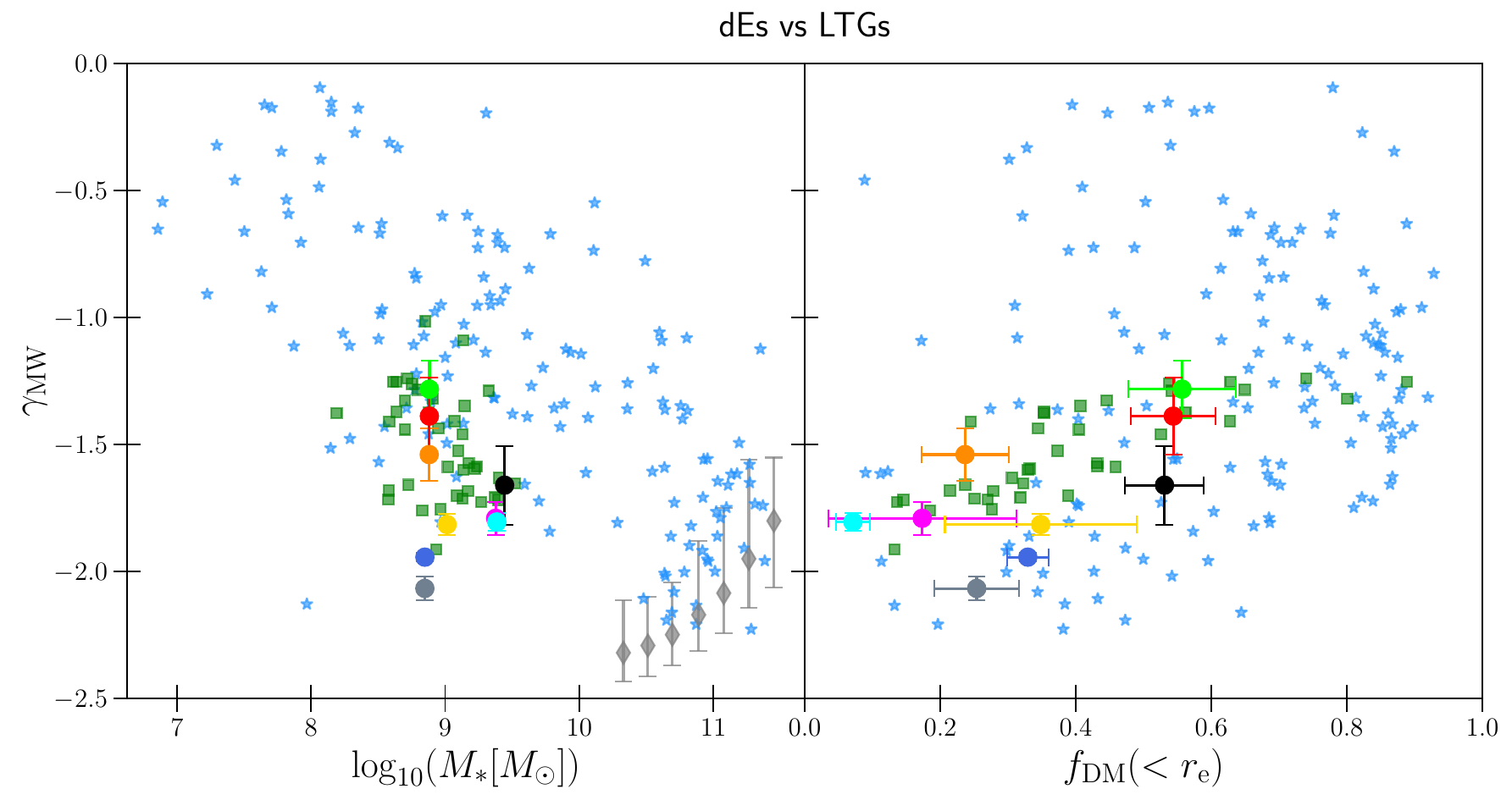}
    \caption{The mass-weighted total density slope $\gamma_{\mathrm{MW}}$ (eq.~\ref{eq:MW_slope}) of dEs and LTGs within $1\reff$ vs stellar mass $M_{*}$ (\textit{left panel}) and dark matter fraction $\dmfrac$ (\textit{right panel}). \textit{Colored Dots:} Our dE sample. All the literature data shown here are from \citet{Tortora_2019} who analyzed the slopes of the following 3 galaxy samples. \textit{Blue stars:} The LTG data originate from a subsample of the SPARC survey \citep[][]{Lelli_2016}, it includes LTGs ranging from S0 to Im. \textit{Green squares:} The dEs of the SMAKCED sample \citep[][]{Toloba_2014,Tortora_2016} which are also part of Virgo cluster. As a reference we also include the location of `ordinary' ETGs in the left panel (\textit{gray diamonds}), they stem from the ETGs of the SPIDER survey \citep[][]{LaBarbera_2010,Tortora_2012}. For these ETGs only the sample medians binned in different mass bins are shown, the error-bars indicate the $1\sigma$ percentiles of the more than 4000 ETGs \citep[cf.][]{Tortora_2019}.} 
    \label{fig:MW_Total_slope_vs_stellar mass}
\end{figure*}

The above anti-correlation of total density slope and stellar dispersion (or stellar mass) has also been noted in LTGs, albeit with some offset/modification \citep[][]{Li_2019,Tortora_2019}. In Fig.~\ref{fig:MW_Total_slope_vs_stellar mass} we compare our results to those of \citet{Tortora_2019}. They analyzed and compared the total density slopes of 3 different galaxy samples: Another Virgo dE sample (\citealt{Toloba_2014}; see also \paperrefeSTARS for a comparison), a LTG sample which includes star-forming galaxies ranging from S0 to Im \citep[][]{Lelli_2016}, and `ordinary' ETGs \citep[][]{LaBarbera_2010,Tortora_2012}. They measured the total density slope (within the one effective radius) using the mass-weighted density slope $\gamma_{\mathrm{MW}}$ \citep[cf.][]{Koopmans_2009,Dutton_2014,Tortora_2014} which is defined as:
\begin{equation}
\gamma_{\mathrm{MW}}=\frac{1}{M(r)}\int_{0}^r -\gamma(x) \rho(x) 4\pi x^2dx 
	\label{eq:MW_slope}
\end{equation}
where $M(r)$ is the total cumulated mass, $\rho$ the density and $\gamma$ the corresponding logarithmic slope. For a single power-law profile $\gamma_{\mathrm{MW}}$ equals $\gamma_{\mathrm{tot}}$, but generally $\gamma_{\mathrm{MW}}\geq\gamma_{\mathrm{tot}}$ holds true\footnote{The mass-weighted slope is essentially the analog of the volume averaged slope we introduced in eq.~\ref{eq:mean_slope} but weighted by the local density. We argue for the dark matter analyzed on its own the volume averaged slope $\eta$ is the safer choice (cf. Sec.~\ref{subsec:mass_slope_shape}), while for the total density $\gamma_{\mathrm{MW}}$ is more suitable.}.

Similar to ETGs (Fig.~\ref{fig:Total_slope_vs_dispersion}) the star-forming galaxies also become cuspier as the total stellar mass (or dispersion) is increased but, in contrast to the ETGs, they exhibit no clear correlation-break at higher masses. The total slopes and dark matter fractions we recovered for our dEs are in remarkable agreement with the dE sample analyzed by \citet{Tortora_2019}. Consequently, dEs of both samples are systematically offset compared to LTGs of the same stellar mass, with dEs having steeper slopes by about $\sim -0.5$ on average. The total slopes we measure are of course spherically averaged quantities. While we may expect both, ETGs and LTGs, to sit in relatively spherical DM halos (Sec.~\ref{subsec:halo_flattening}) at least the stellar density of LTGs can be expected to be flatter than that of dEs. Still, if we assume the worst case where both stars \textit{and DM} are distributed in an exponential disk profile, we expect their intrinsic circular speed curves $v_{c}(r)=\sqrt{GM/r}$ to always stay within 15\% of the purely spherical equivalent \citep[cf.][]{Binney_2008,Tortora_2019}. Compared to the large $\gamma_{\mathrm{MW}}$ difference between LTGs and ETG this effect is miniscule. 

dEs are also offset to LTGs in their dark matter fractions. Most dEs have a dark matter fraction around $30-40\%$. While the LTGs explore the full range of $\dmfrac$ their distribution is lopsided towards higher dark matter fractions. Particularly LTGs in the mass regime of the dEs have typical dark matter fractions of around $70-80\%$ \citep[see also][]{Sharma_2023_B}. In contrast to the LTGs, the two dE samples occupy a narrow range in $\gamma_{\mathrm{MW}}-\dmfrac$-space with both properties slightly correlated.

Considering DM halos are generally thought to be shallower than the stellar distribution (e.g. Sec.~\ref{subsec:mass_slope_shape}) one could argue that the shallower \textit{total} slopes of LTGs compared to dEs are a result of their higher dark matter fraction, whereas stellar differences in two morphologies play a subordinate role. However, it may not be as straightforward to draw conclusions from the total slopes alone. Both, $\dmfrac$ and $\gamma_{\mathrm{MW}}$, are measured within one stellar effective radius and are therefore aperture dependent, i.e. they depend on the concentration of the stellar distribution. However, at the same stellar mass the effective radii of LTGs may well be systematically larger than those of the dEs \citep[][]{Tortora_2019,Li_2019}, which may impede a consistent $\dmfrac$ and $\gamma_{\mathrm{MW}}$ comparison of dEs and LTGs at the same \textit{stellar} mass if the dark matter halo in LTGs is not similarly extended.

\subsubsection{The dichotomy in total slopes - A product of baryonic in- and out-flows?}
\label{subsubsec:tot_slope_formation}
As discussed in Sec.~\ref{subsubsec:tot_slope_massive_ETGs} the total slopes of our dEs are not in tension with the existing evidence \citep[][]{Poci_2017,Tortora_2019,Li_2019} for a \textit{dichotomy} in the early-type sequence below/above $M_{*}\approx10^{10}M_{\sun}$. It appears that below this mass threshold the slopes become continuously shallower at least until $M_{*}\approx10^{8.5}M_{\sun}$, whether and how this changes for even lower mass remains open for now. The breaking point of this dichotomy in the ETG sequence has long been known to be a crucial point of change in many other galaxy properties as well \citep[e.g.][]{Bender_1988_B,Bender_1989,Kormendy_1996,Kormendy_1999,Dekel_2006,Tortora_2010,Kormendy_2013,Cappellari_2013_B,Nelson_2018}. 

This dichotomy in the total slopes of early-type sequence and the offset (in slope and $\dmfrac$) of dEs compared to late-type dwarfs can be interpreted and explained by a coherent (but not necessarily comprehensive) picture of galaxy formation and processing. Depending on the galaxy's total (stellar) mass, several processes could lose or gain in significance such as to produce the observed dichotomy in density distributions. In the following we briefly summarize important effects for different masses with an emphasis on the low-mass regime occupied by the dEs: 

i) The \textit{high-mass regime} $M\gtrsim5\cdot 10^{10}M_{\sun}$ at the turnover/breaking point the ETGs are slightly sub-isothermal ($\gamma \sim -2.3$). ETGs with higher mass are increasingly a product of (multiple) \textit{dry} mergers, and as these dry merger stack up, the slope of the merger product gets more shallow and close to isothermal with $\gamma \sim -2.0$ \citep[][]{Remus_2013,Remus_2017}. This reverts the correlation of total slope and stellar mass and breaks the correlation that is found for lower masses. In parallel to this merger effect, the feedback of AGNs could play an additional driving factor in making the giant ETGs shallower than ETGs at the breaking point. This is because AGN feedback becomes more efficient in suppressing star formation at higher masses \citep[][]{Tortora_2010}.

ii) In the \textit{intermediate mass-regime}, near and around the turn-over point, the situation is likely complex as different mechanism are superimposed and possibly counter-acting each other. Nonetheless, we want to highlight a process that may be specifically crucial in this regime alone. This transition region of the above correlations is located where the \textit{most massive} of LTGs are found. These massive LTGs likely experienced continuous gas supply from their environment. This continued dissipative infall of baryons could be accompanied by an adiabatic \textit{contraction} of the dark matter halo \citep[e.g.][]{Blumenthal_1986,Gnedin_2004,Li_2022} which would lead to a cuspier halo and possibly also to an overall smaller $\gamma_{\mathrm{tot}}$ within $1\reff$. Since the total stellar mass traces not only a galaxy's ability to accrete and hold on to new gas but also its past accretion history we may expect the halo contraction to be more important for massive systems, which could support the anti-correlation of the total slope and $M_{*}$ in LTGs that was observed by \citet{Tortora_2019,Li_2019}.

iii) In the \textit{low-mass regime} (i.e. $M\sim 10^{9}M_{\sun}$), both dEs and LTGs, have similar light profiles and slopes (e.g. dEs have nearly exponential Sersic indices \citep[e.g.][]{Ferrarese_2006}, see also stellar distributions in \paperrefeSTARS). The similarity in stellar profiles implies that the offset in total slope between the early- and late-type dwarfs stems from a difference in their dark matter distributions. And this does indeed seem to be supported by the dynamically \textit{decomposed} halo densities: the core cusp tension usually is milder in early-type dwarfs than for late-types counterparts (cf. Sec.~\ref{subsec:core-cusp}). 

In opposition to the galaxies in the intermediate mass-regime the effect of adiabatic halo \textit{contraction} could be significantly diminished for the dwarfs. This is because as the total mass and, thus the potential well, is decreased continued gas accretion is dampened or shut-off early (e.g. by their environment) hindering the dark matter halo to become more cuspy. As a result, the dwarfs may not be able to reach the steep total slopes of $\approx-2.3$.

Conversely, to the reduced contraction adiabatic halo \textit{expansion} may be an important factor for dwarfs that drives the anti-correlation of mass and total slope in the low-mass regime: any outflow of mass, such as a gas outflow caused by supernova feedback by newly formed stars could force an initially cuspy DM halo to become more cored as time passes \citep[][]{Read_2005,Governato_2010,Pontzen_2012,de_Souza_2011,Pontzen_2014,Madau_2014,Read_2016}. The efficiency of this expansion mechanism is expected to decrease as the total mass of the systems is increased, again due to the ability of the potential well to hold mass outflows. Consequently, lower mass galaxies would become shallower than higher mass galaxies, which could produce the $\gamma_{\mathrm{MW}}$-$M_{*}$ correlation that is seen in the low-mass regime. 

In principle, this mechanism of adiabatic expansion could also explain the offset between quiescent and star-forming dwarfs that is observed in their slopes and dark matter fractions (Fig.~\ref{fig:MW_Total_slope_vs_stellar mass}). The star-forming LTGs may have had a more gradual, but prolonged, star formation history as they continuously processed their surrounding gas reservoir unaffected by the low-density environments these galaxies typically inhabit. The continued supernova outflows that result from this star formation could have gradually and gently expanded the galaxy's size (reducing its surface density, increasing $\reff$ and therefore also $\dmfrac$), while overall making the density distributions shallower (directly for baryons and indirectly for the DM distribution). 

In contrast, for the quiescent dEs adiabatic expansion may have played less of a role because their star formation was being quenched at some point in their past. A manifold of potential quenching mechanisms responsible for this have been suggested (see \paperrefeSTARSSP for more discussion). Among them are \textit{external} processes induced by the environment (e.g. ram-pressure stripping, galaxy starvation, and harassment \citep[e.g.][]{Gunn_1972,Larson_1980,Lin_1983,Moore_1998}) and \textit{internal} processes like outflows caused by rapid supernovae or AGN feedback \citep[e.g.][]{Dekel_1986,Silk_2017,sharma_2023}. These mechanisms can be expected to act within different epochs and on very different timescales, for example ram-pressure stripping is expected to act on a few hundreds of Myrs to a Gyr \citep[][]{Quilis_2000,Steinhauser_2016}. 
If the quenching mechanism that is responsible for the dEs was \textit{fast} acting and \textit{shortly} after galaxy formation then adiabatic expansion could have been suppressed for the dEs leaving their density distributions more cuspy and their effective radii smaller. 

However, as established and discussed in \paperrefeSTARSSP (unlike the more luminous, `ordinary' ETGs) many of the dEs likely had a complex star formation history, possibly involving multiple rapid star formation bursts. Depending on their initial conditions (environment and total mass) some dEs were quenched shortly after the reionization epoch while others were able to form stars up until just a few Gyrs ago. Therefore, if adiabatic expansion is responsible for the shallower densities of star-forming dwarfs compared to the dEs, we may also expect that (within our sample) those dEs that have had prolonged SFH also are more shallow than dEs that were quenched prematurely (e.g. by a single burst). This could reveal itself in a correlation of density slopes with markers of SFH (see next section).

Given galaxies ab initio reside in halos with identical profiles and shapes, but the amount of sustained star formation evolves them in different ways we expect to find the clearest imprint of this correlation in the slopes of the dark matter (not the total density) as the halo's slope traces the gravitational impact of mass outflows largely independent of the stellar morphology. 

The dark matter fraction $\dmfrac$ and total slopes as defined above are aperture dependent measurement because they are measured at $1\reff$. This links stellar and DM slopes indirectly and impairs an objective comparison of their halos. For example, as stated above LTGs tend to have higher $\reff$ at the same mass, consequently, for halos of LTGs the dark matter fraction is measured at larger radii. Therefore, even if star-forming and quiescent dwarfs \textit{currently} reside in \textit{identical} halos, one may measure very different slopes and dark matter fractions. To eliminate these aperture-effect from the analysis, it may be preferable to analyze the dark matter halos within a physically constant aperture, which of course is only sensible if the galaxies are roughly in the same mass regime. Therefore in the following section we will again analyze the halo in the distance-independent $0.8 \rm kpc$ aperture we introduced in Sec.~\ref{sec:CDM-Discussion}. This allows us to assess the dark matter densities independent of the stellar distribution they host and enable a more objective evaluation of the halos formation and subsequent evolution.

\subsection{What makes dE halos cored or cuspy?}
\label{subsec:DM_correlations}
The last sections showed that dEs tend to have shallower total slopes than `ordinary' ETGs, but still steeper ones than star-forming dwarfs of the same total stellar mass. To entangle whether these disparities are due to differences in the DM distributions or differences in the stellar distributions, we analyze the dynamically decomposed DM halos of our dE sample in an aperture that is independent of the stellar extent. One of the goals of this section is to explore whether the halos are initially universal but evolve over time due to their interactions with baryons and environment (Sec.~\ref{subsubsec:DM_evolves}) or, conversely, whether the initial condition of the DM halo determine the morphological evolution of the stars/baryons while the halo remains largely unchanged (Sec.~\ref{subsubsec:DM_init}). 

\subsubsection{Do the halos evolve due to internal feedback?}
\label{subsubsec:DM_evolves}
As considered in Sec.~\ref{subsubsec:tot_slope_formation} a plausible explanation for the observed slopes are baryonic in- and out-flows (e.g. by supernova feedback). In Sec.~\ref{subsec:core-cusp} we discussed that the DM slopes of quiescent dwarf samples (dEs/dSphs) show considerable diversity: some are relatively cored, some are cuspier. Similarly, the periods of active star-formation are expected to vary significantly across different dEs. Depending on their initial conditions, some dEs were quenched shortly after halo formation while others likely had a complex, prolonged SFH (see \paperrefeSTARSSP or \citealt{Seo_2023,Romero_Gomez_2023_B}). If adiabatic expansion by supernova feedback is the predominant factor in evolving the DM slopes of the dEs, we expect to see a correlation of the halo properties with their stellar population properties. Assuming the dEs started out in an initially universal halo density, then the DM halos that host young, metal-rich populations (implying they were quenched recently) with a low $\alpha$-abundance\footnote{A lower $\alpha$-abundance (or equivalently the [Mg/Fe] we measured) is a proxy for a prolonged SFH \citep[e.g.][]{Romero_Gomez_2023_A,Romero_Gomez_2023_B}.} should be more cored and less dense in the center.

\begin{figure*}
	\centering
	\includegraphics[width=1.0\textwidth]{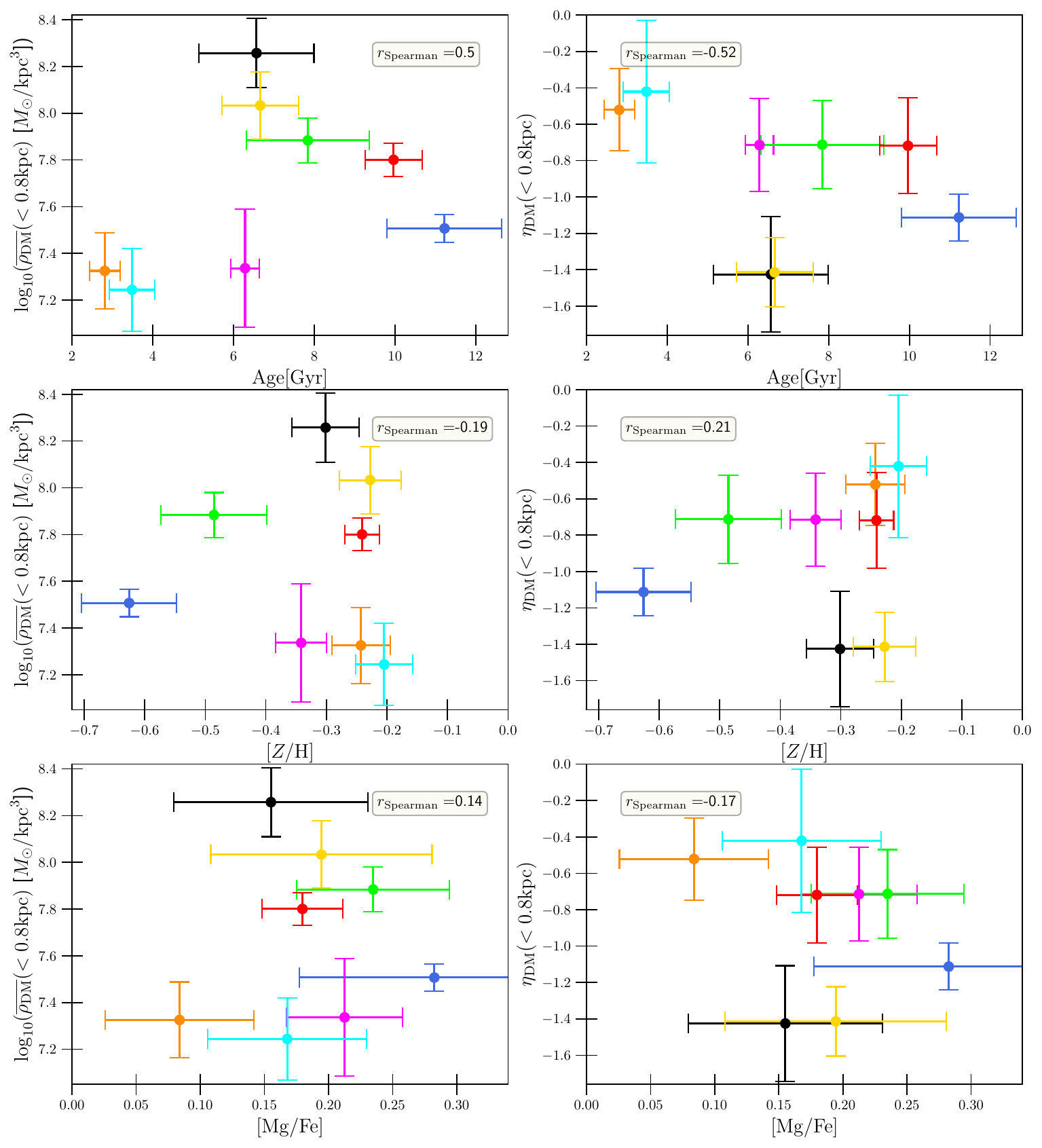}
    \caption{Correlations of the DM halos with stellar population parameters. In the small boxes of each panel, we show the corresponding Spearman correlation coefficient. \textit{Left panels:} The average DM density $\overline{\rhodm}$ (eq.~\ref{eq:mean_density}) inside the sphere of radius $0.8 \rm kpc$. \textit{Right panels:} The average DM slope (eq.~\ref{eq:mean_slope}) inside the same sphere. \textit{Top} to \textit{Bottom}: The age of the stellar population. The metallicity $[Z/\rm H]$. A low abundance ratio [Mg/Fe] is an indicator for a prolonged star formation history (cf. \paperrefeSTARS).}
    \label{fig:correlation_DM_POPULATION}
\end{figure*}

However, as evident in Fig.~\ref{fig:correlation_DM_POPULATION}, we find only little to no correlation of the volume averaged DM densities and slopes with the stellar population parameters we derived in \paperrefeSTARSSP \footnote{In \paperrefeSTARSSP we derived spatially resolved populations from spectra that were binned in two annuli centered around $r=2.5\arcsec$ and $r=7.5\arcsec$. Shown in Fig.~\ref{fig:correlation_DM_POPULATION} are the results for the central aperture at $r=2.5\arcsec \approx 0.2 \rm kpc$. We omitted VCC 1910 because its SSP properties are not very trustworthy due to the badly constrained $H\beta$ feature (cf. \paperrefeSTARS).}. The dEs exhibit no correlation with metallicity and $\alpha$-abundance, and therefore an extended SFH. Only the stellar age is tentatively anti-correlated with the DM slope $\eta_{\mathrm{DM}}$. Qualitatively this age--slope anti-correlation is in agreement with the supernova feedback scenario (younger dEs were quenched recently and possibly were able to form more stars before that), however, the lack of correlation with the chemical enrichment parameters seems to indicate that adiabatic expansion by supernovae feedback only plays a minor role for the DM density distributions. Star formation seemingly does not affect the average halo density and, if any, only very moderately makes them more shallow over time. This may not be a surprise because dEs are among the most massive `dwarf' galaxies which could make their halos more resilient to internal feedback, however, even the much smaller Milky Way dSphs exhibit no clear correlation with SFH either \citep[][]{Hayashi_2020}. In conclusion, if differences in SFH can not explain the diversity of DM slopes that observed in quiescent dwarfs then it may be the case that the halos are not universal from the start (or very early on).

\subsubsection{Are halo profiles and shapes universal or a product of the primordial conditions?}
\label{subsubsec:DM_init}
If the variance of DM slopes in dEs is not a product of fluctuating degrees of internal feedback, then the halos may be different ab initio \citep[e.g.][]{Ascasibar_2004}, i.e. shortly after the gravitational halo collapse but before the majority of their stars has formed. While we lack direct information about the halo structure of the dEs at high redshifts (i.e. shortly after formation) we can investigate correlations with properties that we expect to still hold information about the initial collapse conditions. 

We may assume that the current position the dEs inhabit in Virgo is a reflection of their environment at halo formation. Halos that formed in the cluster outskirts may have only recently fallen into the cluster as they formed later, and/or formed in isolation within a lower background density. 

Similarly, if stellar orbits in dEs were only mildly heated during their evolution (see \paperrefeSTARS), then the \textit{stellar} angular momentum measured today could still hold information about the momentum of the primordial gas disk the stars formed in. And this itself is a proxy for the angular momentum\footnote{While we measured most dEs to have near spherical halo shapes (Sec.~\ref{fig:dE_flattening}) $\qdm$ is likely not a good gauge for the non-radial motions of DM particles in the primordial stage: We measure $\qdm$ `today' (i.e. after baryonic feedback from AGNs or SF) and only within the central one effective stellar radius (the halo is much more extended and could change shape).} of the dark matter halo \citep[][]{Fall_1980,Fall_1983,Romanowsky_2012}.

\begin{figure*}
	\centering
	\includegraphics[width=1.0\textwidth]{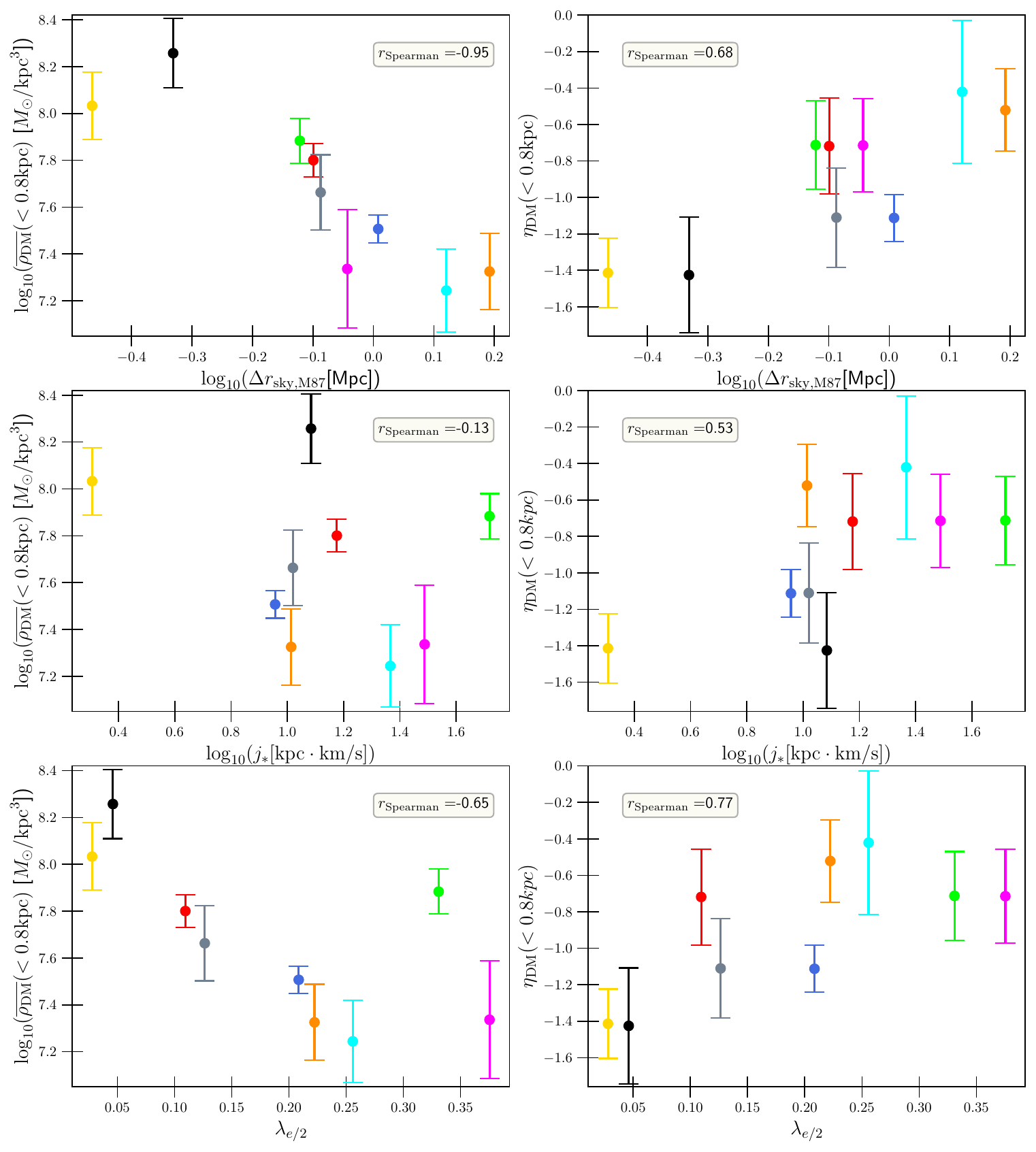}
    \caption{As Fig.~\ref{fig:correlation_DM_POPULATION} but for correlation of DM with cluster environment and \textit{stellar} angular momentum. \textit{Top} to \textit{Bottom}: The projected distance to the central cluster galaxy M87 $\centredis$ (a proxy for the environment the galaxies inhabit, see \paperrefeSTARS). The total stellar angular momentum $j_{*}$. The stellar angular momentum parameter $\lambda_{e/2}$ (see \paperrefeSTARSSP for details) within the half the effective radius, which is not to be confused with the often used halo spin parameter \citep[][]{Peebles_1969}.}
    \label{fig:correlation_DM}
\end{figure*}

In Fig.~\ref{fig:correlation_DM} we show the correlations of the DM with the cluster environment and stellar angular momentum that we obtained in \paperrefeSTARS. The environment is described using a proxy parameter: the projected distance to Virgo's central cluster galaxy M87. Galaxies that are in Virgo's center (roughly where M87 is located) are expected to have experienced longer and stronger interactions with their environment. We either expect them to have formed very early on and enter the cluster, or they formed in-situ directly within the high density background of the cluster's halo. We describe the stellar angular momentum using the approximate total specific angular momentum $j_{*}=J_{*}/M_{*}$ and the stellar angular momentum parameter $\lambda$ within $0.5\reff$ (see \paperrefeSTARSSP for details on the calculation).

Conversely to the stellar population constraints, we find moderate correlations with the angular momentum and strong correlations with the environment. Galaxies in the cluster center have about $1\rm dex$ higher DM densities than galaxies in the Virgo outskirts. Furthermore central dEs are super-NFW ($\eta\sim -1.4$) whereas galaxies in the low-density outskirts of Virgo are almost as cored ($\eta\sim -0.5$) as what is found for LTGs (typically $\eta\sim -0.3$). Similarly, the halos of galaxies that have low stellar angular momentum tend to be more dense and cuspy\footnote{This correlation does not necessarily mean causation, in \paperrefeSTARSSP we discussed the correlation of angular momentum with environment due to dynamical heating. Perhaps the halo profiles solely depend on the environment, but the baryons were also dynamically heated by interaction with the environment, resulting in the observed correlation of angular momentum and DM.}. Only the correlation of average DM density with $j_{*}$ is relatively ambiguous, but this is mostly driven by VCC 856 (green) and VCC 1261 (black)\footnote{The classification of VCC 856 as a dE may be debatable as it could be a face-on spiral resulting in higher $j_{*}$ (cf. \paperrefeSTARS). VCC 1261 is actually known to have very little to no rotation within the FoV in our data or in literature \citep[][]{Geha_2003,van_Zee_2004,Chilingarian_2009,Rys_2013,Toloba_2014,Toloba_2015,Sybilska_2017,Sen_2018}, but it is also the brightest dE ($j_{*}$ increases scales magnitude) and has the largest effective radius which could bias the coarse approximation of the total $j_{*}$ high compared to the other dEs (cf. \paperrefeSTARS).} 

These angular momentum correlations may be understood as an imprint of the joint gravitational collapse of the DM and baryons. When a DM overdensity forms it acquires baryons due to its gravitational pull, these baryons condensate and infall towards the halo center, as a response the halo undergoes contraction \citep[][]{Blumenthal_1986,Gnedin_2004}. However, this steepening of the halo profile during the formation phase is counter-acted by processes that make the halo more cored and less dense: i) Baryons may not be acquired smoothly, particularly in the primordial gas-rich disks the baryons fragment and form larger clumps \citep[][]{Immeli_2004,Aumer_2010,Ceverino_2010}. These clumps undergo dynamical friction and, in the process, heat up and expand the DM halo \citep[][]{El-Zant_2001,Del_Popolo_2009,Inoue_2011}. ii) DM and baryons are not accreted radially but obtain angular momentum from tidal torques \citep[][]{Peebles_1969}. Depending on the strength of these torques, a halo and its primordial gas disk will have high or low angular momentum. Halos with a higher angular momentum have been linked to a flatter density profile, lower surface density and larger extent/size in both, baryons and DM \citep[e.g.][]{Ascasibar_2004,Williams_2004,Del_Popolo_2009,Del_Popolo_2012,Kim_2013}. 

These mechanisms may be the cause for the correlations we found for the halo slopes with angular momentum (Fig.~\ref{fig:correlation_DM}). They could also explain the cuspier slopes (total+DM) and smaller effective radii of the dEs when compared to LTGs (Fig.~\ref{fig:MW_Total_slope_vs_stellar mass}) since the angular momentum of dEs is suppressed in comparison to LTGs (see \paperrefeSTARS). Furthermore the reduced baryonic surface density that comes along with the higher halo spin could be linked to a higher observed dark matter fraction \citep[][]{Sharma_2023_B}.

At first glance, this argument seems counter-intuitive when combined with the observed environment correlation. If the angular momentum grows via tidal torques then one may expect that halos in the cluster center have (on average) experienced more tidal interactions with neighboring structures and, as such, have higher angular momentum and shallower DM slopes (i.e. opposite to the observed $\eta_{\mathrm{DM}}-\centredis$ correlation). But perhaps the assumption that dwarfs in the cluster center should have experienced more tidal torques is an oversimplification and misses the complexity of the problem.

DM halos acquire angular momentum from tidal torques mostly \textit{during} the linear phase of gravitational assembly \citep[][]{Porciani_2002,Lopez_2019}. During the comparatively short time of a halo's assembly epoch (i.e. at a higher redshift) the environmental circumstances may have been very different when compared to the state of Virgo at $z=0$. If the halos assembled at very different formation epochs, their environment and, thus, their ability to acquire angular momentum may change as well. Further complications arise because halos that assembled at early epochs are expected to be denser on average due to a higher background density (see below) and because the geometry of the large-scale surroundings is anisotropic (structures grow along filaments) which affects the angular momentum acquisition of the galaxy halos \citep[e.g.][]{Codis_2015}. Perhaps the duration of halo assembly also varies with the environment and formation epoch. In a low density background DM halos may collapse on different timescales, allowing the halos to gradually acquire angular momentum. 

In conclusion, while currently not obvious if and by how much, we may presume that angular momentum acquisition depends in a complex manner on a halo's assembly epoch and the geometry of its surroundings at the time. As discussed in \paperrefeSTARSSP and Sec.~\ref{subsubsec:DM_evolves}, for the dEs the derived stellar population ages are likely not a good estimator of the halo's formation epoch because many of them probably had extended periods of star formation. Therefore we rely on alternatives to gauge when the dE halos have assembled. Such an alternative is the average DM density we derived above: $\overline{\rho_{\mathrm{DM}}}$ is a proxy for the assembly epoch of a halo because it is partly inherited from the average density of the Universe at the time of assembly, as shown with collapse models and N-body simulations \citep{Gunn_1972,Wechsler_2002,Springel_2005,Gao_2005}. Consequently, galaxies that have been assembled in an early epoch where the average background density of the Universe was higher are also expected to have denser halos than galaxies that formed during later stages of the Universe. While the proxy $\overline{\rho_{\mathrm{DM}}}$ may be superimposed and affected by the manifold of processes (e.g. feedback from prolonged SFH, adiabatic contraction, etc.) the above discussion suggests these effects play a secondary role. They may affect the slopes to some degree, but they are unlikely to be able to reduce the DM density by $\sim1 dex$.

As evident from Fig.~\ref{fig:correlation_DM} the average density is most strongly correlated with the environment, which is very compatible with the assumption that the assembly epoch (or more generally the average background density at formation) is the primary determinant of $\overline{\rho_{\mathrm{DM}}}$. Therefore dEs with a higher $\overline{\rho_{\mathrm{DM}}}$ have formed in a higher density background, e.g. `in-situ' within Virgos halos and/or when the Universe was young and more dense in general which gave them enough time to sink to Virgo center as seen today. Conversely, the dEs currently located in the outskirts have lower $\overline{\rho_{\mathrm{DM}}}$ as they have formed in a lower density background (in isolation) and/or in a more recent epoch such that they only now enter Virgo proper.

In this context, the correlations in Fig.~\ref{fig:correlation_DM} suggest that dEs which have formed early (or in-situ) were less efficient in acquiring angular momentum than dEs that have formed more recently or in isolation\footnote{In terms of stellar population age VCC 200 (dark blue in Fig.~\ref{fig:correlation_DM_POPULATION}) has formed shortly after reionization (12 Gyrs) yet has moderate $\overline{\rho_{\mathrm{DM}}}$, therefore we may suspect that the local background at formation (in-situ or isolation) and not epoch plays the leading role.}. 

\begin{figure}
	\centering
	\includegraphics[width=1.0\columnwidth]{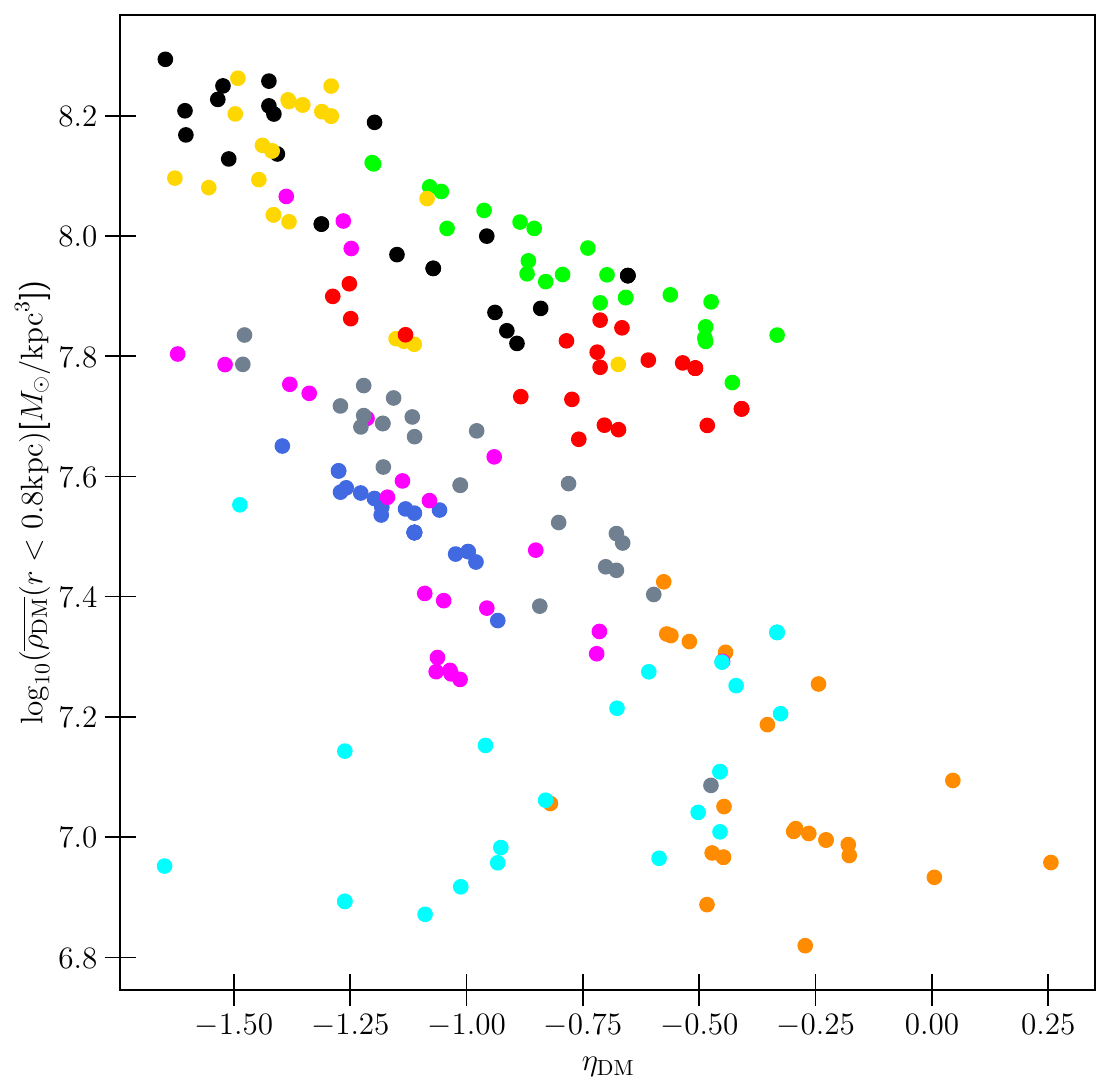}
    \caption{Correlation uncertainty of the average enclosed DM density $\overline{\rho_{\mathrm{DM}}}$ with slope $\eta_{\mathrm{DM}}$. Shown are the 25 best dynamical models we found for each dE.}
    \label{fig:rho_vs_eta}
\end{figure}

Fig.~\ref{fig:correlation_DM} suggests that the average central DM density \textit{and} the central slope are both correlated with the environment. However, $\overline{\rho_{\mathrm{DM}}}$ and $\eta_{\mathrm{DM}}$ are not entirely \textit{independent} properties: a steeper central slope also entails higher average central DM density and vice versa. Therefore it is not obvious that the two environment correlations with $\overline{\rho_{\mathrm{DM}}}$ and $\eta_{\mathrm{DM}}$ are in fact two independent correlations or just an imprint of one and the same correlation combined with a correlated measurement uncertainty. We explore this possibility in Fig.~\ref{fig:rho_vs_eta} which shows $\overline{\rho_{\mathrm{DM}}}$ vs $\eta_{\mathrm{DM}}$ for the 25 best dynamical models for each dE (i.e. the number of models we used to gauge the errors). As expected, the errors in $\overline{\rho_{\mathrm{DM}}}$ and $\eta_{\mathrm{DM}}$ are correlated: if the central slope scatters to steep values, the average central density also often tends to be higher. However, overall, the scatter for a given dE is significantly smaller than the differences between the dEs in our sample. In other words, the dynamical measurements uncertainties are not enough to explain the correlations we find in Fig.~\ref{fig:correlation_DM_POPULATION} and Fig.~\ref{fig:correlation_DM}. And we conclude that the correlations of $\overline{\rho_{\mathrm{DM}}}$ and $\eta_{\mathrm{DM}}$ are two independent and physically meaningful findings.

\subsubsection{Has the environment disturbed the halos after assembly?}
\label{subsubsec:environment}
In Sec.~\ref{subsubsec:DM_evolves} we discussed the effects of \textit{internal feedback}, but we only found tentative trends with SFH indicators. The halos may become slightly more shallow if SF is active for a long time, but this supernova feedback scenario is unable to explain the larger differences in $\overline{\rho_{\mathrm{DM}}}$ and, consequently, the strong $\overline{\rho_{\mathrm{DM}}}$-$\centredis$ correlation. Instead of a variability in primordial conditions (Sec.~\ref{subsubsec:DM_init}) an alternative cause for the DM-environment correlations could be tidal interactions with other cluster members and interaction with the intra-cluster medium that changed the halos \textit{after} they assembled. This could happen by regulating star-formation (and therefore internal feedback) or directly by tidally induced mass in- and out-flows.  

In principle, the $\overline{\rho_{\mathrm{DM}}}$-$\centredis$ and $\eta_{\mathrm{DM}}$-$\centredis$ correlations could be a result of quenching via ram-pressure stripping (RMS) due to it regulating SF. We expect prolonged SFH to gradually make halos more cored and reduce $\overline{\rho_{\mathrm{DM}}}$ (Sec.~\ref{subsubsec:DM_evolves}), but RMS stops this process. If the dEs in Virgo's outskirts were only recently being quenched by RMS as they entered Virgo they may have had more time between assembly and quenching to hollow out their halo, producing the correlations with $\centredis$. In this scenario it is not the process itself (RMS) that drives this but the ceasing of internal feedback \footnote{While RMS may displace dark matter \citep[][]{Smith_2012_B} in the initial quenching phase as the outflowing gas is being stripped, its long-term impact on the slopes and halo density is probably low. Since all dEs are quenched by now, RMS would have affected all dEs equally and an environment correlation wouldn't remain.}. However, we found little evidence that internal feedback is particularly important (Sec.~\ref{subsubsec:DM_evolves}). Furthermore we may expect a significant fraction of dEs to be quiescent even before entering Virgo for the first time due to pre-processing in the groups they arrive in \citep[][]{Bidaran_2022,Romero_Gomez_2024}. Therefore, if SFHs impacts the dE halos, we would expect milder halo-environment correlations as those group dEs would have retained their DM distribution.

Unlike ram-pressure stripping, we may suspect other environment processes to affect the slopes of the dark matter more directly. Examples include tidal harassment by larger galaxies, and mergers with galaxies of smaller or similar size. Galaxy harassment removes both stars and dark matter from the potential well and dynamically heats their orbits. However, the dark matter correlations (Fig.~\ref{fig:correlation_DM}) stand in exact opposition to this scenario, since tidal removal would be expected to reduce the dark matter density and flatten the halo. 

Galaxy mergers violently change the mass structure and orbits of dark and baryonic matter. Wet mergers are unlikely in the cluster's center, but could play a role before the progenitors of the dEs enter Virgo by inducing increased star-formation and rapid bursts. As is the case with tidal harassment though, this effect would be converse to our finding of more dense/cuspy halos in Virgo's center where the dry merger-rate is expected to be higher \footnote{There are some simulations (for less massive dwarfs) that suggest that dry \textit{major} mergers can sometimes make halos \textit{cuspy} by importing additional dark matter to the center \citep[][]{Laporte_2015,Orkeny_2021} but this depends on initial conditions and would likely be a stochastic effect unable to produce the strong environment correlations.}.  

Altogether the direct impact of the environment on the dEs halos \textit{after} assembly should be negligible, if anything it affects the halos indirectly by regulating their star formation and as a consequence internal feedback. 

\subsection{The formation of dEs - Are dEs transformed remnants of spirals?}
\label{subsec:DM_formation}
The synthesis of Sec.~\ref{subsec:DM_correlations} is that the diversity in the halo distribution of dEs are largely a result of the primordial conditions during halo assembly, but modified (to a lesser degree) by the secular evolution caused by the internal SF-feedback the galaxy was able to maintain before it was quenched. By comparing the halos of different morphological types with one another, we may be able to infer how dEs have arrived in their current quiescent and homogeneous stellar state (cf. \paperrefeSTARS). 

The prevalent formation scenario is that dEs are the remnants of late-type progenitors that were, at some point in their evolution, quenched by some internal or external process (see discussion Sec.~\ref{sec:intro} and \paperrefeSTARS). However, the results discussed in Sec.~\ref{subsec:DM_correlations} suggest that the DM (especially $\overline{\rho_{\mathrm{DM}}}$) is unaffected by the quenching mechanism itself and only mildly affected by feedback from its SFH. Consequently, if dEs have formed from late-type progenitors, we would expect their DM structure to be relatively robust and comparable to that of spiral dwarfs that were able to avoid quenching. In other words, differences in SFH may have made the LTG dwarfs slightly more cored on average (see Sec.~\ref{subsec:core-cusp}) but we would expect $\overline{\rho_{\mathrm{DM}}}$ to be similar as it is mostly determined by the background density during halo assembly.

\begin{figure*}
	\centering
	\includegraphics[width=1.0\textwidth]{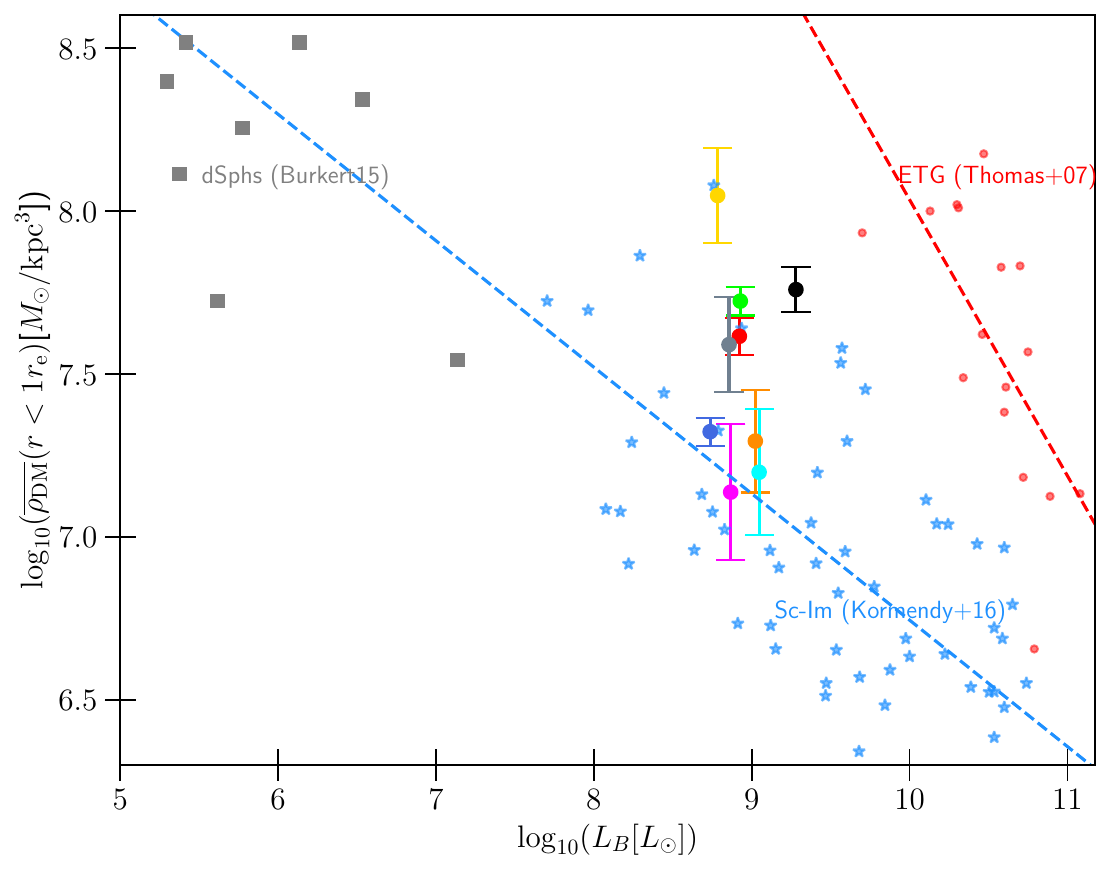}
    \caption{The B-band Luminosity vs average DM density within an aperture of 1 effective radius for different galaxy morphologies. \textit{Colored with errorbars:} Our dE sample. The effective radii we used to calculate $\overline{\rhodm}$ are in the z-Band and tabulated in \paperrefeSTARS. \textit{Red points:} Large ETGs in the Coma cluster as measured by \citet{Thomas_2009}. Following \citet{Thomas_2009} we obtained their effective radii from \citet{Jorgensen_1995,Mehlert_2000} and assumed a Coma distance of 100 Mpc. \textit{Yellow points:} Sc-Im galaxies from \citet{Kormendy_2016}, this sample includes galaxies in different groups within the local volume of 84 Mpc. \textit{Gray squares:} The dSphs of the Milky Way from \citet{Burkert_2015}. \textit{Dashed Lines:} For the LTG and ETG samples we also show fitted scaling relations in the same color. For the samples of \citet{Burkert_2015} and \citet{Kormendy_2016} we plot the \textit{core densities} as they are stated in these studies instead of recalculating the averaged density within the $1\reff$ aperture. This is a good approximation of $\overline{\rhodm}$ because these studies all use halo models with a central core (with minor differences in the model definitions) but find core radii that are considerably larger than one stellar effective radius. Therefore, if we were to correct for the small radial decrease within $1\reff$ it would shift the density only slightly to lower values which would not change our conclusions (see text). Between different studies we expect the systematic differences/uncertainties in the measurement of $\reff$, and in the probed candidate halo models (Sec.~\ref{sec:technique}) to have more of an impact. The dEs do not follow the scaling relations suggested by large ETGs. While they are closer to the DM densities of Sc-Im galaxies, which (presumably) are the progenitors of dEs, they are offset to higher densities. This could indicate that the progenitors of dEs have assembled earlier and/or in a higher density background.}
    \label{fig:scaling_B_band_1reff}
\end{figure*}

Fig.~\ref{fig:scaling_B_band_1reff} shows the average DM density of our dEs compared to galaxies of various morphological types: `ordinary' ETGs in the Coma cluster \citep[][]{Thomas_2009}, dSphs of the Milky Way \citep[][]{Burkert_2015}, and the presumed progenitors of dEs: Sc-Im galaxies \citep{Kormendy_2016}. Since we now compare galaxies spanning a large range of magnitudes and effective radii, we have opted to calculate $\overline{\rhodm}$ within $1\reff$ instead of the size-independent aperture we used in the last sections. To compare the different samples consistently, we decided to use the total, extinction-corrected luminosities in the B band throughout. 

Similar scaling relations of the average (central) DM density with luminosity have been investigated before and both, ETGs and LTGs, were found to have DM densities that are anti-correlated with luminosity, though the scaling relation for the ETG population is offset by $\sim+1 \rm dex$ suggesting an earlier halo assembly consistent with their older stellar populations \citep[cf.][]{Gerhard_2001,Thomas_2009}. In contrast, the quiescent dSphs of the Milky Way fall on the extension of the scaling relations of larger LTGs \citep[][]{Kormendy_2016}, i.e. they are compliant with the formation scenario that they are the remnants of dIm galaxies. These previous results are reproduced in Fig.~\ref{fig:scaling_B_band_1reff} as indicated by the dashed lines that are fitted scaling relation to the ETG and Sc-Im galaxies. While the luminosity range of our small dE sample is too narrow to establish a similar scaling relation for dEs we can compare our galaxies to the existing scaling relations.

The $\overline{\rhodm}$ scatter within our dE sample stays within $1 \rm dex$ which is comparable to the scatter that was found within the individual literature samples. As argued in Sec.~\ref{subsec:DM_correlations} we deem this scatter to be the genuine variety in halo distributions with assembly environment, age and feedback. The dEs follow neither the LTG nor the ETG scaling relations well. The extension of `ordinary' ETGs would predict much higher densities for the dEs. The comparison with the presumed progenitors of the dEs, the Sc-Im galaxies, is less obvious. While they all fall inside the distribution of LTGs of similar luminosity, the dE sample average is offset higher by $\sim+0.5\rm dex$. This $0.5\rm dex$ offset is based on a comparison of LTGs and ETGs at the \textit{same} luminosity $L_{B}$. But this does not account for the fact that LTGs will also be offset towards lower stellar mass-to-light ratios due to their younger populations. Consequently, if we were to compare the LTGs at the \textit{same} stellar mass, this offset between dEs and Sc-Im would only increase further. Therefore a comparison at the same stellar mass would increase the following conclusions, however, considering the luminosity is a more straight-forward measure and not depend on correct mass-to-light ratios, we decided to compare versus the luminosity in Fig.~\ref{fig:scaling_B_band_1reff}.

Generally speaking, we expect the position of a galaxy in the $\overline{\rhodm}-L_{B}$-space to be determined by the assembly conditions (i.e. formation epoch, environment, angular momentum, ...) and the subsequent evolution (e.g. SF-feedback, tidal interaction, ...). Since baryons and dark matter are usually lost/expelled by the latter \citep[e.g.][]{Dekel_1986} such evolutionary processes shift galaxies diagonally to lower $\overline{\rhodm}$ and $L_{B}$, i.e. it could move dEs closer to the LTG scaling relations. However, our results suggest that, at least in the mass regime of our dEs, $\overline{\rhodm}$ is only marginally affected by the latter (Sec.~\ref{subsec:DM_correlations}) despite the fact that some of the dEs seemed to have formed stars actively over a very prolonged period. For the dEs to fall onto LTG relations much more feedback would be needed and it is not obvious why Sc-Im galaxies in the same luminosity/mass regime should not be affected similarly to the dEs. 

\begin{figure*}
	\centering
	\includegraphics[width=1.0\textwidth]{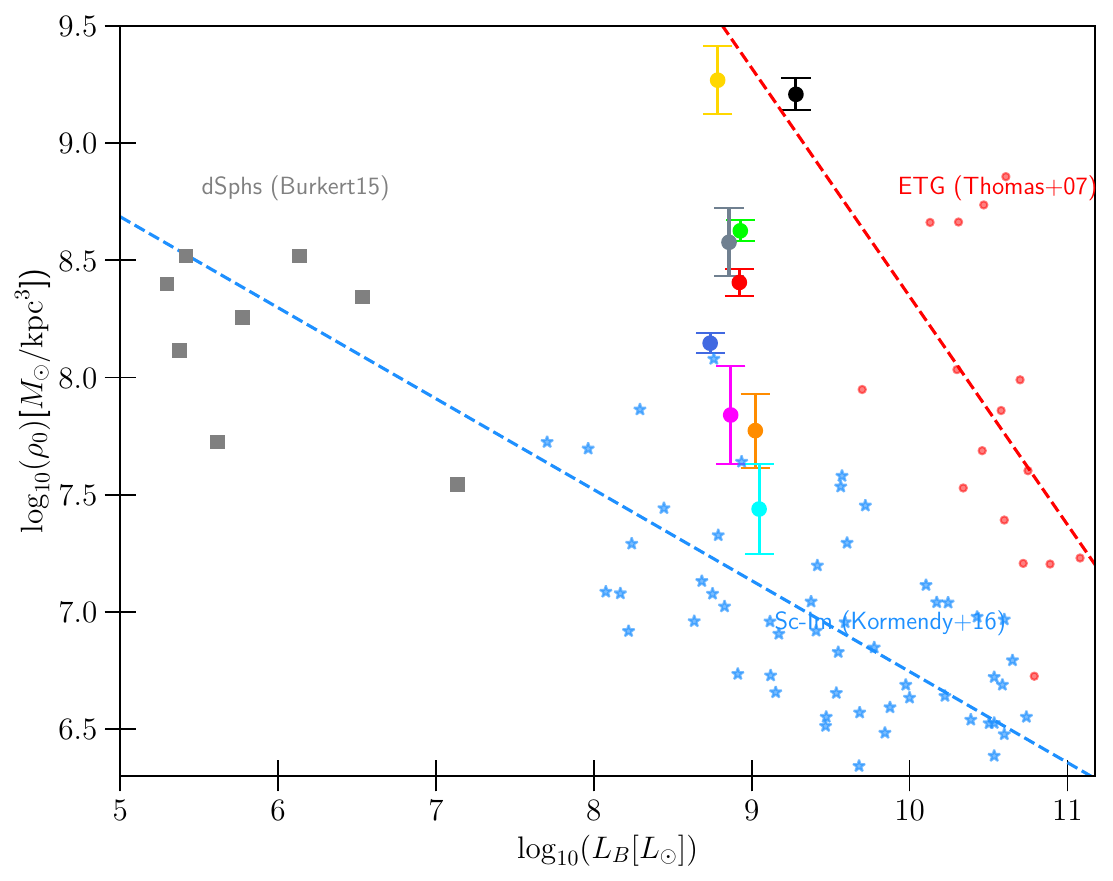}
    \caption{As Fig.~\ref{fig:scaling_B_band_1reff} but for the \textit{central} DM density $\rho_{0}$ (i.e. the maximum of the DM density). For the dEs $\rho_{0}$ is measured as the average density within the central $100\rm pc$, for the massive ETGs \citep[][]{Thomas_2009} within $1 \rm kpc$. The dSphs and Sc-Im galaxies remain unchanged relative to Fig.~\ref{fig:scaling_B_band_1reff} because in both figures we show the parameter $\rho_{0}=\rho(r=0)$ of the cored halo profiles that these studies employ to model the halos.}
    \label{fig:scaling_central_DM}
\end{figure*}

The stripping scenario of dEs can, in principle, `fake' higher average DM densities for dEs within $1\reff$ as we may expect it to change the stellar distribution (and indirectly the DM distribution) and thus affect the aperture we measure $\rho_{\mathrm{DM}}$ in. To exclude this possibility, we can compare the DM density in the very center of the halo $\rho_{0}=\rho_{\mathrm{DM}}\left(r=0\right)$. This is because we expect the central density to be less susceptible to external influences and systematic aperture differences between the LTG and dEs galaxy types. As Fig.~\ref{fig:scaling_central_DM} shows the offset of the dEs becomes even larger when comparing the central DM densities which is no surprise because comparatively the Sc-Im have relatively cored halo profiles. This confirms external influences like stripping are not able to explain the \textit{higher} DM density of the dEs. Therefore it is not obvious how an Sc-Im galaxy with a lower $\rho_{0}$ could be transformed to a dE with a higher $\rho_0$.

Considering all of the above reasons, we suspect the offset between Sc-Im and dE galaxies is a genuine difference in their DM density. In that case, the offset suggests two physical (and likely correlated) reasons: i) dEs have assembled in a higher density background, because they were formed at a higher redshift and/or in-situ in a cluster's dense environment. ii) Internal feedback (Supernovae, AGNs, etc.) was different (stronger) in the LTGs we see today which has shifted them away from the dEs\footnote{The direct impact of external/tidal effects must also be negligible because it would have affected the dEs more than the more isolated LTGs \citep[][]{Dressler_1980}. This is congruent with our findings in Sec.~\ref{subsubsec:environment}.}. If LTGs were progenitors of dEs this would imply that the strength of internal feedback must have increased the later a galaxy has assembled, which poses the question of why LTGs seen today (which reside in lower density halos) were not self-quenched yet\footnote{If dwarfs in this luminosity regime are generally not self-quenched, the dEs we observe must have been quenched solely by the environment (cf. \paperrefeSTARS) but this is challenged by the existence of isolated dEs in the field \citep[e.g.][]{Janz_2017,Paudel_2023}.}. 

\subsection{Distinct formation channels of today's quiescent and late-type dwarfs.}
\label{subsec:formation_channels}
In either case, i) or ii), the Sc-Im galaxies observed today are not directly comparable to the supposed progenitors of dEs because $\overline{\rhodm}$ can not be increased after assembly. Many theories of galaxy evolution attempt to explain the links/similarities between late-type dwarfs and dEs using processes (e.g. RMS stripping or SF feedback) that transform $z=0$ late-types into $z=0$ dEs. But as already stated by \citet{Skillman_1995}, perhaps this is "asking the wrong question" as such theories assumes that the progenitors of dEs are comparable to today's Sc-Im galaxies. Often this makes these theories struggle to explain the co-existence of dEs and Sc-Im within the local Universe. Combined with the findings of \paperrefeSTARS, which suggest that dEs have a suppressed stellar angular momentum compared to $z=0$ LTGs in the same mass regime, we hypothesize two distinct formation channels for dEs and late-type dwarfs observed in the $z=0$ Universe. 

The progenitors of present-day dEs assembled at a higher redshift within a much denser environment in general. In these extreme conditions, angular momentum acquisition may have been suppressed and/or star formation was very different from today (e.g. higher efficiency and very bursty because of the higher avg. density). The former could have made the total density slopes (Fig.~\ref{fig:MW_Total_slope_vs_stellar mass}) steeper, as baryonic contraction is more efficient. The latter could have led to some of the dEs being self-quenched after a single rapid SF burst (e.g. VCC 200) while others were just massive and isolated enough to be able to re-accrete some of the expelled gas (rejuvenating SF) until they were eventually quenched when they first entered the cluster (see \paperrefeSTARS). Additionally, under the extreme early assembly conditions we may expect AGN activity to become more important than in dwarfs that formed recently. AGNs could quench galaxies, affect star-formation and affect the dark matter dynamically \citep[e.g.][]{Koudmani_2022,arjonagalvez_2024}.

Conversely, the $z=0$ late-type dwarfs have all assembled more recently in an epoch where halo assembly happens under different circumstances and SF is prolonged but less bursty (avoiding self-quenching). Their lower average DM density makes them more susceptible to external forces like harassment, which could explain the lack of dwarf late-types in the cluster centers. These distinct formation channels of today's quiescent and star-forming dwarfs could also resolve why the latter have higher DM fractions (Sec.~\ref{fig:MW_Total_slope_vs_stellar mass}) than the dEs. The first JWST observations of high-redshift dwarfs ($z\gtrsim 6$, i.e. the progenitors of dEs in this scenario) and corresponding simulations \citep[][]{dE_Graaff_2023,dE_Graaff2024} suggest that galaxies in this mass-regime have initially high dark matter fractions $\dmfrac\sim 0.8$ but evolve to $\dmfrac\sim 0.4$ at cosmic noon as they form more and more stars. Perhaps the star-forming galaxies at $z=0$ are also still within a dark matter dominated phase, as they are still in the process of converting their gas into stars.

Still, there are caveats to the conclusions drawn from Fig.~\ref{fig:scaling_B_band_1reff}: i) In our work, we have probed halo models that are very flexible (Sec.~\ref{sec:technique}) allowing for a variety of profiles and shapes. Many studies use more restricting assumption about the halo profiles (e.g. spherical NFW- or cored profiles) which could bias $\rhodm$. We plan to explore the effects of model choice in \paperrefe. ii) $\overline{\rhodm}(<\reff)$ is not an aperture independent quantity. At the same stellar mass the effective radii of LTGs may be systematically larger than that of a dEs with the same stellar mass. Then, if we assume an LTG and a dE sit in an identical halo one measures a smaller $\overline{\rhodm}$ for the former. We estimate we would have to increase the effective radii (i.e. the aperture) of our dEs by a factor of $\sim 2.1$ without changing the halo densities in order for them to fall nicely on the LTG scaling relation. iii) We can not easily distinguish between a difference in environment and assembly epoch yet compare galaxies of different groups and clusters, e.g. one cluster/group could have formed later than another. 

Virgo is believed to be a dynamically young cluster. It has a loose, cross-shaped structure with several intact sub-clumps \citep[][]{Bineggli_1987} and an irregular distribution of its intra-cluster medium \citep[][]{Boehringer_1994}. Furthermore Virgo's members exhibit a non-gaussian, unrelaxed velocity distribution \citep[][]{Conselice_2001} and a surprisingly high fraction of LTGs for a cluster \citep[][]{Sandage_1984}. The strong $\overline{\rho_{\mathrm{DM}}}$-$\centredis$ correlation is also telling about the dynamical state of the Virgo cluster. As a cluster relaxes and aggregates nearby galaxy groups, these infalling groups are expected to lose most of their smaller members within a few Gyrs to the larger cluster body \citep[cf.][]{Choque-Challapa_2019}. This process would gradually break down any $\overline{\rho_{\mathrm{DM}}}$-$\centredis$ correlation as the smaller members mix with the cluster. The fact that the $\overline{\rho_{\mathrm{DM}}}$-$\centredis$ correlation is still this strong supports this view of Virgo as a dynamically \textit{young} cluster. 

\section{Summary and Conclusions}
\label{sec:conclusions}
This work presents the first dynamical analysis of a dE sample ($\log_{10}(M_{*}/M_{\sun})\in [8.5,9.5]$) which employs orbit-superposition modeling and resolved, higher-order LOSVD information to constrain the detailed structure of their DM halos. One of the advantages of this approach is that it does not impose a priori restrictions on the stellar anisotropy structure, allowing us to lift the mass-anisotropy degeneracy and measure the halo profiles and shapes unbiased. The models incorporate stellar mass-to-light ratio gradients (i.e. allowing for spatial variances of stellar populations) and explore a large variety of possible halo profiles. For the first time, we also constrain the flattening of dark matter in dEs. We investigate whether the observational constraints are in tension with the standard $\Lambda$CDM paradigm (cusp-core problem and sphericity) and, if not, what they imply regarding halo assembly, star formation, environment feedback, and the progenitors of dEs. Our main conclusions are:

\textbf{dE halos are only mildly cored:} 
We do not find evidence for a strong tension of the DM slopes of dEs with $\Lambda$CDM predictions. The average DM slope of our sample is $\sim -0.9$. While the dEs exhibit some diversity with slopes $\in [-1.4,-0.5]$ this is still within the range of uncertainties in baryonic physics. These observations are formally inconsistent with DMO $\Lambda$CDM simulations, but we do not necessarily need to invoke exotic physics as simulations that include baryonic feedback appropriately may be able to explain the diversity and modest preference to cored profiles.

\textbf{Halos are very round:} 
The tension with $\Lambda$CDM may be more troubling in terms of the sphericity of halos. Similar to spherical shape constraints found for the Milky Way's halo, the dark matter of dEs is distributed nearly spherical ($\qdm \sim 0.9-1.0$) independently of the shape of the stellar distribution. While the inclusion of baryonic feedback in $\Lambda$CDM simulations is known to make halos more spherical, easing some of the tension, the constraints we measured still exceed expectation values from simulations. Still, unlike slope measurements, few constraints on the flattening exist, and the field is still in its infancy. More galaxies will need to be investigated.

\textbf{Is baryonic feedback enough to explain the remaining tension?} The jury is still out. It may be surprising that the dE halos seem more in tension in terms of their sphericity than they are in terms of slopes, considering the same baryonic mechanisms are used to explain both: halos becoming more spherical and cored. Simulations ($\Lambda$CDM and more exotic physics) often attempt and are successful in explaining individual aspects in isolation (e.g. why are halos cored). The challenge remains if simulations can explain multiple aspects at once. Only then we may hope to narrow down the underlying mechanisms if we consider all/more observational constraints simultaneously. 

\textbf{dE centers are dominated by luminous matter:} 
We find upper limits for their central supermassive black hole mass ($\mbh \lesssim 10^{6}M_{\sun}$) and dark matter fractions within one stellar effective radii that are moderate. The bulk of dEs have $\dmfrac \in [0.2,0.5]$ which is slightly more than what is typical for massive ETGs but considerably less than what is found for most LTGs.  

\textbf{Total density slopes are shallower than those of massive ETGs but still steeper than those of LTG dwarfs:} 
The slopes of the total density (baryons+dark matter) in dEs are relatively shallow with $\gamma \sim -1.5$. This places them on the extended anti-correlation found for stellar mass and total slope of `ordinary' ETGs with $M_{\sun}\lesssim 5\cdot 10^{10}M_{\sun}$. Therefore dEs are shallower than `ordinary' ETGs who have approximately isothermal slopes, but still steeper than LTGs of the same total stellar mass.

\textbf{Halo densities and profiles are strongly correlated with environment, and moderately with stellar angular momentum:}
In the cluster center the DM halos are denser and cuspier than an NFW profile. Conversely, in Virgo's outskirts, the average halo density is reduced by $\sim 1\rm dex$ and the profiles are moderately cored ($\sim -0.5$). Similarly, dEs with lower stellar angular momentum (and presumably low DM angular momentum) have cuspier halos. 

\textbf{Halos of dEs evolve only mildly, and are not universal:} 
We find these correlations are difficult to be explained by tidal interactions and quenching mechanisms like supernova feedback. While tidal interactions and ram-pressure stripping are likely important quenching mechanisms (cf. \paperrefeSTARS), they have little effect on the halos. Tidal interactions with the environment are incongruent with the observed correlations. Internal feedback by supernovae winds is a second order effect that may change the slopes of those dEs that have an extended SFH, but even then only moderately. In conclusion, the dE halos have evolved only mildly after their gravitational assembly. 

This lack of DM evolution due to internal or external feedback, the diversity in measured DM slopes $[-1.4,-0.5]$, and the correlations with environment/angular momentum suggest that halo profiles do not have a universal shape after assembly. Depending on the conditions during their halo formation epoch (environment, angular momentum, baryon clumpiness, AGN activity, etc.) some halos will turn out to be more cuspy than NFW if baryonic contraction prevails, or more cored if counter-acting process (e.g. dynamical friction, tidal torques) dominate.  

\textbf{A different formation? dEs may have assembled at higher redshift in more extreme conditions than local star-forming dwarfs:}
Our results suggest that the dE halos in Virgo have formed at a higher redshift than LTG dwarfs of similar mass. During this early assembly epoch star-formation, AGN activity, and environment/formation conditions may have been much more rapid and extreme than for the LTGs that formed more recently. This may have led to additional mechanisms that were able to quench some of the dEs shortly after formation. Considering these extreme early stages, local spirals which seemed to have formed in a much more quiescent epoch of the Universe are not necessarily representative of the star-forming progenitors of dEs. 

Our stellar population analysis in \paperrefeSTARSSP has left open two plausible scenarios: i) dEs have been continuously produced throughout Virgo's history, but the IMF varies with the formation age. ii) dEs have been formed very early in Virgo's history but depending on their mass and environment some of them were quenched early while others have been able to hold onto (or re-accrete) their gas, thus, experiencing a more complex SFH. The results presented here make scenario ii) more likely. This means that if the IMF varies, then it probably varies less than implied by taking the stellar masses of single-stellar-population models at face value (see \paperrefeSTARS). However, whether an early formation epoch is enough to make the dEs' dynamical stellar masses compatible with a Kroupa or even sub-Kroupa IMF remains to be checked with more sophisticated stellar population models.

Future observational constraints for local LTGs could aid in further narrowing down the formation scenarios. For example, we expect that assembly at higher redshift resulted in rounder halos \citep[][]{Chua_2019} than at $z=0$. While our dEs are indeed surprisingly round, not enough constraints on the flattening of local LTGs dwarfs exist to compare the dEs to. 

Our comprehension of small-scale cosmological problems like the cusp-core problem is inextricably linked to our understanding of galaxy formation and evolution. Galaxies are unique objects with their own history of assembly and evolution, resulting in different DM distributions. Our results suggest that there are considerable differences between dEs and comparable late-type dwarfs in the local Universe, but many more galaxy classes exist which may even be more divergent. Ultra-diffuse galaxies, for example, are also quiescent and close in magnitude to dEs, but they are very extended, have a lower surface brightness, and some of them appear to have essentially no dark matter while others have much higher dark matter fractions than the dEs \citep[e.g.][]{van_Dokkum_2016,Danieli_2019,Bar_2022,Zoeller_2023}. Precise and accurate dynamical decompositions (App.~\ref{append:simulation}) of dark matter and baryons for a manifold of different galaxy types is needed, and a consistent comparison between them (e.g. in physical units instead of aperture-dependent) will be essential to investigate small-scale cosmology in the future.

\begin{acknowledgments}
\section*{ACKNOWLEDGEMENTS}
This work is based on observations obtained with the Harlan J. Smith Telescope at the McDonald Observatory, Texas. Computing has been carried out on the COBRA and RAVEN HPC systems at the Max Planck Computing and Data Facility (MPCDF), Germany. We also made frequent use of the NASA/IPAC Extragalactic Database (NED), operated by the Jet Propulsion Laboratory and the California Institute of Technology, NASA’s Astrophysics Data System bibliographic services, and the HyperLeda database \citep{Paturel_2003}. 
\end{acknowledgments}

\vspace{5mm}
\facilities{Smith, HST}
\software{astropy \citep{astropy_2022}}

\appendix

\section{How well can we recover dark components? - A stress test}
\label{append:simulation}
The primary goal of this paper is to investigate the shapes, slopes, and DM fractions of the halos of dEs by use of dynamical orbit modeling. This requires a reliable decomposition of the dark components (black hole and DM halo) from the stars. As preparation for this study, we stress-tested our observational and modeling setup on an N-body simulation. The goal of this test was three-fold: i) Explore the general ability of the orbit models and our observational setup to constrain the mass and kinematic distribution of the dE sample. ii) Investigate how reliable one can decompose the individual matter components. iii) Test whether we can constrain the \textit{flattening} of dark matter halos and gauge how much the assumption of axisymmetry could bias our results. 

In Sec.~\ref{subsec:simulation_setup} we explain how we generated the mock observations for our modeling, then we continue with a general evaluation of the mass recovery and decomposition quality (Sec.~\ref{subsec:results_mass}). We also gauge how much a possible triaxiality of the galaxies could distort the results derived with axisymmetric modeling, specifically the flattening of the dark matter halo (Sec.~\ref{subsec:results_flattening}). 

We expect two circumstances to negatively affect the dark matter recovery: i) If the \textit{local} dark matter contribution to the total mass is negligible, the recovered halo distribution is poorly constrained in these regions because it does change the gravitational potential significantly. ii) If the dark matter follows the exact same distribution as the luminous mass distribution (a mass-follows-light distribution) the two components are degenerate since it's impossible to dynamically differentiate between dark stellar remnants and genuine dark matter. From an orbit's perspective, one component can be absorbed by the other (e.g. by up-scaling the stellar mass-to-light ratio) without changing the combined potential.

Point i) will be explored in detail in \paperrefeSP but we may account for it for now by trusting the dark matter recovery in regions where the models suggest a significant dark matter fraction more than where it doesn't. While a reliable decomposition is completely impossible in the extreme case of point ii) we stress tested our modeling procedure by applying it on an N-body simulation that represents a tough case where at least some of the dark matter distribution follows the luminous matter rather closely within the center of the simulated galaxy. At larger radii the two mass components start to differ in their spatial distributions and amount, which, in principle, should allow the models to break the degeneracy.

\subsection{The simulation setup}
\label{subsec:simulation_setup}
As a test for our modeling routines, we use the results from an idealized N-body simulation of merging low mass galaxies similar to \citet{Partmann_2023}, taking into account the gravitational dynamics of dark matter, stars and black holes.

The N-body simulations are run with the Ketju code \citep[][]{Rantala_2017,Rantala_2018,Mannerkoski_2021}, a combination of the Gadget-3 tree gravity solver with accurate regularized integration \citep[][]{Springel_2005}. This method allows resolving the unsoftened forces between the star/DM components and black holes for an accurate treatment of dynamical friction and scattering of DM and stars by black hole binaries or multiples. The coalescence of black holes by gravitational wave emission is followed with Post-Newtonian corrections up to order 3.5.

The simulations follow the merger of five small galaxies (each with DM mass $4\cdot10^8 \rm M_{\sun}$ and stellar mass $4\cdot10^7 \rm M_{\sun}$) with a five times more massive central halo (DM mass $2\cdot10^9 \rm M_{\sun}$ and stellar mass $2\cdot10^8 \rm M_{\sun}$). Each of the galaxies consists of a DM halo (Hernquist profile with $r_{1/2} = 7.1\mathrm{kpc}$), a stellar component (Hernquist profile with $r_{1/2} = 1\mathrm{kpc}$) and a black hole. By construction, the six galaxies merge within the first few Gyr and result in a galaxy with a total mass of $M_{\mathrm{tot}} = 4.4 \cdot 10^9 \rm M_{\sun}$, where $M_{\rm DM}=4 \cdot 10^9 \, \rm M_\odot$ are contributed by dark matter and $M_{\rm star}=4 \cdot 10^8 \, \rm M_\odot$ by the stellar component. In the state we ‘observe’ the system, the central galaxy has experienced several mergers in its past (7-10 Gyrs ago) and is now de facto in an equilibrium state again.

Each of the merger progenitors carries a black hole. While \citet{Partmann_2023} explores various black hole masses and merger orbits, for this study we choose an extreme case with a central black hole mass of $10^7 \, \rm M_\odot$ and $2\cdot 10^6 \, \rm M_\odot$ black holes in the five infalling smaller galaxies. For reasons explained later, highly overmassive black holes can produce density cores with similar dark matter and stellar densities in the galactic center. We choose this scenario because it is expected to be the most challenging for our modeling pipeline. The extent and mass are typical for the dEs we observed in our Virgo sample. However, beyond that we do not expect the N-body simulation to be a very physical representation of the circumstances in real dEs, i.e. we do not expect a mass-follows-light distribution in dEs nor that they have experienced several dry mergers with very massive black holes.

During the merger process, most black holes sink to the center of the central halo where they have complex evolution paths that result in the formation of black hole binaries or triples, the dynamical ejection of black holes or black hole mergers. In the simulation considered here, three among the five black holes that were brought into the central galaxy with the infalling smaller galaxies were ejected by dynamical interactions. As a result, black holes with a total mass of $1.4\cdot10^{7} \rm M_{\sun}$ remain in the center of the central galaxy. The system of black holes in the center has a sphere of influence of $\rsoi$$\sim$$0.3 \rm kpc$, where $\rsoi$ is defined as the radius at which $M_{*}(r<\rsoi)=\mbh$. This $\mbh$ is barely resolved as $\rsoi$ is close to the spatial resolution ($\sim$$0.25 \rm kpc$) of the observational setup of the real dE sample that we adopted for this simulated test. As such we do not necessarily expect to find strong constraints towards lower black hole mass and possibly only an upper limit for $\mbh$.

In massive early type galaxies “black hole binary scouring”, i.e. the ejection of stars from the galactic center through slingshots with a black hole binary is an important process that can convert an initially cuspy stellar density profile into a profile with a flat density core \citep[e.g.][]{Kormendy_2009_B,Kormendy_2009,Kormendy_2013,Thomas_2014,Rantala_2018,Mehrgan_2019}. As discussed in \citet{Partmann_2023}, even in low mass galaxies the combined effect of black hole sinking through dynamical friction, black hole binary scouring and black holes ejections can lead to the formation of large stellar and dark matter density cores if the black hole masses are large enough. For the large black hole mass used here, this effect leads to the formation of a large core with a size $r_{c}\sim0.4 \rm kpc$ (or about 5\arcsec in our mock setup) and an \textit{assimilation} of the \textit{central} dark and baryonic matter distributions where both matter components follow approximately the same density distribution with a ratio of $\sim 0.5$ of dark and baryonic matter. This means the two components are approximately indistinguishable within the core. Only at larger radii where the scouring had less impact on the mass distribution the two matter components begin to diverge with the dark matter starting to dominate as the stellar density declines more steeply than the dark matter. The mass-to-light ratio of the single stellar population is a spatially constant $\Upsilon_{*}=1.0$, which means the total \textit{dynamical} mass-to-light ratio is $\Upsilon_{dyn} \sim 2$ within the scouring core. $\Upsilon_{dyn}$ is reaching $\sim 5$ (or a dark matter fraction of $\sim 70\%$) near the edge of the mock field of fiew (FoV) at $r_{FoV}=20\arcsec$.

\begin{figure}
	\centering
	\includegraphics[width=1.0\columnwidth]{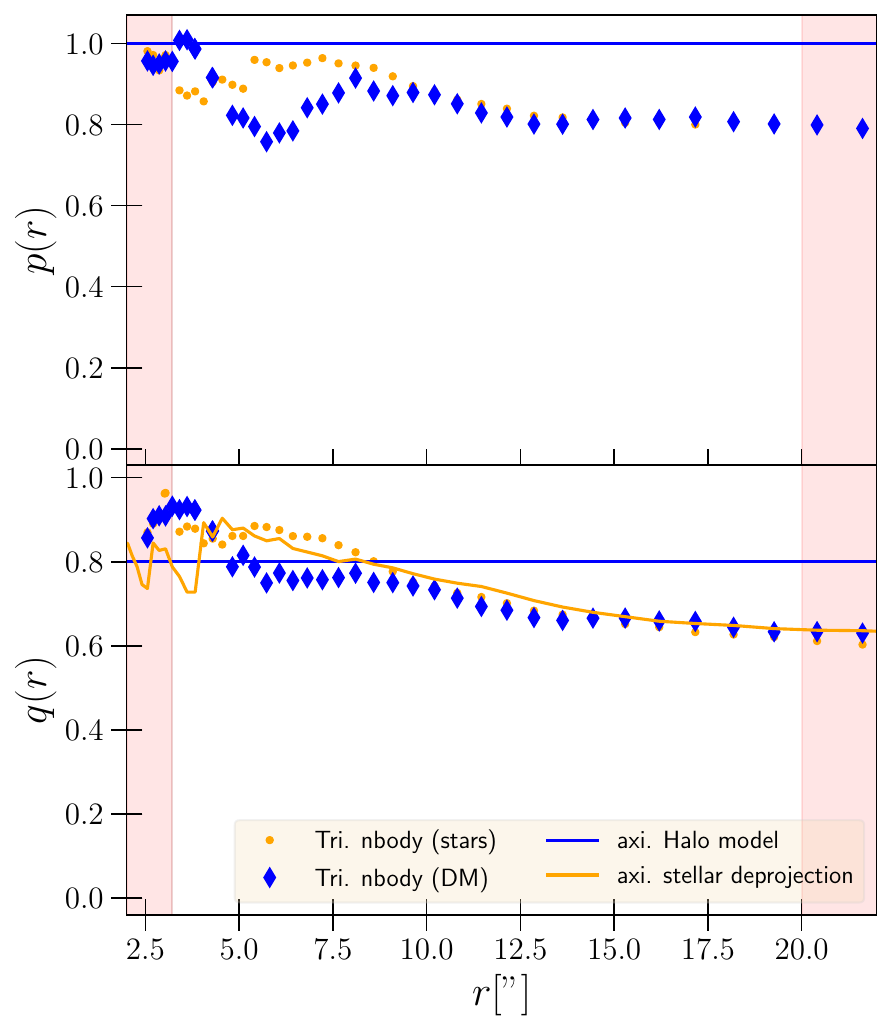}
    \caption{The shape of the N-body simulation's density distribution approximated by an ellipsoid. \textit{Top:} $p=b/a$, \textit{Bottom:} $q=c/a$ where $a$ is the semi-major axis, $b$ the semi-intermediate axis, and $c$ the semi-minor axis of the triaxial ellipsoid. The \textit{Orange Dots} indicate the triaxial axis ratios of the stellar particle distribution, the \textit{Orange Line} shows the corresponding \textit{axisymmetric} approximation, which is also the edge-on `pseudo'-deprojection that we used for the dynamical modeling (but scaled by a variable mass-to-light ratio of the stellar population). The \textit{blue diamonds} show the triaxial ratios of the dark matter particle distribution, it follows the shape of the stellar particles very closely even outside the scouring core. The \textit{blue Line} shows the spatially constant Dark matter flattening $\qdm$ of the \textit{best} axisymmetric (oblate) orbit model we found. We sampled the axisymmetric flattening within $\qdm \in [0.7,0.8,0.9,1.0]$. Similar to Fig.~\ref{fig:nbody_results} we mark regions smaller than the central spatial resolution and outside the FoV in \textit{red}.} 
    \label{fig:nbody_flattening}
\end{figure}

Since the galaxy in the N-body simulation is a results of several mergers, the density distribution is not expected to follow a particular symmetry, hence the axisymmetric orbit models we employ are unlikely to be a perfect fit. If we approximate the particle distributions using a \textit{triaxial} ellipsoid we find that, both the baryonic and dark matter component, follow a very similar shape profile with both components being triaxial and oblate outside the core. However, within the core the shape of both mass components becomes near spherical. Within the core the particle distribution is also more complex and neither an axisymmetric nor a triaxial approximation is ideal. We plot the triaxial axis ratios $p$ and $q$ versus radius for both matter components in Fig.~\ref{fig:nbody_flattening}. 

\begin{figure}
	\centering
	\includegraphics[width=1.0\columnwidth]{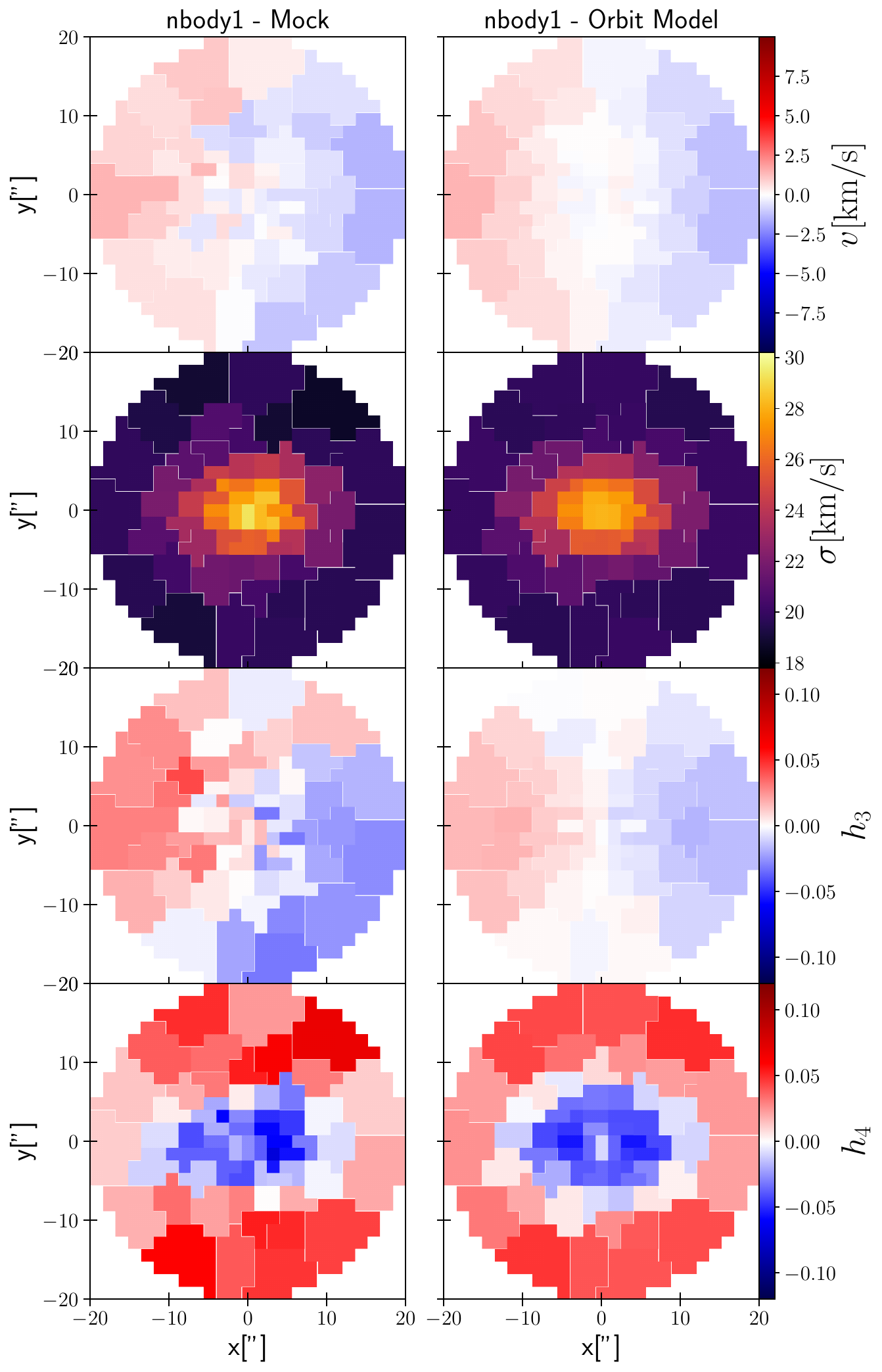}
    \caption{\textit{Left:} The mock observations for the triaxial N-body simulation. \textit{Right:} The corresponding best axisymmetric model we found using our orbit modeling setup. Both, mock and model LOSVDs are characterized here by a Gauss--Hermite series truncated at $h_{4}$, higher order deviations in the actual binned LOSVDs (that are still being modelled dynamically) are not shown.}
    \label{fig:nbody_mock_1}
\end{figure}

To obtain realistic mock observations of the N-body simulation that emulates our observational setup used for the real dEs we place the simulated galaxy at the average distance of the Virgo cluster $d = 16.5 \, \rm Mpc$ and projected its stellar kinematics along its \textit{intermediate} axis into a realistic Voronoi grid with a FoV of $20\arcsec$. The number of bins and their resolution are typical for one of the higher $S/N$ observations we obtained in the real sample. The central Voronoi bins are typically $3\arcsec$ large. This means our spatial resolution is just below the sphere of influence $\rsoi \sim 4\arcsec$ and the size of the scoured core $\sim 5\arcsec$. The resulting mock observations are shown on the left side of Fig.~\ref{fig:nbody_mock_1}. These kinematic maps are illustrated by a truncated Gauss--Hermite series up to $h_{4}$. However, the input for the dynamical models are actually the fully non-parametric descriptions of the LOSVDs as is the case for the dE modeling. The kinematic maps of the simulation show some interesting features that are usually not observed in real galaxies: i) A positive $v$-$h_3$ correlation ii) and a negative $h_4$ within the scoured core. This discrepancy may tell us something about the significance of black hole scouring in real galaxies and assumptions in N-body simulations, however, an investigation is beyond the scope of this paper. 

Like we did with the dE sample we use the search algorithm NOMAD and probe a total of approximately 5000 orbit models using the halo parametrization we also used for the dEs (see Sec.~\ref{sec:technique}). However, we only probed a single deprojection (or inclination) for the \textit{stellar} component. As mentioned above the N-body simulation is observed along its intermediate axis, therefore this deprojection is assumed to be edge-on. The right hand side panels in Fig.~\ref{fig:nbody_mock_1} illustrate the projected kinematics of the single best orbit model we were able to find to represent the simulated mock observation shown on the left side of Fig.~\ref{fig:nbody_mock_1}. As is the case for the modeling of the Virgo dEs we opted to model the entire FoV with the orbit models, i.e. differences in individual quadrants can not be reflected as easily by the axisymmetric model. Nevertheless, the orbit model is able to reproduce all important features of the mock observations well, giving confidence that the orbit setup (e.g. number of orbits) and mass model are adequate and flexible enough to emulate the mass and kinematic of the N-body simulation. 

\subsection{Mass recovery and decomposition}
\label{subsec:results_mass}
\begin{figure}
	\centering
	\includegraphics[width=1.0\columnwidth]{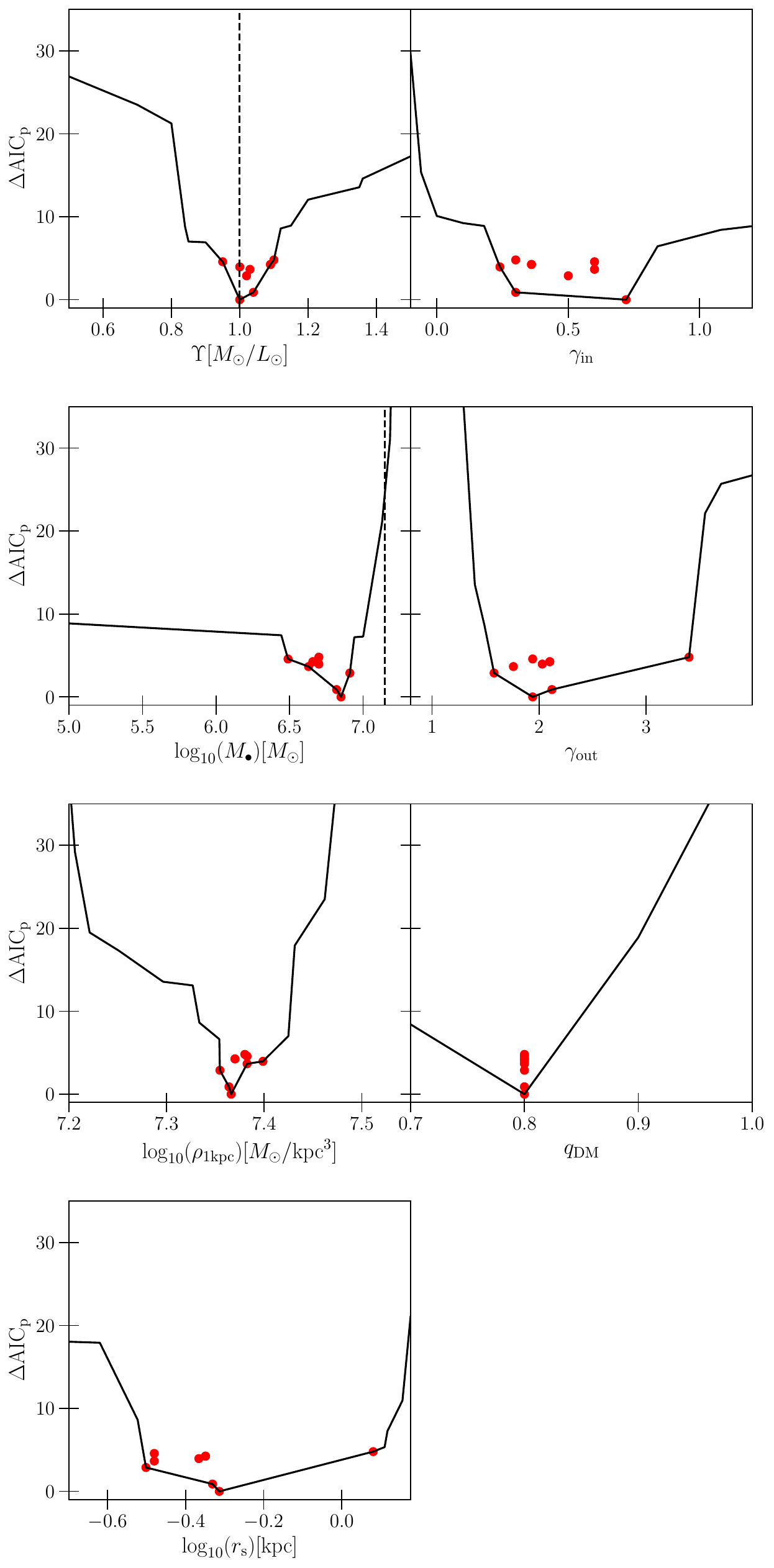}
    \caption{The minimum subtracted $\aicmod$-constraints for the axisymmetric orbit models of the N-body simulation. In total we calculated in the order of $\sim 5\cdot10^{3}$ orbit models. The red dots indicate the parameters of the best few models (ranked in $\aicmod$). The vertical dashed lines indicate the true stellar mass-to-light ratio and black hole mass.}
    \label{fig:nbody_envelope}
\end{figure}

In Fig.~\ref{fig:nbody_envelope} we show the $\Delta \aicmod$-curves for each of the nuisance parameters that generate the mass model. The stellar mass-to-light ratio $\Upsilon_{*}=1.0$ is recovered by the single best orbit model. The next best orbit models $\aicmod$ tend to have slightly larger $\Upsilon_{*}$ but the scatter is in the single digit percentage range. If one was to apply the error estimation approach we employ for the dEs sample, i.e. by calculating the spread of the 25 best orbit models we estimate an uncertainty in $\Upsilon_{*}$ of $6\%$. This implies the stellar mass component is well decomposed by the dynamical modeling even when the dark matter is distributed similarly as the stars over some regions of the galaxy. As long as there is a detectable difference between stars and DM in some parts of the FoV (in this case between $10\arcsec$ to $20\arcsec$) the models are able to `recognize' that a stellar component that is simply scaled up or down is an insufficient description. 

The black hole mass of the best model on the other hand is underestimated with the best model only having $0.7\cdot 10^7$ instead of the actual SMBH mass of $1.4\cdot 10^7$. The $\Delta \aicmod$-curves for the black hole are very a-symmetric indicating that models with essentially no black hole and models with $\mbh\leq10^7$ only show very little difference. Without prior knowledge one would likely conclude an upper limit at $\mbh$$\sim$$1\cdot10^7$ due to the sphere of influence of the best-fitting models being close to the central resolution limit. 

The parameters of the Zhao-profile suggest a scale radius $r_{s}$$\sim$$0.5 \rm kpc$, i.e. the profile transitions its slope in the vicinity of the scouring core radius. However, models with a lower or larger scale radius (e.g. $0.3 \rm kpc$ and $1 \rm kpc$) are not much worse ($\Delta \aicmod \lesssim 5$). Only for the extreme values $\aicmod$ begins to rise more rapidly. We find similarly broad $\aicmod$ valleys for the inner and outer slopes $\gamma_{\mathrm{in}}$ and $\gamma_{\mathrm{out}}$ of the Zhao parametrization. 

\begin{figure}
	\centering
	\includegraphics[width=1.0\columnwidth]{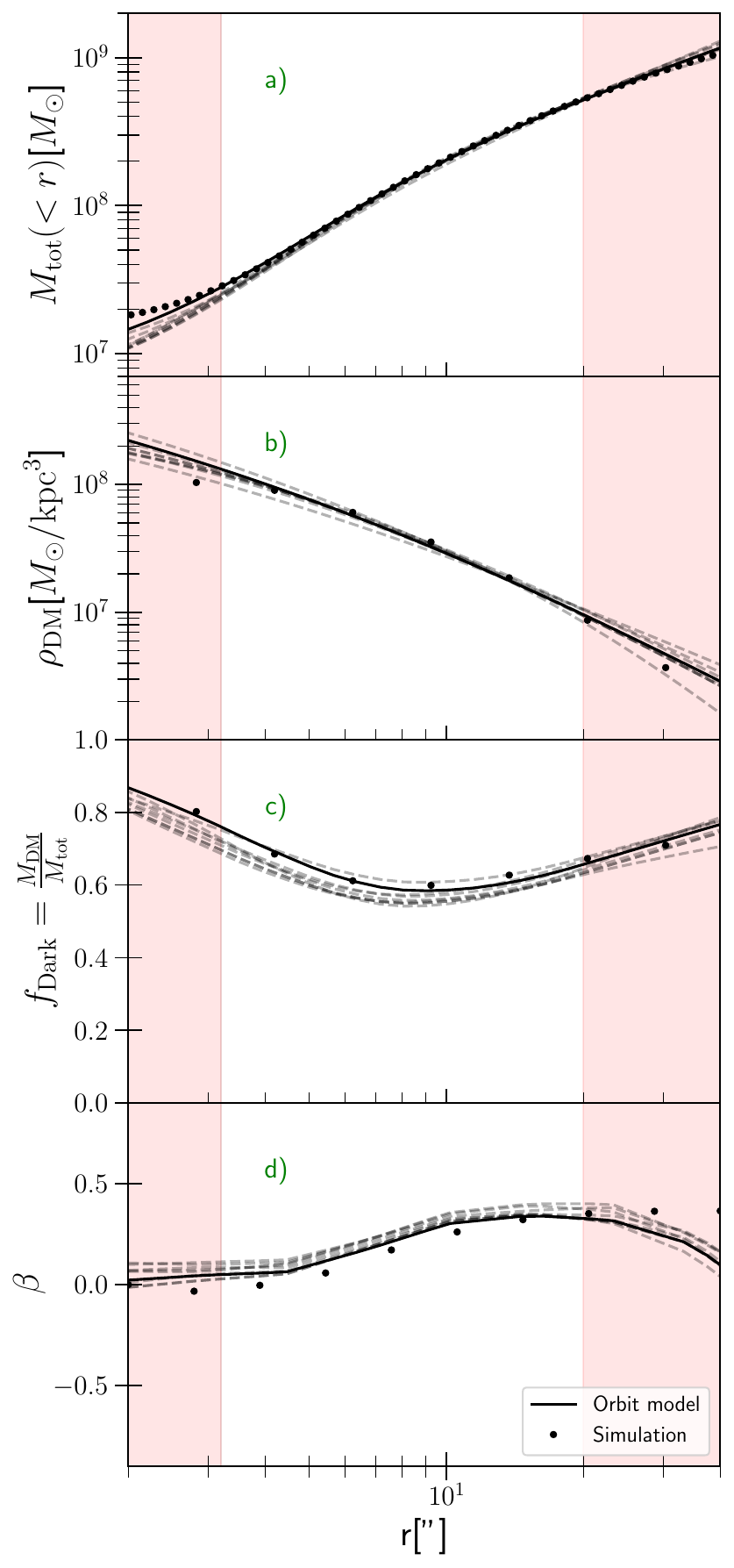}
    \caption{Mass and and anisotropy structure (spherically averaged) of the simulation (\textit{dots}) and the recovered profiles of the best \textit{axisymmetric} orbit model we were able to find (\textit{solid line}). The \textit{dashed} lines show the next best models when ranking all sampled models by their $\aicmod$, their nuisance parameters are indicated by the red dots in Fig.~\ref{fig:nbody_envelope}. The \textit{inner} red region indicates the typical size of our resolution limit. The \textit{outer} red region marks areas outside the FoV. Panel a) the total enclosed mass within radius r, b) the local DM density, c) the (cumulated) DM fraction (including the black hole), d) the anisotropy $\beta$ of the stars.}
    \label{fig:nbody_results}
\end{figure}

Such broad $\aicmod$ valleys in the parameters are not surprising (particularly in the more correlated parameters) and possibly even desirable if one wants to minimize biases because of erroneous halo parametrization. The dynamical models are sensitive to the actual mass distribution that is described by the above set of parameters and not the specific set of parameters or the halo parametrization itself. Therefore one should investigate how well the actual mass distributions are recovered to gauge how well the recovery worked. This is illustrated in Fig.~\ref{fig:nbody_results} where we plot the recovered mass and anisotropy profiles of the best axisymmetric orbit model and the actual N-body simulation. We color radial regions that are within our spatial resolution or outside the FoV in red to highlight the regions where we expect the data do \textit{not} impose strong constraints. We also plot the next few best $\aicmod$ model we found as dashed lines and indicate their corresponding halo parameters in Fig.~\ref{fig:nbody_envelope}. Even though the nominal values of the strongly inter-correlated shape parameters ($\slpin$,$\slpout$,$\sclrad$) can differ significantly the actual mass distributions they describe are very similar. Again we conclude the parameters that `generate' the halo density should be treated as nuisance parameters.  

We find that the total enclosed mass $\MTOT$ is the property constrained the best as it's linked very directly to the gravitational potential $\Phi$ which itself determines all the orbits that constitute the orbit-superposition model. In other words, the model selection is, first and foremost, a predictor of the gravitational potential and thus the total mass (regardless of composition). One also sees that the next best few $\aicmod$ models have very similar enclosed masses to the single best $\aicmod$ model, though they are generally slightly further away from the true N-body mass than the single best $\aicmod$ model which suggests that $\aicmod$ is a consistent predictor of the enclosed mass. 

However, when plotting $\MTOT$ as a function of radius and comparing it to the actual enclosed mass of the simulation we also observe that the systematic deviation between the two varies with the radius it is measured at: Within the unresolved center (inner red region) and outside the FoV (outer red region) the differences between the `truth' and the model predictions start to flare-up and diverge. The next best $\aicmod$ models also seem to scatter \textit{a-symmetrically} in these regions: Preferably underestimating $\MTOT$ in the center and overestimating it at large radii outside the FoV (the latter was also noted in \citealt{Gerhard_1998} and \citealt{Thomas_2005}). It seems the exact matter distribution far inside the spatial resolution (or central Voronoi elements) and far outside the FoV seems to be constrained less by the kinematic data, as several models can be constructed that differ significantly in these regions but produce essentially the same model LOSVDs (Fig.~\ref{fig:nbody_mock_1}) and $\aicmod$. 

Considering $\MTOT$ is fairly well constrained within the radial range covered by the data and the fact that the models found the correct stellar mass-to-light ratios $\Upsilon_{*}=1$ (cf. Fig.~\ref{fig:nbody_envelope}) it is not surprising that the recovery of the DM distribution (second panel) also performs similarly well. Even though the underlying dark matter distribution is more complex than the Zhao-model and quasi-degenerate with the stellar matter within $5\arcsec$ due to the mass-follows-light core, both, the dark matter density itself and its local slope $\dmgrad$ is recovered well within the white regions where $\MTOT$ is constraint.

The dark matter inside the unresolved center does not become uncertain in a random fashion: Within the unresolved center it systematically scatters above the true cored profile of the simulation. This can be explained by a degeneracy of the \textit{two} dark components of the model: The dark matter halo and the central black hole. As seen in Fig.~\ref{fig:nbody_envelope} we find a sharp upper limit for the black hole mass of $\sim1\cdot10^7$ but only very little difference in $\aicmod$ for models with a lower black hole mass.

This underestimation of the black hole is complemented by an overestimation of the central DM density with the sphere of influence. Both these errors cancel each other if we examine the (enclosed) fraction $f_{Dark}$ of the \textit{combined} dark components (i.e. DM halo + black hole) versus the total enclosed mass (third panel in Fig.~\ref{fig:nbody_results}). Even in the very center the best orbit model traces the true dark fraction of the simulation well, suggesting that the separation of dark and luminous matter actually works well, merely the separation of the two dark components proves difficult. A model with an underestimated black hole mass can essentially achieve the same LOSVD predictions by assigning the `missing' mass to its central dark matter halo. On the other hand, models with a much larger black hole than the true mass $\mbh=1.4\cdot10^{7}M_{\sun}$ are effectively ruled out because the halo can only add additional mass in the center but not subtract it.

We do not believe that previous black hole measurements in real galaxies were affected significantly by such a degeneracy, at least as long as their sphere of influence was resolved. This is because most evidence and models of real galaxies suggest that the dark matter halo fraction is almost 0 in the center, i.e. the dark matter contribution to $\MTOT$ in unresolved scales is negligible. Which indicates that the black hole is the only significant mass in the resolved bins. Only in the cases where the dark matter fraction within the central scales is significant or if there is suspicion of distinct unresolved stellar population (i.e. variable $\Upsilon_{*}$) it might be worth to revisit existing black hole measurements as they may have been affected by a decomposition degeneracy. For example, \citet{Mehrgan_2023_b} have found black-hole masses to reduced by $25$\% when allowing for central mass-to-light ratio gradients in a sample of massive early-type galaxies.

Assuming the two dark components are indeed degenerate it nonetheless surprising that the $\aicmod$-envelopes (Fig.~\ref{fig:nbody_envelope}) rapidly rise as $\mbh>1\cdot10^{7}M_{\sun}$, effectively ruling out the true mass $\mbh=1.4\cdot10^{7}M_{\sun}$ as too high. If the black hole constraints are degenerate but unbiased $\aicmod$ should only start to rise rapidly for masses beyond $1.4\cdot10^{7}M_{\sun}$. In \paperrefeSP we plan to identify and discuss possible reasons for this discrepancy in the upper limit for the black hole mass. As it turns out better but more computationally expensive model parametrizations can resolve this tension resulting in a better upper limit for the black hole mass $\sim 1.5\cdot10^{7}M_{\sun}$.

While there are no obvious signs that the dEs are triaxial (see also \paperrefeSTARS) the real dE galaxy could at least to some degree be triaxial like the N-body simulation. To gauge how deviations from axisymmetry could affect the results for the dE sample we analyse the effects the axisymmetry assumption had on the modeling of the triaxial simulation. Two properties that we expect to be very affected by such an erroneous symmetry assumption are the intrinsic 3D \textit{kinematics} of the stars and the \textit{shape} (flattening) of the mass distribution. The former because the orbits of the model are restricted to an axisymmetric potential, thus, its orbits may not be as representative of the plethora of different orbits that are possible in a more general potential, and the latter because it is impossible for the model to emulate the non-axisymmetric shape of the mass distribution. This symmetry mismatch in the orbit structure and shape may in turn then negatively affect the recovered mass distributions as well \citep[][]{Thomas_2007_b}.

We can examine the quality of the kinematic recovery using the orbital anisotropy parameter $\beta$: 
\begin{equation}
\beta=1-\frac{\sigma_{\phi}^2+\sigma_{\theta}^2}{2\sigma_{r}^2}
	\label{eq:aniso_parameter}
\end{equation}
where the $\sigma_{i}$ are the velocity dispersion in spherical coordinates \citep{Binney_2008}. 

The spherically averaged $\beta$ of the N-body simulation, together with the $\beta$ of the best orbit models, is shown in the bottom panel of Fig.~\ref{fig:nbody_results}. Within the radial range covered by the data the anisotropy structure is recovered well within a few percent. Albeit the best $\aicmod$ models appear to be biased slightly more radial than the truth. This could be a an expression of the incomplete orbit representation forced by the axisymmetry assumption. Similar to the mass recovery the $\beta$ structure of the dynamical models starts to deviate significantly outside the FoV, again implying it is not important for a model's ability to fit the LOSVD data what exactly the mass/kinematic structure is in unconstrained regions outside the FoV. 

The simulated tests show that the mass decomposition and distribution at or below unresolved scales can be misleading. In the N-body simulation this shows itself in the correlation of $\mbh$ and the central dark matter excess. Density gradients (i.e. slopes) are recovered best for radii larger than the resolution and smaller than the FoV (white regions). The tests also shows that we can expect to constrain the kinematic structure of the dE sample well even if they are slightly triaxial. However, as was the case with $\MTOT$, the kinematic recovery suffers in areas where we have sparse or no data coverage.

\subsection{Shape recovery}
\label{subsec:results_flattening}
To evaluate the model's ability to recover the intrinsic shape of a non-axisymmetric mass distribution we approximate the N-body's particle distribution directly with \textit{triaxial} ellipsoids (which is still a symmetry assumption but less restrictive than the axisymmetry) and determine its semi axis ratios $p=b/a$ and $q=c/a$. Where $a$,$b$, and $c$ are the semi major-, intermediate- and minor axis of the ellipsoid. We compare this to the flattening $q$ of the axisymmetric mass distributions which have $p=1$ by definition since $b=a$ for the oblate, axisymmetric models.

We compare the triaxial flattening of the simulation and the dynamically recovered axisymmetric shapes of the orbit models in Fig.~\ref{fig:nbody_flattening}. The true triaxial shapes of the dark and baryonic components trace each other closely for the entire radial range with minor differences at intermediate radii just outside the scouring core ($5-8\arcsec$). As mentioned above the N-body simulation is viewed along its intermediate axis. As such, the flattening of the edge-on axisymmetric deprojection that represents the stellar component of the orbit models, is essentially a direct approximation of the minor axis flattening $q(r)$. However, the information about the additional flattening along the line of sight ($p\sim0.8$) is lost due to the axisymmetry assumption. 

While the dynamical models `know' the axisymmetric flattening of the stellar distribution from the deprojection they have no a priori knowledge of neither the $q$ nor $p$ axis ratios of the dark matter distribution. In the past a long-standing point of discussion was whether the viewing angles of galaxies can be accurately recovered using only dynamical constraints. However, recently we have demonstrated that viewing angles, and consequently the intrinsic shapes, of the \textit{luminous} component can in fact be accurately measured with dynamical models if one accounts for variation in model flexibility \citep[e.g.][]{Lipka_2021,de_Nicola_2022}. Since dynamical models trace the gravitational potential, there is no reason to assume that a flattening of the \textit{dark} component should not be detectable as well.

Despite this though, the shapes of the dark matter component of the model are often simply assumed to be spherical even when the stellar component is obviously flattened. This assumption could heavily bias the entire mass recovery (in particular stellar mass-to-light ratios and shapes) as the model would attempt to compensate for the non-spherical halo by adjusting other components in order to approximate the total gravitational potential better. For this reason we equipped the Zhao halos we test in this study with a (radially constant) axis ratio $\qdm$ as an additional free parameter. In the case of the N-body simulation we probed orbit models with four different axisymmetric flattening $\qdm \in [0.7,0.8,0.9,1.0]$. As the $\aicmod$-curves in Fig.~\ref{fig:nbody_envelope} suggest the orbit models show a strong preference towards an intermediate flattening of $\qdm=0.8$, essentially ruling out a spherical halo. In Fig.~\ref{fig:nbody_flattening} we plot the (constant) axisymmetric flattening $\qdm=0.8$ of the best orbit model together with the true triaxial shapes of the simulation. While it's obvious that the halo model is not sophisticated enough to describe the halo's radial variation in $q$ or its triaxiality ($p\neq1$) one can see that a $\qdm=0.8$ is the closest approximation for the average flattening in $q$ \textit{and} $p$ within the radial range that is constrained by the data (white region). This suggests that we can place \textit{unbiased} dynamical constraints on the average halo shapes of galaxies, even if the true halo is slightly triaxial or has variable flattening. In the future, a more sophisticated halo description may allow an even more accurate halo recovery. At present, however, this appears computationally unfeasible. Nevertheless, the simulated test implies that we can at least expect to infer whether the halos of the dE sample are spherical or show signs of flattening, even if they are triaxial. 

\bibliography{literature}{}

\begin{thebibliography}{}
\expandafter\ifx\csname natexlab\endcsname\relax\def\natexlab#1{#1}\fi
\providecommand{\url}[1]{\href{#1}{#1}}
\providecommand{\dodoi}[1]{doi:~\href{http://doi.org/#1}{\nolinkurl{#1}}}
\providecommand{\doeprint}[1]{\href{http://ascl.net/#1}{\nolinkurl{http://ascl.net/#1}}}
\providecommand{\doarXiv}[1]{\href{https://arxiv.org/abs/#1}{\nolinkurl{https://arxiv.org/abs/#1}}}

\bibitem[{{Abadi} {et~al.}(2010){Abadi}, {Navarro}, {Fardal}, {Babul}, \& {Steinmetz}}]{Abadi_2010}
{Abadi}, M.~G., {Navarro}, J.~F., {Fardal}, M., {Babul}, A., \& {Steinmetz}, M. 2010, \mnras, 407, 435, \dodoi{10.1111/j.1365-2966.2010.16912.x}

\bibitem[{{Adams} {et~al.}(2014){Adams}, {Simon}, {Fabricius}, {van den Bosch}, {Barentine}, {Bender}, {Gebhardt}, {Hill}, {Murphy}, {Swaters}, {Thomas}, \& {van de Ven}}]{Adams_2014}
{Adams}, J.~J., {Simon}, J.~D., {Fabricius}, M.~H., {et~al.} 2014, \apj, 789, 63, \dodoi{10.1088/0004-637X/789/1/63}

\bibitem[{Akaike(1973)}]{Akaike_73}
Akaike, H. 1973, Information Theory and an Extension of the Maximum Likelihood Principle (New York, NY: Springer New York), 199--213

\bibitem[{{Akaike}(1974)}]{Akaike_74}
{Akaike}, H. 1974, IEEE Transactions on Automatic Control, 19, 716

\bibitem[{{Allgood} {et~al.}(2006){Allgood}, {Flores}, {Primack}, {Kravtsov}, {Wechsler}, {Faltenbacher}, \& {Bullock}}]{Allgood_2006}
{Allgood}, B., {Flores}, R.~A., {Primack}, J.~R., {et~al.} 2006, \mnras, 367, 1781, \dodoi{10.1111/j.1365-2966.2006.10094.x}

\bibitem[{{Arjona-Galvez} {et~al.}(2024){Arjona-Galvez}, {Di Cintio}, \& {Grand}}]{arjonagalvez_2024}
{Arjona-Galvez}, E., {Di Cintio}, A., \& {Grand}, R. J.~J. 2024, arXiv e-prints, arXiv:2402.00929, \dodoi{10.48550/arXiv.2402.00929}

\bibitem[{{Ascasibar} {et~al.}(2004){Ascasibar}, {Yepes}, {Gottl{\"o}ber}, \& {M{\"u}ller}}]{Ascasibar_2004}
{Ascasibar}, Y., {Yepes}, G., {Gottl{\"o}ber}, S., \& {M{\"u}ller}, V. 2004, \mnras, 352, 1109, \dodoi{10.1111/j.1365-2966.2004.08005.x}

\bibitem[{{Astropy Collaboration} {et~al.}(2022){Astropy Collaboration}, {Price-Whelan}, {Lim}, {Earl}, {Starkman}, {Bradley}, {Shupe}, {Patil}, {Corrales}, {Brasseur}, {N{\"o}the}, {Donath}, {Tollerud}, {Morris}, {Ginsburg}, {Vaher}, {Weaver}, {Tocknell}, {Jamieson}, {van Kerkwijk}, {Robitaille}, {Merry}, {Bachetti}, {G{\"u}nther}, {Aldcroft}, {Alvarado-Montes}, {Archibald}, {B{\'o}di}, {Bapat}, {Barentsen}, {Baz{\'a}n}, {Biswas}, {Boquien}, {Burke}, {Cara}, {Cara}, {Conroy}, {Conseil}, {Craig}, {Cross}, {Cruz}, {D'Eugenio}, {Dencheva}, {Devillepoix}, {Dietrich}, {Eigenbrot}, {Erben}, {Ferreira}, {Foreman-Mackey}, {Fox}, {Freij}, {Garg}, {Geda}, {Glattly}, {Gondhalekar}, {Gordon}, {Grant}, {Greenfield}, {Groener}, {Guest}, {Gurovich}, {Handberg}, {Hart}, {Hatfield-Dodds}, {Homeier}, {Hosseinzadeh}, {Jenness}, {Jones}, {Joseph}, {Kalmbach}, {Karamehmetoglu}, {Ka{\l}uszy{\'n}ski}, {Kelley}, {Kern}, {Kerzendorf}, {Koch}, {Kulumani}, {Lee}, {Ly}, {Ma}, {MacBride}, {Maljaars}, {Muna}, {Murphy}, {Norman},
  {O'Steen}, {Oman}, {Pacifici}, {Pascual}, {Pascual-Granado}, {Patil}, {Perren}, {Pickering}, {Rastogi}, {Roulston}, {Ryan}, {Rykoff}, {Sabater}, {Sakurikar}, {Salgado}, {Sanghi}, {Saunders}, {Savchenko}, {Schwardt}, {Seifert-Eckert}, {Shih}, {Jain}, {Shukla}, {Sick}, {Simpson}, {Singanamalla}, {Singer}, {Singhal}, {Sinha}, {Sip{\H{o}}cz}, {Spitler}, {Stansby}, {Streicher}, {{\v{S}}umak}, {Swinbank}, {Taranu}, {Tewary}, {Tremblay}, {de Val-Borro}, {Van Kooten}, {Vasovi{\'c}}, {Verma}, {de Miranda Cardoso}, {Williams}, {Wilson}, {Winkel}, {Wood-Vasey}, {Xue}, {Yoachim}, {Zhang}, {Zonca}, \& {Astropy Project Contributors}}]{astropy_2022}
{Astropy Collaboration}, {Price-Whelan}, A.~M., {Lim}, P.~L., {et~al.} 2022, \apj, 935, 167, \dodoi{10.3847/1538-4357/ac7c74}

\bibitem[{Audet \& Dennis(2006)}]{Audet_2006}
Audet, C., \& Dennis, J.~E. 2006, SIAM Journal on Optimization, 17, 188, \dodoi{10.1137/040603371}

\bibitem[{{Auger} {et~al.}(2009){Auger}, {Treu}, {Bolton}, {Gavazzi}, {Koopmans}, {Marshall}, {Bundy}, \& {Moustakas}}]{Auger_2009}
{Auger}, M.~W., {Treu}, T., {Bolton}, A.~S., {et~al.} 2009, \apj, 705, 1099, \dodoi{10.1088/0004-637X/705/2/1099}

\bibitem[{{Auger} {et~al.}(2010){Auger}, {Treu}, {Bolton}, {Gavazzi}, {Koopmans}, {Marshall}, {Moustakas}, \& {Burles}}]{Auger_2010}
---. 2010, \apj, 724, 511, \dodoi{10.1088/0004-637X/724/1/511}

\bibitem[{{Aumer} {et~al.}(2010){Aumer}, {Burkert}, {Johansson}, \& {Genzel}}]{Aumer_2010}
{Aumer}, M., {Burkert}, A., {Johansson}, P.~H., \& {Genzel}, R. 2010, \apj, 719, 1230, \dodoi{10.1088/0004-637X/719/2/1230}

\bibitem[{{Bar} {et~al.}(2022){Bar}, {Danieli}, \& {Blum}}]{Bar_2022}
{Bar}, N., {Danieli}, S., \& {Blum}, K. 2022, \apjl, 932, L10, \dodoi{10.3847/2041-8213/ac70df}

\bibitem[{{Bender}(1988)}]{Bender_1988_B}
{Bender}, R. 1988, \aap, 193, L7

\bibitem[{{Bender} {et~al.}(1989){Bender}, {Surma}, {Doebereiner}, {Moellenhoff}, \& {Madejsky}}]{Bender_1989}
{Bender}, R., {Surma}, P., {Doebereiner}, S., {Moellenhoff}, C., \& {Madejsky}, R. 1989, \aap, 217, 35

\bibitem[{{Benetti} {et~al.}(2023){Benetti}, {Lapi}, {Gandolfi}, {Salucci}, \& {Danese}}]{Benetti_2023}
{Benetti}, F., {Lapi}, A., {Gandolfi}, G., {Salucci}, P., \& {Danese}, L. 2023, \apj, 949, 65, \dodoi{10.3847/1538-4357/acc8ca}

\bibitem[{{Bidaran} {et~al.}(2022){Bidaran}, {La Barbera}, {Pasquali}, {Peletier}, {van de Ven}, {Grebel}, {Falc{\'o}n-Barroso}, {Sybilska}, {Gadotti}, \& {Coccato}}]{Bidaran_2022}
{Bidaran}, B., {La Barbera}, F., {Pasquali}, A., {et~al.} 2022, \mnras, 515, 4622, \dodoi{10.1093/mnras/stac2005}

\bibitem[{{Binggeli} {et~al.}(1987){Binggeli}, {Tammann}, \& {Sandage}}]{Bineggli_1987}
{Binggeli}, B., {Tammann}, G.~A., \& {Sandage}, A. 1987, \aj, 94, 251, \dodoi{10.1086/114467}

\bibitem[{{Binney} \& {Tremaine}(2008)}]{Binney_2008}
{Binney}, J., \& {Tremaine}, S. 2008, {Galactic Dynamics: Second Edition} ({Princeton University Press})

\bibitem[{{Blumenthal} {et~al.}(1986){Blumenthal}, {Faber}, {Flores}, \& {Primack}}]{Blumenthal_1986}
{Blumenthal}, G.~R., {Faber}, S.~M., {Flores}, R., \& {Primack}, J.~R. 1986, \apj, 301, 27, \dodoi{10.1086/163867}

\bibitem[{{B{\"o}hringer} {et~al.}(1994){B{\"o}hringer}, {Briel}, {Schwarz}, {Voges}, {Hartner}, \& {Tr{\"u}mper}}]{Boehringer_1994}
{B{\"o}hringer}, H., {Briel}, U.~G., {Schwarz}, R.~A., {et~al.} 1994, \nat, 368, 828, \dodoi{10.1038/368828a0}

\bibitem[{{Bolton} {et~al.}(2008){Bolton}, {Burles}, {Koopmans}, {Treu}, {Gavazzi}, {Moustakas}, {Wayth}, \& {Schlegel}}]{Bolton_2008}
{Bolton}, A.~S., {Burles}, S., {Koopmans}, L. V.~E., {et~al.} 2008, \apj, 682, 964, \dodoi{10.1086/589327}

\bibitem[{{Bovy} {et~al.}(2016){Bovy}, {Bahmanyar}, {Fritz}, \& {Kallivayalil}}]{Bovy_2016}
{Bovy}, J., {Bahmanyar}, A., {Fritz}, T.~K., \& {Kallivayalil}, N. 2016, \apj, 833, 31, \dodoi{10.3847/1538-4357/833/1/31}

\bibitem[{{Boylan-Kolchin} {et~al.}(2011){Boylan-Kolchin}, {Bullock}, \& {Kaplinghat}}]{Boylan_Kolchin_2011}
{Boylan-Kolchin}, M., {Bullock}, J.~S., \& {Kaplinghat}, M. 2011, \mnras, 415, L40, \dodoi{10.1111/j.1745-3933.2011.01074.x}

\bibitem[{{Boylan-Kolchin} {et~al.}(2012){Boylan-Kolchin}, {Bullock}, \& {Kaplinghat}}]{Boylan_Kolchin_2012}
---. 2012, \mnras, 422, 1203, \dodoi{10.1111/j.1365-2966.2012.20695.x}

\bibitem[{{Brinckmann} {et~al.}(2018){Brinckmann}, {Zavala}, {Rapetti}, {Hansen}, \& {Vogelsberger}}]{Brinckmann_2018}
{Brinckmann}, T., {Zavala}, J., {Rapetti}, D., {Hansen}, S.~H., \& {Vogelsberger}, M. 2018, \mnras, 474, 746, \dodoi{10.1093/mnras/stx2782}

\bibitem[{{Burkert}(2015)}]{Burkert_2015}
{Burkert}, A. 2015, \apj, 808, 158, \dodoi{10.1088/0004-637X/808/2/158}

\bibitem[{Burnham \& Anderson(2002)}]{burnham_2002}
Burnham, K., \& Anderson, D. 2002, Model selection and multimodel inference: a practical information-theoretic approach (Springer Verlag)

\bibitem[{{Bustamante-Rosell} {et~al.}(2021){Bustamante-Rosell}, {Noyola}, {Gebhardt}, {Fabricius}, {Mazzalay}, {Thomas}, \& {Zeimann}}]{Bustamante-Rosell_2021}
{Bustamante-Rosell}, M.~J., {Noyola}, E., {Gebhardt}, K., {et~al.} 2021, arXiv e-prints, arXiv:2111.04770.
\newblock \doarXiv{2111.04770}

\bibitem[{{Cappellari}(2008)}]{Cappellari_2008}
{Cappellari}, M. 2008, \mnras, 390, 71, \dodoi{10.1111/j.1365-2966.2008.13754.x}

\bibitem[{{Cappellari} \& {Copin}(2003)}]{Cappellari_2003}
{Cappellari}, M., \& {Copin}, Y. 2003, \mnras, 342, 345, \dodoi{10.1046/j.1365-8711.2003.06541.x}

\bibitem[{{Cappellari} {et~al.}(2011){Cappellari}, {Emsellem}, {Krajnovi{\'c}}, {McDermid}, {Scott}, {Verdoes Kleijn}, {Young}, {Alatalo}, {Bacon}, {Blitz}, {Bois}, {Bournaud}, {Bureau}, {Davies}, {Davis}, {de Zeeuw}, {Duc}, {Khochfar}, {Kuntschner}, {Lablanche}, {Morganti}, {Naab}, {Oosterloo}, {Sarzi}, {Serra}, \& {Weijmans}}]{Cappellari_2011}
{Cappellari}, M., {Emsellem}, E., {Krajnovi{\'c}}, D., {et~al.} 2011, \mnras, 413, 813, \dodoi{10.1111/j.1365-2966.2010.18174.x}

\bibitem[{{Cappellari} {et~al.}(2013{\natexlab{a}}){Cappellari}, {Scott}, {Alatalo}, {Blitz}, {Bois}, {Bournaud}, {Bureau}, {Crocker}, {Davies}, {Davis}, {de Zeeuw}, {Duc}, {Emsellem}, {Khochfar}, {Krajnovi{\'c}}, {Kuntschner}, {McDermid}, {Morganti}, {Naab}, {Oosterloo}, {Sarzi}, {Serra}, {Weijmans}, \& {Young}}]{Cappellari_2013}
{Cappellari}, M., {Scott}, N., {Alatalo}, K., {et~al.} 2013{\natexlab{a}}, \mnras, 432, 1709, \dodoi{10.1093/mnras/stt562}

\bibitem[{{Cappellari} {et~al.}(2013{\natexlab{b}}){Cappellari}, {McDermid}, {Alatalo}, {Blitz}, {Bois}, {Bournaud}, {Bureau}, {Crocker}, {Davies}, {Davis}, {de Zeeuw}, {Duc}, {Emsellem}, {Khochfar}, {Krajnovi{\'c}}, {Kuntschner}, {Morganti}, {Naab}, {Oosterloo}, {Sarzi}, {Scott}, {Serra}, {Weijmans}, \& {Young}}]{Cappellari_2013_B}
{Cappellari}, M., {McDermid}, R.~M., {Alatalo}, K., {et~al.} 2013{\natexlab{b}}, \mnras, 432, 1862, \dodoi{10.1093/mnras/stt644}

\bibitem[{{Cataldi} {et~al.}(2023){Cataldi}, {Pedrosa}, {Tissera}, {Artale}, {Padilla}, {Dominguez-Tenreiro}, {Bignone}, {Gonzalez}, \& {Pellizza}}]{Cataldi_2023}
{Cataldi}, P., {Pedrosa}, S.~E., {Tissera}, P.~B., {et~al.} 2023, \mnras, 523, 1919, \dodoi{10.1093/mnras/stad1601}

\bibitem[{{Ceverino} {et~al.}(2010){Ceverino}, {Dekel}, \& {Bournaud}}]{Ceverino_2010}
{Ceverino}, D., {Dekel}, A., \& {Bournaud}, F. 2010, \mnras, 404, 2151, \dodoi{10.1111/j.1365-2966.2010.16433.x}

\bibitem[{Chilingarian(2009)}]{Chilingarian_2009}
Chilingarian, I.~V. 2009, Monthly Notices of the Royal Astronomical Society, 394, 1229, \dodoi{10.1111/j.1365-2966.2009.14450.x}

\bibitem[{{Chisari} {et~al.}(2017){Chisari}, {Koukoufilippas}, {Jindal}, {Peirani}, {Beckmann}, {Codis}, {Devriendt}, {Miller}, {Dubois}, {Laigle}, {Slyz}, \& {Pichon}}]{Chisari_2017}
{Chisari}, N.~E., {Koukoufilippas}, N., {Jindal}, A., {et~al.} 2017, \mnras, 472, 1163, \dodoi{10.1093/mnras/stx1998}

\bibitem[{{Choque-Challapa} {et~al.}(2019){Choque-Challapa}, {Smith}, {Candlish}, {Peletier}, \& {Shin}}]{Choque-Challapa_2019}
{Choque-Challapa}, N., {Smith}, R., {Candlish}, G., {Peletier}, R., \& {Shin}, J. 2019, \mnras, 490, 3654, \dodoi{10.1093/mnras/stz2829}

\bibitem[{Chua {et~al.}(2019)Chua, Pillepich, Vogelsberger, \& Hernquist}]{Chua_2019}
Chua, K. T.~E., Pillepich, A., Vogelsberger, M., \& Hernquist, L. 2019, Monthly Notices of the Royal Astronomical Society, 484, 476, \dodoi{10.1093/mnras/sty3531}

\bibitem[{{Chua} {et~al.}(2022){Chua}, {Vogelsberger}, {Pillepich}, \& {Hernquist}}]{Chua_2022}
{Chua}, K. T.~E., {Vogelsberger}, M., {Pillepich}, A., \& {Hernquist}, L. 2022, \mnras, 515, 2681, \dodoi{10.1093/mnras/stac1897}

\bibitem[{{Codis} {et~al.}(2015){Codis}, {Pichon}, \& {Pogosyan}}]{Codis_2015}
{Codis}, S., {Pichon}, C., \& {Pogosyan}, D. 2015, \mnras, 452, 3369, \dodoi{10.1093/mnras/stv1570}

\bibitem[{{Cole} {et~al.}(2011){Cole}, {Dehnen}, \& {Wilkinson}}]{Cole_2011}
{Cole}, D.~R., {Dehnen}, W., \& {Wilkinson}, M.~I. 2011, \mnras, 416, 1118, \dodoi{10.1111/j.1365-2966.2011.19110.x}

\bibitem[{Comerón {et~al.}(2023)Comerón, Trujillo, Cappellari, Buitrago, Garduño, Zaragoza-Cardiel, Zinchenko, Lara-López, Ferré-Mateu, \& Dib}]{comeron_2023_arxiv}
Comerón, S., Trujillo, I., Cappellari, M., {et~al.} 2023, The massive relic galaxy NGC 1277 is dark matter deficient. From dynamical models of integral-field stellar kinematics out to five effective radii.
\newblock \doarXiv{2303.11360}

\bibitem[{{Conselice} {et~al.}(2001){Conselice}, {Gallagher}, \& {Wyse}}]{Conselice_2001}
{Conselice}, C.~J., {Gallagher}, John~S., I., \& {Wyse}, R. F.~G. 2001, \apj, 559, 791, \dodoi{10.1086/322373}

\bibitem[{{C{\^o}t{\'e}} {et~al.}(2000){C{\^o}t{\'e}}, {Carignan}, \& {Freeman}}]{Cote_2000}
{C{\^o}t{\'e}}, S., {Carignan}, C., \& {Freeman}, K.~C. 2000, \aj, 120, 3027, \dodoi{10.1086/316883}

\bibitem[{{Croft} {et~al.}(2002){Croft}, {Weinberg}, {Bolte}, {Burles}, {Hernquist}, {Katz}, {Kirkman}, \& {Tytler}}]{Croft_2002}
{Croft}, R. A.~C., {Weinberg}, D.~H., {Bolte}, M., {et~al.} 2002, \apj, 581, 20, \dodoi{10.1086/344099}

\bibitem[{{{\c{S}}en} {et~al.}(2018){{\c{S}}en}, {Peletier}, {Boselli}, {den Brok}, {Falc{\'o}n-Barroso}, {Hensler}, {Janz}, {Laurikainen}, {Lisker}, {Mentz}, {Paudel}, {Salo}, {Sybilska}, {Toloba}, {van de Ven}, {Vazdekis}, \& {Yesilyaprak}}]{Sen_2018}
{{\c{S}}en}, {\c{S}}., {Peletier}, R.~F., {Boselli}, A., {et~al.} 2018, \mnras, 475, 3453, \dodoi{10.1093/mnras/stx3254}

\bibitem[{{Danieli} {et~al.}(2019){Danieli}, {van Dokkum}, {Conroy}, {Abraham}, \& {Romanowsky}}]{Danieli_2019}
{Danieli}, S., {van Dokkum}, P., {Conroy}, C., {Abraham}, R., \& {Romanowsky}, A.~J. 2019, \apjl, 874, L12, \dodoi{10.3847/2041-8213/ab0e8c}

\bibitem[{{de Blok}(2010)}]{de_Blok2010}
{de Blok}, W.~J.~G. 2010, Advances in Astronomy, 2010, 789293, \dodoi{10.1155/2010/789293}

\bibitem[{{de Blok} \& {Bosma}(2002)}]{de_Blok_2002}
{de Blok}, W.~J.~G., \& {Bosma}, A. 2002, \aap, 385, 816, \dodoi{10.1051/0004-6361:20020080}

\bibitem[{{de Blok} {et~al.}(2008){de Blok}, {Walter}, {Brinks}, {Trachternach}, {Oh}, \& {Kennicutt}}]{de_Blok_2008}
{de Blok}, W.~J.~G., {Walter}, F., {Brinks}, E., {et~al.} 2008, \aj, 136, 2648, \dodoi{10.1088/0004-6256/136/6/2648}

\bibitem[{{de Graaff} {et~al.}(2024){de Graaff}, {Pillepich}, \& {Rix}}]{dE_Graaff2024}
{de Graaff}, A., {Pillepich}, A., \& {Rix}, H.-W. 2024, arXiv e-prints, arXiv:2403.00907.
\newblock \doarXiv{2403.00907}

\bibitem[{{de Graaff} {et~al.}(2023){de Graaff}, {Rix}, {Carniani}, {Suess}, {Charlot}, {Curtis-Lake}, {Arribas}, {Baker}, {Boyett}, {Bunker}, {Cameron}, {Chevallard}, {Curti}, {Eisenstein}, {Franx}, {Hainline}, {Hausen}, {Ji}, {Johnson}, {Jones}, {Maiolino}, {Maseda}, {Nelson}, {Parlanti}, {Rawle}, {Robertson}, {Tacchella}, {{\"U}bler}, {Williams}, {Willmer}, \& {Willott}}]{dE_Graaff_2023}
{de Graaff}, A., {Rix}, H.-W., {Carniani}, S., {et~al.} 2023, arXiv e-prints, arXiv:2308.09742, \dodoi{10.48550/arXiv.2308.09742}

\bibitem[{{De Leo} {et~al.}(2023){De Leo}, {Read}, {Noel}, {Erkal}, {Massana}, \& {Carrera}}]{De_Leo_2023}
{De Leo}, M., {Read}, J.~I., {Noel}, N. E.~D., {et~al.} 2023, arXiv e-prints, arXiv:2303.08838, \dodoi{10.48550/arXiv.2303.08838}

\bibitem[{{de Nicola} {et~al.}(2022){de Nicola}, {Neureiter}, {Thomas}, {Saglia}, \& {Bender}}]{de_Nicola_2022}
{de Nicola}, S., {Neureiter}, B., {Thomas}, J., {Saglia}, R.~P., \& {Bender}, R. 2022, \mnras, 517, 3445, \dodoi{10.1093/mnras/stac2852}

\bibitem[{{de Souza} {et~al.}(2011){de Souza}, {Rodrigues}, {Ishida}, \& {Opher}}]{de_Souza_2011}
{de Souza}, R.~S., {Rodrigues}, L.~F.~S., {Ishida}, E.~E.~O., \& {Opher}, R. 2011, \mnras, 415, 2969, \dodoi{10.1111/j.1365-2966.2011.18916.x}

\bibitem[{{Debattista} {et~al.}(2008){Debattista}, {Moore}, {Quinn}, {Kazantzidis}, {Maas}, {Mayer}, {Read}, \& {Stadel}}]{Debattista_2008}
{Debattista}, V.~P., {Moore}, B., {Quinn}, T., {et~al.} 2008, \apj, 681, 1076, \dodoi{10.1086/587977}

\bibitem[{{Dekel} \& {Birnboim}(2006)}]{Dekel_2006}
{Dekel}, A., \& {Birnboim}, Y. 2006, \mnras, 368, 2, \dodoi{10.1111/j.1365-2966.2006.10145.x}

\bibitem[{{Dekel} \& {Silk}(1986)}]{Dekel_1986}
{Dekel}, A., \& {Silk}, J. 1986, \apj, 303, 39, \dodoi{10.1086/164050}

\bibitem[{{Del Popolo}(2009)}]{Del_Popolo_2009}
{Del Popolo}, A. 2009, \apj, 698, 2093, \dodoi{10.1088/0004-637X/698/2/2093}

\bibitem[{{Del Popolo}(2012)}]{Del_Popolo_2012}
---. 2012, \mnras, 419, 971, \dodoi{10.1111/j.1365-2966.2011.19754.x}

\bibitem[{{Del Popolo} \& {Le Delliou}(2021)}]{Del_Popolo_2021}
{Del Popolo}, A., \& {Le Delliou}, M. 2021, Galaxies, 9, 123, \dodoi{10.3390/galaxies9040123}

\bibitem[{{Derkenne} {et~al.}(2023){Derkenne}, {McDermid}, {Poci}, {Mendel}, {D'Eugenio}, {Jeon}, {Remus}, {Bellstedt}, {Battisti}, {Bland-Hawthorn}, {Ferr{\'e}-Mateu}, {Foster}, {Harborne}, {Lagos}, {Peng}, {Sharda}, {Sharma}, {Sweet}, {Tran}, {Valenzuela}, {Vaughan}, {Wisnioski}, \& {Yi}}]{Derkenne_2023}
{Derkenne}, C., {McDermid}, R.~M., {Poci}, A., {et~al.} 2023, \mnras, 522, 3602, \dodoi{10.1093/mnras/stad1079}

\bibitem[{{Diemand} \& {Moore}(2011)}]{Diemand_2011}
{Diemand}, J., \& {Moore}, B. 2011, Advanced Science Letters, 4, 297, \dodoi{10.1166/asl.2011.1211}

\bibitem[{{Diemand} {et~al.}(2004){Diemand}, {Moore}, \& {Stadel}}]{Diemand_2004}
{Diemand}, J., {Moore}, B., \& {Stadel}, J. 2004, \mnras, 353, 624, \dodoi{10.1111/j.1365-2966.2004.08094.x}

\bibitem[{{Donato} {et~al.}(2009){Donato}, {Gentile}, {Salucci}, {Frigerio Martins}, {Wilkinson}, {Gilmore}, {Grebel}, {Koch}, \& {Wyse}}]{Donato_2009}
{Donato}, F., {Gentile}, G., {Salucci}, P., {et~al.} 2009, \mnras, 397, 1169, \dodoi{10.1111/j.1365-2966.2009.15004.x}

\bibitem[{{Dressler}(1980)}]{Dressler_1980}
{Dressler}, A. 1980, \apj, 236, 351, \dodoi{10.1086/157753}

\bibitem[{{Dutta Chowdhury} {et~al.}(2023){Dutta Chowdhury}, {van den Bosch}, {van Dokkum}, {Robles}, {Schive}, \& {Chiueh}}]{Chowdhury_2023}
{Dutta Chowdhury}, D., {van den Bosch}, F.~C., {van Dokkum}, P., {et~al.} 2023, \apj, 949, 68, \dodoi{10.3847/1538-4357/acc73d}

\bibitem[{{Dutton} \& {Treu}(2014)}]{Dutton_2014}
{Dutton}, A.~A., \& {Treu}, T. 2014, \mnras, 438, 3594, \dodoi{10.1093/mnras/stt2489}

\bibitem[{{Eftekhari} {et~al.}(2022){Eftekhari}, {Peletier}, {Scott}, {Mieske}, {Bland-Hawthorn}, {Bryant}, {Cantiello}, {Croom}, {Drinkwater}, {Falc{\'o}n-Barroso}, {Hilker}, {Iodice}, {Napolitano}, {Spavone}, {Valentijn}, {van de Ven}, \& {Venhola}}]{Eftekhari_2022}
{Eftekhari}, F.~S., {Peletier}, R.~F., {Scott}, N., {et~al.} 2022, \mnras, 517, 4714, \dodoi{10.1093/mnras/stac2606}

\bibitem[{{El-Zant} {et~al.}(2001){El-Zant}, {Shlosman}, \& {Hoffman}}]{El-Zant_2001}
{El-Zant}, A., {Shlosman}, I., \& {Hoffman}, Y. 2001, \apj, 560, 636, \dodoi{10.1086/322516}

\bibitem[{{Elbert} {et~al.}(2015){Elbert}, {Bullock}, {Garrison-Kimmel}, {Rocha}, {O{\~n}orbe}, \& {Peter}}]{Elbert_2015}
{Elbert}, O.~D., {Bullock}, J.~S., {Garrison-Kimmel}, S., {et~al.} 2015, \mnras, 453, 29, \dodoi{10.1093/mnras/stv1470}

\bibitem[{{Fabricius} {et~al.}(2008){Fabricius}, {Barnes}, {Bender}, {Drory}, {Grupp}, {Hill}, {Hopp}, \& {MacQueen}}]{Fabricius_2008}
{Fabricius}, M.~H., {Barnes}, S., {Bender}, R., {et~al.} 2008, in Society of Photo-Optical Instrumentation Engineers (SPIE) Conference Series, Vol. 7014, Ground-based and Airborne Instrumentation for Astronomy II, ed. I.~S. {McLean} \& M.~M. {Casali}, 701473, \dodoi{10.1117/12.787204}

\bibitem[{{Fabricius} {et~al.}(2012){Fabricius}, {Grupp}, {Bender}, {Drory}, {Arns}, {Barnes}, {G{\"o}ssl}, {Snigula}, {Hill}, {Hopp}, {Lang-Bardl}, {MacQueen}, {Saglia}, \& {Wullstein}}]{Fabricius_2012}
{Fabricius}, M.~H., {Grupp}, F., {Bender}, R., {et~al.} 2012, in Society of Photo-Optical Instrumentation Engineers (SPIE) Conference Series, Vol. 8446, Ground-based and Airborne Instrumentation for Astronomy IV, ed. I.~S. {McLean}, S.~K. {Ramsay}, \& H.~{Takami}, 84465K, \dodoi{10.1117/12.925177}

\bibitem[{{Fall}(1983)}]{Fall_1983}
{Fall}, S.~M. 1983, in Internal Kinematics and Dynamics of Galaxies, ed. E.~{Athanassoula}, Vol. 100, 391--398

\bibitem[{{Fall} \& {Efstathiou}(1980)}]{Fall_1980}
{Fall}, S.~M., \& {Efstathiou}, G. 1980, \mnras, 193, 189, \dodoi{10.1093/mnras/193.2.189}

\bibitem[{{Ferrarese} \& {Merritt}(2000)}]{Ferrarese_2000}
{Ferrarese}, L., \& {Merritt}, D. 2000, \apjl, 539, L9, \dodoi{10.1086/312838}

\bibitem[{{Ferrarese} {et~al.}(2006){Ferrarese}, {C{\^o}t{\'e}}, {Jord{\'a}n}, {Peng}, {Blakeslee}, {Piatek}, {Mei}, {Merritt}, {Milosavljevi{\'c}}, {Tonry}, \& {West}}]{Ferrarese_2006}
{Ferrarese}, L., {C{\^o}t{\'e}}, P., {Jord{\'a}n}, A., {et~al.} 2006, \apjs, 164, 334, \dodoi{10.1086/501350}

\bibitem[{{Frenk} {et~al.}(1988){Frenk}, {White}, {Davis}, \& {Efstathiou}}]{Frenk_1988}
{Frenk}, C.~S., {White}, S. D.~M., {Davis}, M., \& {Efstathiou}, G. 1988, \apj, 327, 507, \dodoi{10.1086/166213}

\bibitem[{{Fukushige} \& {Makino}(2001)}]{Fukushige_2001}
{Fukushige}, T., \& {Makino}, J. 2001, \apj, 557, 533, \dodoi{10.1086/321666}

\bibitem[{{Gao} {et~al.}(2008){Gao}, {Navarro}, {Cole}, {Frenk}, {White}, {Springel}, {Jenkins}, \& {Neto}}]{Gao_2008}
{Gao}, L., {Navarro}, J.~F., {Cole}, S., {et~al.} 2008, \mnras, 387, 536, \dodoi{10.1111/j.1365-2966.2008.13277.x}

\bibitem[{{Gao} {et~al.}(2005){Gao}, {Springel}, \& {White}}]{Gao_2005}
{Gao}, L., {Springel}, V., \& {White}, S. D.~M. 2005, \mnras, 363, L66, \dodoi{10.1111/j.1745-3933.2005.00084.x}

\bibitem[{{Gebhardt} {et~al.}(2000){Gebhardt}, {Bender}, {Bower}, {Dressler}, {Faber}, {Filippenko}, {Green}, {Grillmair}, {Ho}, {Kormendy}, {Lauer}, {Magorrian}, {Pinkney}, {Richstone}, \& {Tremaine}}]{Gebhardt_2000}
{Gebhardt}, K., {Bender}, R., {Bower}, G., {et~al.} 2000, \apjl, 539, L13, \dodoi{10.1086/312840}

\bibitem[{{Geha} {et~al.}(2003){Geha}, {Guhathakurta}, \& {van der Marel}}]{Geha_2003}
{Geha}, M., {Guhathakurta}, P., \& {van der Marel}, R.~P. 2003, \aj, 126, 1794, \dodoi{10.1086/377624}

\bibitem[{{Genina} {et~al.}(2018){Genina}, {Ben{\'\i}tez-Llambay}, {Frenk}, {Cole}, {Fattahi}, {Navarro}, {Oman}, {Sawala}, \& {Theuns}}]{Genina_2018}
{Genina}, A., {Ben{\'\i}tez-Llambay}, A., {Frenk}, C.~S., {et~al.} 2018, \mnras, 474, 1398, \dodoi{10.1093/mnras/stx2855}

\bibitem[{{Gentile} {et~al.}(2004){Gentile}, {Salucci}, {Klein}, {Vergani}, \& {Kalberla}}]{Gentile_2004}
{Gentile}, G., {Salucci}, P., {Klein}, U., {Vergani}, D., \& {Kalberla}, P. 2004, \mnras, 351, 903, \dodoi{10.1111/j.1365-2966.2004.07836.x}

\bibitem[{Gerhard {et~al.}(1998)Gerhard, Jeske, Saglia, \& Bender}]{Gerhard_1998}
Gerhard, O., Jeske, G., Saglia, R.~P., \& Bender, R. 1998, Monthly Notices of the Royal Astronomical Society, 295, 197, \dodoi{https://doi.org/10.1046/j.1365-8711.1998.29511341.x}

\bibitem[{{Gerhard} {et~al.}(2001){Gerhard}, {Kronawitter}, {Saglia}, \& {Bender}}]{Gerhard_2001}
{Gerhard}, O., {Kronawitter}, A., {Saglia}, R.~P., \& {Bender}, R. 2001, \aj, 121, 1936, \dodoi{10.1086/319940}

\bibitem[{{Gnedin} {et~al.}(2004){Gnedin}, {Kravtsov}, {Klypin}, \& {Nagai}}]{Gnedin_2004}
{Gnedin}, O.~Y., {Kravtsov}, A.~V., {Klypin}, A.~A., \& {Nagai}, D. 2004, \apj, 616, 16, \dodoi{10.1086/424914}

\bibitem[{{Gnedin} \& {Zhao}(2002)}]{Gnedin_2002}
{Gnedin}, O.~Y., \& {Zhao}, H. 2002, \mnras, 333, 299, \dodoi{10.1046/j.1365-8711.2002.05361.x}

\bibitem[{{Governato} {et~al.}(2010){Governato}, {Brook}, {Mayer}, {Brooks}, {Rhee}, {Wadsley}, {Jonsson}, {Willman}, {Stinson}, {Quinn}, \& {Madau}}]{Governato_2010}
{Governato}, F., {Brook}, C., {Mayer}, L., {et~al.} 2010, \nat, 463, 203, \dodoi{10.1038/nature08640}

\bibitem[{{Governato} {et~al.}(2012){Governato}, {Zolotov}, {Pontzen}, {Christensen}, {Oh}, {Brooks}, {Quinn}, {Shen}, \& {Wadsley}}]{Governato_2012}
{Governato}, F., {Zolotov}, A., {Pontzen}, A., {et~al.} 2012, \mnras, 422, 1231, \dodoi{10.1111/j.1365-2966.2012.20696.x}

\bibitem[{{Gunn} \& {Gott}(1972)}]{Gunn_1972}
{Gunn}, J.~E., \& {Gott}, J.~Richard, I. 1972, \apj, 176, 1, \dodoi{10.1086/151605}

\bibitem[{{Harko}(2011)}]{Harko_2011}
{Harko}, T. 2011, \jcap, 2011, 022, \dodoi{10.1088/1475-7516/2011/05/022}

\bibitem[{{Hayashi} {et~al.}(2007){Hayashi}, {Navarro}, \& {Springel}}]{Hayashi_2007}
{Hayashi}, E., {Navarro}, J.~F., \& {Springel}, V. 2007, \mnras, 377, 50, \dodoi{10.1111/j.1365-2966.2007.11599.x}

\bibitem[{{Hayashi} {et~al.}(2020){Hayashi}, {Chiba}, \& {Ishiyama}}]{Hayashi_2020}
{Hayashi}, K., {Chiba}, M., \& {Ishiyama}, T. 2020, \apj, 904, 45, \dodoi{10.3847/1538-4357/abbe0a}

\bibitem[{{Hu}(2008)}]{Hu_2008}
{Hu}, J. 2008, \mnras, 386, 2242, \dodoi{10.1111/j.1365-2966.2008.13195.x}

\bibitem[{{Ibata} {et~al.}(2001){Ibata}, {Lewis}, {Irwin}, {Totten}, \& {Quinn}}]{Ibata_2001}
{Ibata}, R., {Lewis}, G.~F., {Irwin}, M., {Totten}, E., \& {Quinn}, T. 2001, \apj, 551, 294, \dodoi{10.1086/320060}

\bibitem[{{Immeli} {et~al.}(2004){Immeli}, {Samland}, {Westera}, \& {Gerhard}}]{Immeli_2004}
{Immeli}, A., {Samland}, M., {Westera}, P., \& {Gerhard}, O. 2004, \apj, 611, 20, \dodoi{10.1086/422179}

\bibitem[{{Inoue} \& {Saitoh}(2011)}]{Inoue_2011}
{Inoue}, S., \& {Saitoh}, T.~R. 2011, \mnras, 418, 2527, \dodoi{10.1111/j.1365-2966.2011.19873.x}

\bibitem[{{Janz} {et~al.}(2017){Janz}, {Penny}, {Graham}, {Forbes}, \& {Davies}}]{Janz_2017}
{Janz}, J., {Penny}, S.~J., {Graham}, A.~W., {Forbes}, D.~A., \& {Davies}, R.~L. 2017, \mnras, 468, 2850, \dodoi{10.1093/mnras/stx634}

\bibitem[{{Jardel} \& {Gebhardt}(2012)}]{Jardel_2012}
{Jardel}, J.~R., \& {Gebhardt}, K. 2012, \apj, 746, 89, \dodoi{10.1088/0004-637X/746/1/89}

\bibitem[{{Jardel} \& {Gebhardt}(2013)}]{Jardel_2013_b}
---. 2013, \apjl, 775, L30, \dodoi{10.1088/2041-8205/775/1/L30}

\bibitem[{{Jardel} {et~al.}(2013){Jardel}, {Gebhardt}, {Fabricius}, {Drory}, \& {Williams}}]{Jardel_2013_a}
{Jardel}, J.~R., {Gebhardt}, K., {Fabricius}, M.~H., {Drory}, N., \& {Williams}, M.~J. 2013, \apj, 763, 91, \dodoi{10.1088/0004-637X/763/2/91}

\bibitem[{{Jorgensen} {et~al.}(1995){Jorgensen}, {Franx}, \& {Kjaergaard}}]{Jorgensen_1995}
{Jorgensen}, I., {Franx}, M., \& {Kjaergaard}, P. 1995, \mnras, 273, 1097, \dodoi{10.1093/mnras/273.4.1097}

\bibitem[{{Katz} \& {Gunn}(1991)}]{Katz_1991}
{Katz}, N., \& {Gunn}, J.~E. 1991, \apj, 377, 365, \dodoi{10.1086/170367}

\bibitem[{{Kim} \& {Lee}(2013)}]{Kim_2013}
{Kim}, J.-h., \& {Lee}, J. 2013, \mnras, 432, 1701, \dodoi{10.1093/mnras/stt632}

\bibitem[{{Klypin} {et~al.}(2001){Klypin}, {Kravtsov}, {Bullock}, \& {Primack}}]{Klypin_2001}
{Klypin}, A., {Kravtsov}, A.~V., {Bullock}, J.~S., \& {Primack}, J.~R. 2001, \apj, 554, 903, \dodoi{10.1086/321400}

\bibitem[{{Klypin} {et~al.}(1999){Klypin}, {Kravtsov}, {Valenzuela}, \& {Prada}}]{Klypin_1999}
{Klypin}, A., {Kravtsov}, A.~V., {Valenzuela}, O., \& {Prada}, F. 1999, \apj, 522, 82, \dodoi{10.1086/307643}

\bibitem[{{Klypin} {et~al.}(2011){Klypin}, {Trujillo-Gomez}, \& {Primack}}]{Klypin_2011}
{Klypin}, A.~A., {Trujillo-Gomez}, S., \& {Primack}, J. 2011, \apj, 740, 102, \dodoi{10.1088/0004-637X/740/2/102}

\bibitem[{{Koopmans} {et~al.}(2009){Koopmans}, {Bolton}, {Treu}, {Czoske}, {Auger}, {Barnab{\`e}}, {Vegetti}, {Gavazzi}, {Moustakas}, \& {Burles}}]{Koopmans_2009}
{Koopmans}, L.~V.~E., {Bolton}, A., {Treu}, T., {et~al.} 2009, \apjl, 703, L51, \dodoi{10.1088/0004-637X/703/1/L51}

\bibitem[{{Kormendy}(1999)}]{Kormendy_1999}
{Kormendy}, J. 1999, in Astronomical Society of the Pacific Conference Series, Vol. 182, Galaxy Dynamics - A Rutgers Symposium, ed. D.~R. {Merritt}, M.~{Valluri}, \& J.~A. {Sellwood}, 124

\bibitem[{{Kormendy} \& {Bender}(1996)}]{Kormendy_1996}
{Kormendy}, J., \& {Bender}, R. 1996, \apjl, 464, L119, \dodoi{10.1086/310095}

\bibitem[{{Kormendy} \& {Bender}(2009)}]{Kormendy_2009_B}
---. 2009, \apjl, 691, L142, \dodoi{10.1088/0004-637X/691/2/L142}

\bibitem[{{Kormendy} \& {Bender}(2013)}]{Kormendy_2013}
---. 2013, \apjl, 769, L5, \dodoi{10.1088/2041-8205/769/1/L5}

\bibitem[{{Kormendy} {et~al.}(2009){Kormendy}, {Fisher}, {Cornell}, \& {Bender}}]{Kormendy_2009}
{Kormendy}, J., {Fisher}, D.~B., {Cornell}, M.~E., \& {Bender}, R. 2009, \apjs, 182, 216, \dodoi{10.1088/0067-0049/182/1/216}

\bibitem[{{Kormendy} \& {Freeman}(2016)}]{Kormendy_2016}
{Kormendy}, J., \& {Freeman}, K.~C. 2016, \apj, 817, 84, \dodoi{10.3847/0004-637X/817/2/84}

\bibitem[{{Koudmani} {et~al.}(2022){Koudmani}, {Sijacki}, \& {Smith}}]{Koudmani_2022}
{Koudmani}, S., {Sijacki}, D., \& {Smith}, M.~C. 2022, \mnras, 516, 2112, \dodoi{10.1093/mnras/stac2252}

\bibitem[{{La Barbera} {et~al.}(2010){La Barbera}, {de Carvalho}, {de La Rosa}, {Lopes}, {Kohl-Moreira}, \& {Capelato}}]{LaBarbera_2010}
{La Barbera}, F., {de Carvalho}, R.~R., {de La Rosa}, I.~G., {et~al.} 2010, \mnras, 408, 1313, \dodoi{10.1111/j.1365-2966.2010.16850.x}

\bibitem[{{Laporte} \& {Penarrubia}(2015)}]{Laporte_2015}
{Laporte}, C.~F.~P., \& {Penarrubia}, J. 2015, \mnras, 449, L90, \dodoi{10.1093/mnrasl/slv008}

\bibitem[{{Larson} {et~al.}(1980){Larson}, {Tinsley}, \& {Caldwell}}]{Larson_1980}
{Larson}, R.~B., {Tinsley}, B.~M., \& {Caldwell}, C.~N. 1980, \apj, 237, 692, \dodoi{10.1086/157917}

\bibitem[{{Law} \& {Majewski}(2010)}]{Law_2010}
{Law}, D.~R., \& {Majewski}, S.~R. 2010, \apj, 714, 229, \dodoi{10.1088/0004-637X/714/1/229}

\bibitem[{Le~Digabel(2011)}]{Digabel_2022}
Le~Digabel, S. 2011, ACM Trans. Math. Softw., 37, \dodoi{10.1145/1916461.1916468}

\bibitem[{{Lelli} {et~al.}(2016){Lelli}, {McGaugh}, \& {Schombert}}]{Lelli_2016}
{Lelli}, F., {McGaugh}, S.~S., \& {Schombert}, J.~M. 2016, \aj, 152, 157, \dodoi{10.3847/0004-6256/152/6/157}

\bibitem[{{Li} {et~al.}(2022){Li}, {McGaugh}, {Lelli}, {Schombert}, \& {Pawlowski}}]{Li_2022}
{Li}, P., {McGaugh}, S.~S., {Lelli}, F., {Schombert}, J.~M., \& {Pawlowski}, M.~S. 2022, \aap, 665, A143, \dodoi{10.1051/0004-6361/202243916}

\bibitem[{{Li} {et~al.}(2019){Li}, {Li}, {Shao}, {Lu}, {Zhu}, {Wang}, {Gao}, {Mao}, {Dutton}, {Ge}, {Wang}, {Leauthaud}, {Zheng}, {Bundy}, \& {Brownstein}}]{Li_2019}
{Li}, R., {Li}, H., {Shao}, S., {et~al.} 2019, \mnras, 490, 2124, \dodoi{10.1093/mnras/stz2565}

\bibitem[{{Lin} \& {Faber}(1983)}]{Lin_1983}
{Lin}, D.~N.~C., \& {Faber}, S.~M. 1983, \apjl, 266, L21, \dodoi{10.1086/183971}

\bibitem[{{Lipka} \& {Thomas}(2021)}]{Lipka_2021}
{Lipka}, M., \& {Thomas}, J. 2021, \mnras, 504, 4599, \dodoi{10.1093/mnras/stab1092}

\bibitem[{{L{\'o}pez} {et~al.}(2019){L{\'o}pez}, {Merch{\'a}n}, \& {Paz}}]{Lopez_2019}
{L{\'o}pez}, P., {Merch{\'a}n}, M.~E., \& {Paz}, D.~J. 2019, \mnras, 485, 5244, \dodoi{10.1093/mnras/stz762}

\bibitem[{{Lovell} {et~al.}(2018){Lovell}, {Pillepich}, {Genel}, {Nelson}, {Springel}, {Pakmor}, {Marinacci}, {Weinberger}, {Torrey}, {Vogelsberger}, {Alabi}, \& {Hernquist}}]{Lovell_2018}
{Lovell}, M.~R., {Pillepich}, A., {Genel}, S., {et~al.} 2018, \mnras, 481, 1950, \dodoi{10.1093/mnras/sty2339}

\bibitem[{{Madau} {et~al.}(2014){Madau}, {Shen}, \& {Governato}}]{Madau_2014}
{Madau}, P., {Shen}, S., \& {Governato}, F. 2014, \apjl, 789, L17, \dodoi{10.1088/2041-8205/789/1/L17}

\bibitem[{{Mannerkoski} {et~al.}(2021){Mannerkoski}, {Johansson}, {Rantala}, {Naab}, \& {Liao}}]{Mannerkoski_2021}
{Mannerkoski}, M., {Johansson}, P.~H., {Rantala}, A., {Naab}, T., \& {Liao}, S. 2021, \apjl, 912, L20, \dodoi{10.3847/2041-8213/abf9a5}

\bibitem[{{Marsh} \& {Silk}(2014)}]{Marsh_2014}
{Marsh}, D. J.~E., \& {Silk}, J. 2014, \mnras, 437, 2652, \dodoi{10.1093/mnras/stt2079}

\bibitem[{{Mashchenko} {et~al.}(2006){Mashchenko}, {Couchman}, \& {Wadsley}}]{Mashchenko_2006}
{Mashchenko}, S., {Couchman}, H.~M.~P., \& {Wadsley}, J. 2006, \nat, 442, 539, \dodoi{10.1038/nature04944}

\bibitem[{{McGaugh} \& {de Blok}(1998)}]{McGaugh_1998}
{McGaugh}, S.~S., \& {de Blok}, W.~J.~G. 1998, \apj, 499, 41, \dodoi{10.1086/305612}

\bibitem[{{Mehlert} {et~al.}(2000){Mehlert}, {Saglia}, {Bender}, \& {Wegner}}]{Mehlert_2000}
{Mehlert}, D., {Saglia}, R.~P., {Bender}, R., \& {Wegner}, G. 2000, \aaps, 141, 449, \dodoi{10.1051/aas:2000322}

\bibitem[{{Mehrgan} {et~al.}(2019){Mehrgan}, {Thomas}, {Saglia}, {Mazzalay}, {Erwin}, {Bender}, {Kluge}, \& {Fabricius}}]{Mehrgan_2019}
{Mehrgan}, K., {Thomas}, J., {Saglia}, R., {et~al.} 2019, \apj, 887, 195, \dodoi{10.3847/1538-4357/ab5856}

\bibitem[{{Mehrgan} {et~al.}(2024){Mehrgan}, {Thomas}, {Saglia}, {Parikh}, {Neureiter}, {Erwin}, \& {Bender}}]{Mehrgan_2023_b}
---. 2024, \apj, 961, 127, \dodoi{10.3847/1538-4357/acfe09}

\bibitem[{{Merrifield} \& {Kent}(1990)}]{Merrifield_1990}
{Merrifield}, M.~R., \& {Kent}, S.~M. 1990, \aj, 99, 1548, \dodoi{10.1086/115438}

\bibitem[{{Mezcua} {et~al.}(2019){Mezcua}, {Suh}, \& {Civano}}]{Mezcua_2019}
{Mezcua}, M., {Suh}, H., \& {Civano}, F. 2019, \mnras, 488, 685, \dodoi{10.1093/mnras/stz1760}

\bibitem[{{Moore}(1994)}]{Moore_1994}
{Moore}, B. 1994, \nat, 370, 629, \dodoi{10.1038/370629a0}

\bibitem[{{Moore} {et~al.}(2001){Moore}, {Calc{\'a}neo-Rold{\'a}n}, {Stadel}, {Quinn}, {Lake}, {Ghigna}, \& {Governato}}]{Moore_2001}
{Moore}, B., {Calc{\'a}neo-Rold{\'a}n}, C., {Stadel}, J., {et~al.} 2001, \prd, 64, 063508, \dodoi{10.1103/PhysRevD.64.063508}

\bibitem[{{Moore} {et~al.}(1998){Moore}, {Governato}, {Quinn}, {Stadel}, \& {Lake}}]{Moore_1998}
{Moore}, B., {Governato}, F., {Quinn}, T., {Stadel}, J., \& {Lake}, G. 1998, \apjl, 499, L5, \dodoi{10.1086/311333}

\bibitem[{{Mukherjee} {et~al.}(2022){Mukherjee}, {Koopmans}, {Tortora}, {Schaller}, {Metcalf}, {Schaye}, \& {Vernardos}}]{Mukherjee_2022}
{Mukherjee}, S., {Koopmans}, L. V.~E., {Tortora}, C., {et~al.} 2022, \mnras, 509, 1245, \dodoi{10.1093/mnras/stab3014}

\bibitem[{{Navarro} {et~al.}(1996{\natexlab{a}}){Navarro}, {Eke}, \& {Frenk}}]{Navarro_1996}
{Navarro}, J.~F., {Eke}, V.~R., \& {Frenk}, C.~S. 1996{\natexlab{a}}, \mnras, 283, L72, \dodoi{10.1093/mnras/283.3.L72}

\bibitem[{{Navarro} {et~al.}(1996{\natexlab{b}}){Navarro}, {Frenk}, \& {White}}]{Navarro_1996_B}
{Navarro}, J.~F., {Frenk}, C.~S., \& {White}, S. D.~M. 1996{\natexlab{b}}, \apj, 462, 563, \dodoi{10.1086/177173}

\bibitem[{{Navarro} {et~al.}(1997){Navarro}, {Frenk}, \& {White}}]{Navarro_1997}
---. 1997, \apj, 490, 493, \dodoi{10.1086/304888}

\bibitem[{{Navarro} {et~al.}(2010){Navarro}, {Ludlow}, {Springel}, {Wang}, {Vogelsberger}, {White}, {Jenkins}, {Frenk}, \& {Helmi}}]{Navarro_2010}
{Navarro}, J.~F., {Ludlow}, A., {Springel}, V., {et~al.} 2010, \mnras, 402, 21, \dodoi{10.1111/j.1365-2966.2009.15878.x}

\bibitem[{{Nelson} {et~al.}(2018){Nelson}, {Pillepich}, {Springel}, {Weinberger}, {Hernquist}, {Pakmor}, {Genel}, {Torrey}, {Vogelsberger}, {Kauffmann}, {Marinacci}, \& {Naiman}}]{Nelson_2018}
{Nelson}, D., {Pillepich}, A., {Springel}, V., {et~al.} 2018, \mnras, 475, 624, \dodoi{10.1093/mnras/stx3040}

\bibitem[{{Nipoti} \& {Binney}(2015)}]{Nipoti_2015}
{Nipoti}, C., \& {Binney}, J. 2015, \mnras, 446, 1820, \dodoi{10.1093/mnras/stu2217}

\bibitem[{{Oh} {et~al.}(2011{\natexlab{a}}){Oh}, {Brook}, {Governato}, {Brinks}, {Mayer}, {de Blok}, {Brooks}, \& {Walter}}]{Oh_2011_b}
{Oh}, S.-H., {Brook}, C., {Governato}, F., {et~al.} 2011{\natexlab{a}}, \aj, 142, 24, \dodoi{10.1088/0004-6256/142/1/24}

\bibitem[{{Oh} {et~al.}(2011{\natexlab{b}}){Oh}, {de Blok}, {Brinks}, {Walter}, \& {Kennicutt}}]{Oh_2011_a}
{Oh}, S.-H., {de Blok}, W.~J.~G., {Brinks}, E., {Walter}, F., \& {Kennicutt}, Robert~C., J. 2011{\natexlab{b}}, \aj, 141, 193, \dodoi{10.1088/0004-6256/141/6/193}

\bibitem[{{Oh} {et~al.}(2015){Oh}, {Hunter}, {Brinks}, {Elmegreen}, {Schruba}, {Walter}, {Rupen}, {Young}, {Simpson}, {Johnson}, {Herrmann}, {Ficut-Vicas}, {Cigan}, {Heesen}, {Ashley}, \& {Zhang}}]{Oh_2015}
{Oh}, S.-H., {Hunter}, D.~A., {Brinks}, E., {et~al.} 2015, \aj, 149, 180, \dodoi{10.1088/0004-6256/149/6/180}

\bibitem[{{Orkney} {et~al.}(2023){Orkney}, {Taylor}, {Read}, {Rey}, {Pontzen}, {Agertz}, {Kim}, \& {Delorme}}]{Orkney_2023}
{Orkney}, M. D.~A., {Taylor}, E., {Read}, J.~I., {et~al.} 2023, arXiv e-prints, arXiv:2302.12818, \dodoi{10.48550/arXiv.2302.12818}

\bibitem[{{Orkney} {et~al.}(2021){Orkney}, {Read}, {Rey}, {Nasim}, {Pontzen}, {Agertz}, {Kim}, {Delorme}, \& {Dehnen}}]{Orkeny_2021}
{Orkney}, M. D.~A., {Read}, J.~I., {Rey}, M.~P., {et~al.} 2021, \mnras, 504, 3509, \dodoi{10.1093/mnras/stab1066}

\bibitem[{Pacucci {et~al.}(2023)Pacucci, Ni, \& Loeb}]{pacucci_2023_arxiv}
Pacucci, F., Ni, Y., \& Loeb, A. 2023, Extreme Tidal Stripping May Explain the Overmassive Black Hole in Leo I: a Proof of Concept.
\newblock \doarXiv{2309.02487}

\bibitem[{{Partmann} {et~al.}(2023){Partmann}, {Naab}, {Rantala}, {Genina}, {Mannerkoski}, \& {Johansson}}]{Partmann_2023}
{Partmann}, C., {Naab}, T., {Rantala}, A., {et~al.} 2023, arXiv e-prints, arXiv:2310.08079, \dodoi{10.48550/arXiv.2310.08079}

\bibitem[{{Paturel} {et~al.}(2003){Paturel}, {Petit}, {Prugniel}, {Theureau}, {Rousseau}, {Brouty}, {Dubois}, \& {Cambr{\'e}sy}}]{Paturel_2003}
{Paturel}, G., {Petit}, C., {Prugniel}, P., {et~al.} 2003, \aap, 412, 45, \dodoi{10.1051/0004-6361:20031411}

\bibitem[{{Paudel} {et~al.}(2023){Paudel}, {Yoon}, {Yoo}, {Smith}, {Chhatkuli}, {Kumar Bachchan}, {Aryal}, {Adhikari}, {Adhikari}, {Sedain}, {Sheikh}, {Dhital}, {Giri}, \& {Baral}}]{Paudel_2023}
{Paudel}, S., {Yoon}, S.-J., {Yoo}, J., {et~al.} 2023, \apjs, 265, 57, \dodoi{10.3847/1538-4365/acbfa7}

\bibitem[{{Peebles}(1969)}]{Peebles_1969}
{Peebles}, P.~J.~E. 1969, \apj, 155, 393, \dodoi{10.1086/149876}

\bibitem[{{Peter} {et~al.}(2013){Peter}, {Rocha}, {Bullock}, \& {Kaplinghat}}]{Peter_2013}
{Peter}, A. H.~G., {Rocha}, M., {Bullock}, J.~S., \& {Kaplinghat}, M. 2013, \mnras, 430, 105, \dodoi{10.1093/mnras/sts535}

\bibitem[{{Peters} {et~al.}(2017){Peters}, {van der Kruit}, {Allen}, \& {Freeman}}]{Peters_2017}
{Peters}, S.~P.~C., {van der Kruit}, P.~C., {Allen}, R.~J., \& {Freeman}, K.~C. 2017, \mnras, 464, 65, \dodoi{10.1093/mnras/stw2101}

\bibitem[{{Plana} {et~al.}(2010){Plana}, {Amram}, {Mendes de Oliveira}, \& {Balkowski}}]{Plana_2010}
{Plana}, H., {Amram}, P., {Mendes de Oliveira}, C., \& {Balkowski}, C. 2010, \aj, 139, 1, \dodoi{10.1088/0004-6256/139/1/1}

\bibitem[{{Poci} {et~al.}(2017){Poci}, {Cappellari}, \& {McDermid}}]{Poci_2017}
{Poci}, A., {Cappellari}, M., \& {McDermid}, R.~M. 2017, \mnras, 467, 1397, \dodoi{10.1093/mnras/stx101}

\bibitem[{{Pontzen} \& {Governato}(2012)}]{Pontzen_2012}
{Pontzen}, A., \& {Governato}, F. 2012, \mnras, 421, 3464, \dodoi{10.1111/j.1365-2966.2012.20571.x}

\bibitem[{{Pontzen} \& {Governato}(2014)}]{Pontzen_2014}
---. 2014, \nat, 506, 171, \dodoi{10.1038/nature12953}

\bibitem[{{Porciani} {et~al.}(2002){Porciani}, {Dekel}, \& {Hoffman}}]{Porciani_2002}
{Porciani}, C., {Dekel}, A., \& {Hoffman}, Y. 2002, \mnras, 332, 325, \dodoi{10.1046/j.1365-8711.2002.05305.x}

\bibitem[{Posacki {et~al.}(2014)Posacki, Cappellari, Treu, Pellegrini, \& Ciotti}]{Posacki_2014}
Posacki, S., Cappellari, M., Treu, T., Pellegrini, S., \& Ciotti, L. 2014, Monthly Notices of the Royal Astronomical Society, 446, 493, \dodoi{10.1093/mnras/stu2098}

\bibitem[{{Posti} \& {Helmi}(2019)}]{Posti_2019}
{Posti}, L., \& {Helmi}, A. 2019, \aap, 621, A56, \dodoi{10.1051/0004-6361/201833355}

\bibitem[{{Pulsoni} {et~al.}(2021){Pulsoni}, {Gerhard}, {Arnaboldi}, {Pillepich}, {Rodriguez-Gomez}, {Nelson}, {Hernquist}, \& {Springel}}]{Pulsoni_2021}
{Pulsoni}, C., {Gerhard}, O., {Arnaboldi}, M., {et~al.} 2021, \aap, 647, A95, \dodoi{10.1051/0004-6361/202039166}

\bibitem[{{Quilis} {et~al.}(2000){Quilis}, {Moore}, \& {Bower}}]{Quilis_2000}
{Quilis}, V., {Moore}, B., \& {Bower}, R. 2000, Science, 288, 1617, \dodoi{10.1126/science.288.5471.1617}

\bibitem[{{Rantala} {et~al.}(2018){Rantala}, {Johansson}, {Naab}, {Thomas}, \& {Frigo}}]{Rantala_2018}
{Rantala}, A., {Johansson}, P.~H., {Naab}, T., {Thomas}, J., \& {Frigo}, M. 2018, \apj, 864, 113, \dodoi{10.3847/1538-4357/aada47}

\bibitem[{{Rantala} {et~al.}(2017){Rantala}, {Pihajoki}, {Johansson}, {Naab}, {Lah{\'e}n}, \& {Sawala}}]{Rantala_2017}
{Rantala}, A., {Pihajoki}, P., {Johansson}, P.~H., {et~al.} 2017, \apj, 840, 53, \dodoi{10.3847/1538-4357/aa6d65}

\bibitem[{{Read} {et~al.}(2016){Read}, {Agertz}, \& {Collins}}]{Read_2016}
{Read}, J.~I., {Agertz}, O., \& {Collins}, M.~L.~M. 2016, \mnras, 459, 2573, \dodoi{10.1093/mnras/stw713}

\bibitem[{{Read} \& {Gilmore}(2005)}]{Read_2005}
{Read}, J.~I., \& {Gilmore}, G. 2005, \mnras, 356, 107, \dodoi{10.1111/j.1365-2966.2004.08424.x}

\bibitem[{{Remus} {et~al.}(2013){Remus}, {Burkert}, {Dolag}, {Johansson}, {Naab}, {Oser}, \& {Thomas}}]{Remus_2013}
{Remus}, R.-S., {Burkert}, A., {Dolag}, K., {et~al.} 2013, \apj, 766, 71, \dodoi{10.1088/0004-637X/766/2/71}

\bibitem[{{Remus} {et~al.}(2017){Remus}, {Dolag}, {Naab}, {Burkert}, {Hirschmann}, {Hoffmann}, \& {Johansson}}]{Remus_2017}
{Remus}, R.-S., {Dolag}, K., {Naab}, T., {et~al.} 2017, \mnras, 464, 3742, \dodoi{10.1093/mnras/stw2594}

\bibitem[{{Robles} \& {Matos}(2012)}]{Robles_2012}
{Robles}, V.~H., \& {Matos}, T. 2012, \mnras, 422, 282, \dodoi{10.1111/j.1365-2966.2012.20603.x}

\bibitem[{{Romanowsky} \& {Fall}(2012)}]{Romanowsky_2012}
{Romanowsky}, A.~J., \& {Fall}, S.~M. 2012, \apjs, 203, 17, \dodoi{10.1088/0067-0049/203/2/17}

\bibitem[{{Romero-G{\'o}mez} {et~al.}(2024){Romero-G{\'o}mez}, {J.}, {Peletier}, {Aguerri}, \& {Smith}}]{Romero_Gomez_2024}
{Romero-G{\'o}mez}, {J.}, {Peletier}, R.~F., {Aguerri}, J.~A.~L., \& {Smith}, R. 2024, arXiv e-prints, arXiv:2404.15519, \dodoi{10.48550/arXiv.2404.15519}

\bibitem[{Romero-Gómez {et~al.}(2023{\natexlab{a}})Romero-Gómez, Aguerri, Peletier, Mieske, van~de Ven, \& Falcón-Barroso}]{Romero_Gomez_2023_B}
Romero-Gómez, J., Aguerri, J. A.~L., Peletier, R.~F., {et~al.} 2023{\natexlab{a}}, Monthly Notices of the Royal Astronomical Society, 527, 9715, \dodoi{10.1093/mnras/stad3801}

\bibitem[{Romero-Gómez {et~al.}(2023{\natexlab{b}})Romero-Gómez, Peletier, Aguerri, Mieske, Scott, Bland-Hawthorn, Bryant, Croom, Eftekhari, Falcón-Barroso, Hilker, van~de Ven, \& Venhola}]{Romero_Gomez_2023_A}
Romero-Gómez, J., Peletier, R.~F., Aguerri, J. A.~L., {et~al.} 2023{\natexlab{b}}, Monthly Notices of the Royal Astronomical Society, 522, 130, \dodoi{10.1093/mnras/stad953}

\bibitem[{Rubin \& Ford(1970)}]{Rubin_1970}
Rubin, V.~C., \& Ford, W.~K. 1970, The Astrophysical Journal, 159, 379.
\newblock \url{https://api.semanticscholar.org/CorpusID:122756867}

\bibitem[{{Ry{\'s}} {et~al.}(2013){Ry{\'s}}, {Falc{\'o}n-Barroso}, \& {van de Ven}}]{Rys_2013}
{Ry{\'s}}, A., {Falc{\'o}n-Barroso}, J., \& {van de Ven}, G. 2013, \mnras, 428, 2980, \dodoi{10.1093/mnras/sts245}

\bibitem[{{Sandage} \& {Binggeli}(1984)}]{Sandage_1984}
{Sandage}, A., \& {Binggeli}, B. 1984, \aj, 89, 919, \dodoi{10.1086/113588}

\bibitem[{{Schwarzschild}(1979)}]{Schwarzschild_1979}
{Schwarzschild}, M. 1979, \apj, 232, 236, \dodoi{10.1086/157282}

\bibitem[{{Scott} {et~al.}(2020){Scott}, {Eftekhari}, {Peletier}, {Bryant}, {Bland-Hawthorn}, {Capaccioli}, {Croom}, {Drinkwater}, {Falc{\'o}n-Barroso}, {Hilker}, {Iodice}, {Lorente}, {Mieske}, {Spavone}, {van de Ven}, \& {Venhola}}]{Scott_2020}
{Scott}, N., {Eftekhari}, F.~S., {Peletier}, R.~F., {et~al.} 2020, \mnras, 497, 1571, \dodoi{10.1093/mnras/staa2042}

\bibitem[{{Seo} \& {Ann}(2023)}]{Seo_2023}
{Seo}, M., \& {Ann}, H.~B. 2023, \mnras, 520, 5521, \dodoi{10.1093/mnras/stad425}

\bibitem[{{Sharma} {et~al.}(2023{\natexlab{a}}){Sharma}, {Freundlich}, {van de Ven}, {Famaey}, {Salucci}, {Martorano}, \& {Renaud}}]{Sharma_2023_B}
{Sharma}, G., {Freundlich}, J., {van de Ven}, G., {et~al.} 2023{\natexlab{a}}, arXiv e-prints, arXiv:2309.04541, \dodoi{10.48550/arXiv.2309.04541}

\bibitem[{{Sharma} {et~al.}(2023{\natexlab{b}}){Sharma}, {Brooks}, {Tremmel}, {Bellovary}, \& {Quinn}}]{sharma_2023}
{Sharma}, R.~S., {Brooks}, A.~M., {Tremmel}, M., {Bellovary}, J., \& {Quinn}, T.~R. 2023{\natexlab{b}}, \apj, 957, 16, \dodoi{10.3847/1538-4357/ace046}

\bibitem[{{Sharma} {et~al.}(2022){Sharma}, {Brooks}, {Tremmel}, {Bellovary}, {Ricarte}, \& {Quinn}}]{Sharma_2022}
{Sharma}, R.~S., {Brooks}, A.~M., {Tremmel}, M., {et~al.} 2022, arXiv e-prints, arXiv:2203.05580.
\newblock \doarXiv{2203.05580}

\bibitem[{{Shen} {et~al.}(2021){Shen}, {Danieli}, {van Dokkum}, {Abraham}, {Brodie}, {Conroy}, {Dolphin}, {Romanowsky}, {Kruijssen}, \& {Dutta Chowdhury}}]{Shen_2021}
{Shen}, Z., {Danieli}, S., {van Dokkum}, P., {et~al.} 2021, \apjl, 914, L12, \dodoi{10.3847/2041-8213/ac0335}

\bibitem[{{Silk}(2017)}]{Silk_2017}
{Silk}, J. 2017, \apjl, 839, L13, \dodoi{10.3847/2041-8213/aa67da}

\bibitem[{{Skillman} \& {Bender}(1995)}]{Skillman_1995}
{Skillman}, E.~D., \& {Bender}, R. 1995, in Revista Mexicana de Astronomia y Astrofisica Conference Series, Vol.~3, Revista Mexicana de Astronomia y Astrofisica Conference Series, ed. M.~{Pena} \& S.~{Kurtz}, 25

\bibitem[{{Smith} {et~al.}(2012){Smith}, {Fellhauer}, \& {Assmann}}]{Smith_2012_B}
{Smith}, R., {Fellhauer}, M., \& {Assmann}, P. 2012, \mnras, 420, 1990, \dodoi{10.1111/j.1365-2966.2011.20077.x}

\bibitem[{{Somerville} {et~al.}(2018){Somerville}, {Behroozi}, {Pandya}, {Dekel}, {Faber}, {Fontana}, {Koekemoer}, {Koo}, {P{\'e}rez-Gonz{\'a}lez}, {Primack}, {Santini}, {Taylor}, \& {van der Wel}}]{Somerville_2018}
{Somerville}, R.~S., {Behroozi}, P., {Pandya}, V., {et~al.} 2018, \mnras, 473, 2714, \dodoi{10.1093/mnras/stx2040}

\bibitem[{{Spergel} \& {Steinhardt}(2000)}]{Spergel_2000}
{Spergel}, D.~N., \& {Steinhardt}, P.~J. 2000, \prl, 84, 3760, \dodoi{10.1103/PhysRevLett.84.3760}

\bibitem[{{Spergel} {et~al.}(2003){Spergel}, {Verde}, {Peiris}, {Komatsu}, {Nolta}, {Bennett}, {Halpern}, {Hinshaw}, {Jarosik}, {Kogut}, {Limon}, {Meyer}, {Page}, {Tucker}, {Weiland}, {Wollack}, \& {Wright}}]{Spergel_2003}
{Spergel}, D.~N., {Verde}, L., {Peiris}, H.~V., {et~al.} 2003, \apjs, 148, 175, \dodoi{10.1086/377226}

\bibitem[{{Springel} {et~al.}(2006){Springel}, {Frenk}, \& {White}}]{Springel_2006}
{Springel}, V., {Frenk}, C.~S., \& {White}, S. D.~M. 2006, \nat, 440, 1137, \dodoi{10.1038/nature04805}

\bibitem[{{Springel} {et~al.}(2005){Springel}, {White}, {Jenkins}, {Frenk}, {Yoshida}, {Gao}, {Navarro}, {Thacker}, {Croton}, {Helly}, {Peacock}, {Cole}, {Thomas}, {Couchman}, {Evrard}, {Colberg}, \& {Pearce}}]{Springel_2005}
{Springel}, V., {White}, S. D.~M., {Jenkins}, A., {et~al.} 2005, \nat, 435, 629, \dodoi{10.1038/nature03597}

\bibitem[{{Stadel} {et~al.}(2009){Stadel}, {Potter}, {Moore}, {Diemand}, {Madau}, {Zemp}, {Kuhlen}, \& {Quilis}}]{Stadel_2009}
{Stadel}, J., {Potter}, D., {Moore}, B., {et~al.} 2009, \mnras, 398, L21, \dodoi{10.1111/j.1745-3933.2009.00699.x}

\bibitem[{{Steinhauser} {et~al.}(2016){Steinhauser}, {Schindler}, \& {Springel}}]{Steinhauser_2016}
{Steinhauser}, D., {Schindler}, S., \& {Springel}, V. 2016, \aap, 591, A51, \dodoi{10.1051/0004-6361/201527705}

\bibitem[{{Sybilska} {et~al.}(2017){Sybilska}, {Lisker}, {Kuntschner}, {Vazdekis}, {van de Ven}, {Peletier}, {Falc{\'o}n-Barroso}, {Vijayaraghavan}, \& {Janz}}]{Sybilska_2017}
{Sybilska}, A., {Lisker}, T., {Kuntschner}, H., {et~al.} 2017, \mnras, 470, 815, \dodoi{10.1093/mnras/stx1138}

\bibitem[{{Tenneti} {et~al.}(2014){Tenneti}, {Mandelbaum}, {Di Matteo}, {Feng}, \& {Khandai}}]{Tenneti_2014}
{Tenneti}, A., {Mandelbaum}, R., {Di Matteo}, T., {Feng}, Y., \& {Khandai}, N. 2014, \mnras, 441, 470, \dodoi{10.1093/mnras/stu586}

\bibitem[{{Thomas} {et~al.}(2007{\natexlab{a}}){Thomas}, {Jesseit}, {Naab}, {Saglia}, {Burkert}, \& {Bender}}]{Thomas_2007_b}
{Thomas}, J., {Jesseit}, R., {Naab}, T., {et~al.} 2007{\natexlab{a}}, \mnras, 381, 1672, \dodoi{10.1111/j.1365-2966.2007.12343.x}

\bibitem[{{Thomas} \& {Lipka}(2022)}]{Thomas_2022}
{Thomas}, J., \& {Lipka}, M. 2022, \mnras, 514, 6203, \dodoi{10.1093/mnras/stac1581}

\bibitem[{{Thomas} {et~al.}(2014){Thomas}, {Saglia}, {Bender}, {Erwin}, \& {Fabricius}}]{Thomas_2014}
{Thomas}, J., {Saglia}, R.~P., {Bender}, R., {Erwin}, P., \& {Fabricius}, M. 2014, \apj, 782, 39, \dodoi{10.1088/0004-637X/782/1/39}

\bibitem[{{Thomas} {et~al.}(2005){Thomas}, {Saglia}, {Bender}, {Thomas}, {Gebhardt}, {Magorrian}, {Corsini}, \& {Wegner}}]{Thomas_2005}
{Thomas}, J., {Saglia}, R.~P., {Bender}, R., {et~al.} 2005, \mnras, 360, 1355, \dodoi{10.1111/j.1365-2966.2005.09139.x}

\bibitem[{{Thomas} {et~al.}(2007{\natexlab{b}}){Thomas}, {Saglia}, {Bender}, {Thomas}, {Gebhardt}, {Magorrian}, {Corsini}, \& {Wegner}}]{Thomas_2007}
---. 2007{\natexlab{b}}, \mnras, 382, 657, \dodoi{10.1111/j.1365-2966.2007.12434.x}

\bibitem[{{Thomas} {et~al.}(2009){Thomas}, {Saglia}, {Bender}, {Thomas}, {Gebhardt}, {Magorrian}, {Corsini}, \& {Wegner}}]{Thomas_2009}
---. 2009, \apj, 691, 770, \dodoi{10.1088/0004-637X/691/1/770}

\bibitem[{{Thomas} {et~al.}(2004){Thomas}, {Saglia}, {Bender}, {Thomas}, {Gebhardt}, {Magorrian}, \& {Richstone}}]{Thomas_2004}
---. 2004, \mnras, 353, 391, \dodoi{10.1111/j.1365-2966.2004.08072.x}

\bibitem[{{Toloba} {et~al.}(2014){Toloba}, {Guhathakurta}, {Peletier}, {Boselli}, {Lisker}, {Falc{\'o}n-Barroso}, {Simon}, {van de Ven}, {Paudel}, {Emsellem}, {Janz}, {den Brok}, {Gorgas}, {Hensler}, {Laurikainen}, {Niemi}, {Ry{\'s}}, \& {Salo}}]{Toloba_2014}
{Toloba}, E., {Guhathakurta}, P., {Peletier}, R.~F., {et~al.} 2014, \apjs, 215, 17, \dodoi{10.1088/0067-0049/215/2/17}

\bibitem[{{Toloba} {et~al.}(2015){Toloba}, {Guhathakurta}, {Boselli}, {Peletier}, {Emsellem}, {Lisker}, {van de Ven}, {Simon}, {Falc{\'o}n-Barroso}, {Adams}, {Benson}, {Boissier}, {den Brok}, {Gorgas}, {Hensler}, {Janz}, {Laurikainen}, {Paudel}, {Ry{\'s}}, \& {Salo}}]{Toloba_2015}
{Toloba}, E., {Guhathakurta}, P., {Boselli}, A., {et~al.} 2015, \apj, 799, 172, \dodoi{10.1088/0004-637X/799/2/172}

\bibitem[{{Tortora} {et~al.}(2016){Tortora}, {La Barbera}, \& {Napolitano}}]{Tortora_2016}
{Tortora}, C., {La Barbera}, F., \& {Napolitano}, N.~R. 2016, \mnras, 455, 308, \dodoi{10.1093/mnras/stv2250}

\bibitem[{{Tortora} {et~al.}(2012){Tortora}, {La Barbera}, {Napolitano}, {de Carvalho}, \& {Romanowsky}}]{Tortora_2012}
{Tortora}, C., {La Barbera}, F., {Napolitano}, N.~R., {de Carvalho}, R.~R., \& {Romanowsky}, A.~J. 2012, \mnras, 425, 577, \dodoi{10.1111/j.1365-2966.2012.21506.x}

\bibitem[{{Tortora} {et~al.}(2010){Tortora}, {Napolitano}, {Cardone}, {Capaccioli}, {Jetzer}, \& {Molinaro}}]{Tortora_2010}
{Tortora}, C., {Napolitano}, N.~R., {Cardone}, V.~F., {et~al.} 2010, \mnras, 407, 144, \dodoi{10.1111/j.1365-2966.2010.16938.x}

\bibitem[{{Tortora} {et~al.}(2014){Tortora}, {Napolitano}, {Saglia}, {Romanowsky}, {Covone}, \& {Capaccioli}}]{Tortora_2014}
{Tortora}, C., {Napolitano}, N.~R., {Saglia}, R.~P., {et~al.} 2014, \mnras, 445, 162, \dodoi{10.1093/mnras/stu1712}

\bibitem[{{Tortora} {et~al.}(2019){Tortora}, {Posti}, {Koopmans}, \& {Napolitano}}]{Tortora_2019}
{Tortora}, C., {Posti}, L., {Koopmans}, L.~V.~E., \& {Napolitano}, N.~R. 2019, \mnras, 489, 5483, \dodoi{10.1093/mnras/stz2320}

\bibitem[{{van der Marel} \& {Franx}(1993)}]{Marel_1993}
{van der Marel}, R.~P., \& {Franx}, M. 1993, \apj, 407, 525, \dodoi{10.1086/172534}

\bibitem[{{van Dokkum} {et~al.}(2016){van Dokkum}, {Abraham}, {Brodie}, {Conroy}, {Danieli}, {Merritt}, {Mowla}, {Romanowsky}, \& {Zhang}}]{van_Dokkum_2016}
{van Dokkum}, P., {Abraham}, R., {Brodie}, J., {et~al.} 2016, \apjl, 828, L6, \dodoi{10.3847/2041-8205/828/1/L6}

\bibitem[{{van Dokkum} {et~al.}(2018){van Dokkum}, {Danieli}, {Cohen}, {Merritt}, {Romanowsky}, {Abraham}, {Brodie}, {Conroy}, {Lokhorst}, {Mowla}, {O'Sullivan}, \& {Zhang}}]{van_Dokkum_2018}
{van Dokkum}, P., {Danieli}, S., {Cohen}, Y., {et~al.} 2018, \nat, 555, 629, \dodoi{10.1038/nature25767}

\bibitem[{{van Zee} {et~al.}(2004){van Zee}, {Skillman}, \& {Haynes}}]{van_Zee_2004}
{van Zee}, L., {Skillman}, E.~D., \& {Haynes}, M.~P. 2004, \aj, 128, 121, \dodoi{10.1086/421368}

\bibitem[{{Vera-Ciro} \& {Helmi}(2013)}]{Vera_ciro_2013}
{Vera-Ciro}, C., \& {Helmi}, A. 2013, \apjl, 773, L4, \dodoi{10.1088/2041-8205/773/1/L4}

\bibitem[{{Vogelsberger} {et~al.}(2016){Vogelsberger}, {Zavala}, {Cyr-Racine}, {Pfrommer}, {Bringmann}, \& {Sigurdson}}]{Vogelsberger_2016}
{Vogelsberger}, M., {Zavala}, J., {Cyr-Racine}, F.-Y., {et~al.} 2016, \mnras, 460, 1399, \dodoi{10.1093/mnras/stw1076}

\bibitem[{{Walker} \& {Pe{\~n}arrubia}(2011)}]{Walker_2011}
{Walker}, M.~G., \& {Pe{\~n}arrubia}, J. 2011, \apj, 742, 20, \dodoi{10.1088/0004-637X/742/1/20}

\bibitem[{{Wechsler} {et~al.}(2002){Wechsler}, {Bullock}, {Primack}, {Kravtsov}, \& {Dekel}}]{Wechsler_2002}
{Wechsler}, R.~H., {Bullock}, J.~S., {Primack}, J.~R., {Kravtsov}, A.~V., \& {Dekel}, A. 2002, \apj, 568, 52, \dodoi{10.1086/338765}

\bibitem[{{Wegg} {et~al.}(2019){Wegg}, {Gerhard}, \& {Bieth}}]{Wegg_2019}
{Wegg}, C., {Gerhard}, O., \& {Bieth}, M. 2019, \mnras, 485, 3296, \dodoi{10.1093/mnras/stz572}

\bibitem[{Weller {et~al.}(2023)Weller, Pacucci, Natarajan, \& Matteo}]{weller2023_arxiv}
Weller, E.~J., Pacucci, F., Natarajan, P., \& Matteo, T.~D. 2023, Over-massive Central Black Holes in the Cosmological Simulations ASTRID and Illustris TNG50.
\newblock \doarXiv{2305.02335}

\bibitem[{{Williams} {et~al.}(2004){Williams}, {Babul}, \& {Dalcanton}}]{Williams_2004}
{Williams}, L. L.~R., {Babul}, A., \& {Dalcanton}, J.~J. 2004, \apj, 604, 18, \dodoi{10.1086/381722}

\bibitem[{{Zhao}(1996)}]{Zhao_1996}
{Zhao}, H. 1996, \mnras, 278, 488, \dodoi{10.1093/mnras/278.2.488}

\bibitem[{{Z{\"o}ller} {et~al.}(2023){Z{\"o}ller}, {Kluge}, {Staiger}, \& {Bender}}]{Zoeller_2023}
{Z{\"o}ller}, R., {Kluge}, M., {Staiger}, B., \& {Bender}, R. 2023, arXiv e-prints, arXiv:2310.09330, \dodoi{10.48550/arXiv.2310.09330}

\end{thebibliography}
\bibliographystyle{aasjournal}

\end{document}